\documentclass[twocolumn,trackchanges,]{aastex631}
\hypersetup{linkcolor=blue,colorlinks=true,citecolor=cyan,filecolor=black}

\received{03.10.2023}
\revised{13.12.2023}
\accepted{28.12.2023}

\submitjournal{ApJ}

\shorttitle{Using Empirical Thermal Emission Spectra as an Input for Atmospheric Retrievals of an Earth-Twin Exoplanet}
\shortauthors{Mettler et al.}

\usepackage{graphicx}
\usepackage{placeins}
\usepackage{amsmath}
\usepackage{url}
\usepackage{hyperref} 
\usepackage[T1]{fontenc}
\usepackage{txfonts}
\usepackage{textgreek}
\usepackage{tabularx}
\usepackage{booktabs}
\usepackage{multirow}
\usepackage{makecell}
\usepackage[version=4]{mhchem}
\usepackage{enumitem}

\usepackage{natbib}

\begin{document}
\newcommand{\pt}[0]{\textit{P}$-$\textit{T}}

\newcommand{\R}[0]{\ensuremath{R}}
\newcommand{\Rv}[1]{\ensuremath{R = #1}}
\newcommand{\SN}[0]{\ensuremath{S/N}}
\newcommand{\SNv}[1]{\ensuremath{S/N = #1}}

\newcommand{\mic}[1]{\ensuremath{#1}~\textmu m}

\newcommand{\life}[0]{LIFE}
\newcommand{\lifesim}[0]{LIFE\textsc{sim}}

\newcommand{\Rpl}[0]{\ensuremath{R_{\text{pl}}}}
\newcommand{\Rearth}[0]{\ensuremath{R_\oplus}}   
\newcommand{\Mpl}[0]{\ensuremath{M_{\text{pl}}}}
\newcommand{\Teq}[0]{\ensuremath{T_\mathrm{eq}}}
\newcommand{\Teff}[0]{\ensuremath{T_\mathrm{eff}}}
\newcommand{\Ab}[0]{\ensuremath{A_\mathrm{B}}}
\newcommand{\Ps}[0]{\ensuremath{P_0}}
\newcommand{\Ts}[0]{\ensuremath{T_0}}
\newcommand{\Tct}[0]{\ensuremath{T_\mathrm{cloud-top}}}
\newcommand{\aWA}[0]{\ensuremath{\alpha_{\ce{H2O}}}}
\newcommand{\bWA}[0]{\ensuremath{\beta_{\ce{H2O}}}}

\newcommand{\modelparam}[0]{\ensuremath{\boldsymbol{\checkmark}}}
\newcommand{\notparam}[0]{\ensuremath{\boldsymbol{\times}}}

\newcommand{\lgrt}[1]{\ensuremath{\log_{10}(#1)}}
\newcommand{\lgrtdaj}[1]{\ensuremath{\log_{10}\left(#1\right)}}

\title{Earth as an Exoplanet. III. Using Empirical Thermal Emission Spectra as an Input for Atmospheric Retrieval of an Earth-twin Exoplanet}

\correspondingauthor{Jean-Noël Mettler, Bj\"orn S. Konrad}
\email{jmettler@phys.ethz.ch, konradb@student.ethz.ch}

\author[0000-0002-8653-0226]{Jean-No\"el Mettler}
\altaffiliation{Both authors contributed equally to this publication.}
\affiliation{ETH Zurich, Institute for Particle Physics and Astrophysics, Wolfgang-Pauli-Strasse 27, CH-8093 Zurich, Switzerland}
\affiliation{Center for Theoretical Astrophysics $\&$ Cosmology, Institute for Computational Science, University of Zurich, Switzerland}

\author[0000-0002-9912-8340]{Bj\"orn S. Konrad}
\altaffiliation{Both authors contributed equally to this publication.}
\affiliation{ETH Zurich, Institute for Particle Physics and Astrophysics, Wolfgang-Pauli-Strasse 27, CH-8093 Zurich, Switzerland}

\author[0000-0003-3829-7412]{Sascha P. Quanz}
\affiliation{ETH Zurich, Institute for Particle Physics and Astrophysics, Wolfgang-Pauli-Strasse 27, CH-8093 Zurich, Switzerland}
\affiliation{ETH Zurich, Department of Earth Sciences, Sonneggstrasse 5, 8092 Zurich, Switzerland}

\author[0000-0001-5555-2652]{Ravit Helled}
\affiliation{Center for Theoretical Astrophysics $\&$ Cosmology, Institute for Computational Science, University of Zurich, Switzerland}

\begin{abstract}
   In this study, we treat Earth as an exoplanet and investigate our home planet by means of a potential future mid-infrared (MIR) space mission called the Large Interferometer For Exoplanets (\life{}). We combine thermal spectra from an empirical dataset of disk-integrated Earth observations with a noise model for \life{} to create mock observations. We apply a state-of-the-art atmospheric retrieval framework to characterize the planet, assess the potential for detecting the known bioindicators, and investigate the impact of viewing geometry and seasonality on the characterization. Our key findings reveal that we are observing a temperate habitable planet with significant abundances of \ce{CO2}, \ce{H2O}, \ce{O3}, and \ce{CH4}. Seasonal variations in the surface and equilibrium temperature, as well as in the Bond albedo, are detectable. Furthermore, the viewing geometry and the spatially and temporally unresolved nature of our observations only have a minor impact on the characterization. Additionally, Earth's variable abundance profiles and patchy cloud coverage can bias retrieval results for the atmospheric structure and trace gas abundances. Lastly, the limited extent of Earth's seasonal variations in biosignature abundances makes the direct detection of its biosphere through atmospheric seasonality unlikely. Our results suggest that \life{} could correctly identify Earth as a planet where life could thrive, with detectable levels of bioindicators, a temperate climate, and surface conditions allowing liquid surface water. Even if atmospheric seasonality is not easily observed, our study demonstrates that next generation space missions can assess whether nearby temperate terrestrial exoplanets are habitable or even inhabited. 

\end{abstract}


\section{Introduction}

    The atmospheric characterization of terrestrial exoplanets in the habitable zone \citep[HZ;][]{HZ_Kasting93,HZ_kopparapu13} and the search for life are key endeavors in exoplanet science \citep[e.g., Astrobiology Strategy and Astro 2020 Decadal Survey in the United States:][]{Astrobiology_Strategy_2017, Decadal_Study_2021}. Constraining the composition, structure, and dynamics of exoplanet atmospheres yields valuable insights into planetary habitability and could lead to the detection of life beyond our solar system.

    Terrestrial HZ exoplanets are detectable with current observatories \citep[see, e.g.,][for a catalogue]{Hill_2023}. Exoplanet transit surveys such as the Kepler mission \citep{Borucki2010} and the Transiting Exoplanet Survey Satellite \citep[TESS;][]{TESS_2015} as well as current long-term radial velocity (RV) surveys have revealed that HZ planets with Earth-like radii and masses are abundant in the galaxy \citep[e.g.,][]{Bryson2021}. Such exoplanets have already been detected within 20~pc of the sun with both the transit \citep[e.g.,][]{Bthompson2015,Gillon2017,Vanderspek2019} and the RV \citep[e.g.,][]{Anglada2016,Ribas2016,Zechmeister2019} methods.
    Ongoing observations with the James Webb Space Telescope (JWST) are revealing whether terrestrial HZ exoplanets transiting nearby M dwarfs have significant atmospheres \citep[e.g.,][]{Koll2019,Greene2023,Zieba2023,Lustig2023,Ih2023,Lincowski2023,Madhusudhan2023,Lim2023}. However, performing a detailed atmospheric characterization for such planets with JWST is challenging \citep[e.g.,][]{morley2017,Krissansen-Totton2018_Biosigs_JWST}. Observations with the future 40~m ground-based extremely large telescopes (ELTs) will reach unprecedented spatial resolution and sensitivity. The ELTs will directly detect HZ exoplanets around the nearest stars via their thermal emission \citep[e.g.,][]{quanz2015,Bowens2021} and the reflected stellar light \citep[e.g.,][]{Kasper2021}. However, none of the current or approved future ground- or space-based instruments is capable of performing an in-depth atmosphere characterization for a statistically meaningful sample (dozens) of such exoplanets.
    
    Therefore, the exoplanet community is working toward more capable observatories. LUVOIR \citep{LUVOIR_2019} and HabEx \citep{Gaudi2020} were designed to directly detect the stellar light reflected by terrestrial exoplanets at ultraviolet, optical, and near-infrared (UV/O/NIR) wavelengths. Following the evaluation of both concepts in the Astro 2020 Decadal Survey in the United States \citep{Decadal_Study_2021}, the space-based UV/O/NIR Habitable Worlds Observatory (HWO) was recommended. However, also the mid-infrared (MIR) thermal emission of exoplanets (and its time variability) contains a wealth of unique information about the planetary atmosphere and surface conditions \citep[e.g.,][]{DesMarais2002, Hearty2009, Catling2018, Schwieterman2018, mettler_2020,mettler_2023}. The Large Interferometer For Exoplanets (\life{}), a space-based MIR nulling interferometer concept, aims to directly measure the MIR spectrum of terrestrial HZ exoplanets \citep{K&QLIFE, Quanz:exoplanets_and_atmospheric_characterization, LIFE_I}.

    One key challenge in exoplanet characterization is the correct interpretation of their spectra. Measured exoplanet spectra are global averages (due to the large exoplanet-observer separation). Hence, local variations in the atmospheric composition, pressure-temperature (\pt{}) structure, and clouds are unresolved. Further, since signals from terrestrial exoplanets are faint, the temporal and spectral resolution of observations is limited. Such temporally and spatially unresolved observations can lead to degeneracies, making it  hard to interpret the observations. Finally, the inference of planetary characteristics from such spectra is model-dependent \citep[e.g.,][]{Paradise2021, mettler_2023}. Without thorough exploration and validation of our characterization methods, it will not be possible to accurately infer the wide range of climate states expected for habitable planets. Currently, in-situ data, which are necessary for the validation of our methods, can only be acquired for solar system objects. While the spectral libraries and the knowledge about the formation, composition, and atmospheric properties of solar system planets and their moons is continuously growing, Earth remains the most extensively studied planet and the sole known globally habitable planet harboring life. Therefore, Earth and its unique characteristics remain the key reference point to study the factors required for habitability and (the origin of) life \citep[e.g.,][]{Meadows2018_habitability, Robinson2018}. 

\subsection{Disk-Integrated Earth Spectra Characteristics}
\label{SubSec: Earth's Characteristics}

    From space, Earth’s appearance is dominated by oceans, deserts, vegetation, ice, and clouds. Earth's surface is dominated by oceans ($\approx70\%$ of surface), and the land-to-ocean ratio differs between the hemispheres \citep[Northern Hemisphere $\approx2/3$, Southern Hemisphere $\approx1/4$;][]{pidwirny_2006}. The contribution of different surface types and climate zones to a disk-integrated Earth spectrum (and its seasonal variability) depends on their thermal properties, their fractional contributions, and positions on the observed hemisphere.

    In general, in the thermal emission spectrum of Earth, land-dominated views show not only higher flux readings but also larger flux variations over one full orbit than ocean-dominated views \citep[e.g.,][]{Hearty2009, Gomez2012, mettler_2023}. Specifically, from Table~2 in \citet{mettler_2023}, we see that at Earth’s peak emission wavelength ($\approx$~\mic{10.2}) the disk-integrated Northern Hemisphere pole-on view (NP) and the Africa-centered equatorial view (EqA) show annual flux variations of $33\%$ and $22\%$, respectively. In contrast, the ocean dominated Southern Hemisphere pole-on view (SP) and the Pacific-centered equatorial view (EqP), show smaller annual variations ($\approx11\%$) due to the large thermal inertia of oceans.

    Another distinctive characteristic of Earth is its patchy cloud cover (see also Appendix~\ref{app:patchy_clouds}). Earth's patchy cloud coverage is unique among the three terrestrial planets with significant atmospheres in the Solar System (Venus is completely covered in clouds; Mars has negligible cloud coverage). Using nearly a decade of satellite data, \citet{King2013} show that roughly $67\%$ of Earth's surface is covered by clouds at all times. The cloud fraction over land is approximately $55\%$ and shows a distinct seasonal cycle. Over oceans, cloudiness is significantly higher ($\approx72\%$) and shows smaller seasonal variations. In addition, the cloud fraction is nearly identical during day and night, with only modest diurnal variation. Clouds are particularly abundant in the mid-latitudes (latitudes of $\approx\pm60^{\circ}$), and infrequent at latitudes from $\pm15^{\circ}$ to $\pm30^{\circ}$ (often characterized by arid desert conditions). Thus, there are three bands with a high cloud fraction in Earth’s atmosphere: a narrowband at the equator and two wider mid-latitude bands.

    Atmospheric clouds can significantly impact both the reflected light and thermal emission spectrum of a planet and can reduce or eliminate spectral features \citep[particularly in the UV/O/NIR; e.g.,][]{DesMarais2002, Lu_2023}. Parameters such as cloud fraction, composition, particle size, and altitude as well as multi-layered cloud coverage and cloud seasonality all affect the resulting spectrum significantly \citep[e.g.,][]{DesMarais2002,Tinetti_2006_1, Tinetti_2006_2, Hearty2009, Kitzmann2011, Rugheimer_2013, Vasquez_2013, Komacek2020}. \citet{LIFE_IX} ran retrievals on simulated MIR thermal emission spectra of a Venus-twin exoplanet. They showed that the presence of clouds can be inferred and requires a minimal spectral resolution of $50$ and a signal-to-noise ratio of $20$. Further, clouds inhibit the accurate retrieval of surface conditions, and inadequate cloud treatment in retrievals (i.e., choosing too complex/simple cloud model given the quality of the input spectrum) can bias the estimates for important planetary parameters (e.g., planet radius, equilibrium temperature, and Bond albedo). However, despite recent efforts to understand how patchy clouds could alter the spectra of terrestrial exoplanets \citep[e.g.,][]{May_2021, Windsor_2023}, it remains unclear how they affect the characterization of terrestrial HZ exoplanets through MIR retrievals.

\subsection{MIR Observables of Habitable and Inhabited Worlds}
\label{SubSec: Observable Indicators of Habitability in the MIR}

    Habitability refers to the degree to which a global environment can support life, and depends on a myriad of factors \citep{Meadows2018_habitability}. The characteristics of a planet and its atmosphere, the architecture of the planetary system, the host star, and the galactic environment all affect habitability \citep[for an extensive list, see, e.g.,][]{Meadows2018_habitability}. For exoplanets, which can only be observed via remote sensing, we require observable characteristics to assess their habitability.
    
    Analyzing MIR thermal emission spectra of exoplanets with atmospheric retrievals \citep[see, e.g., Section~\ref{sec:retrievals};][]{Madhusudhan:Atmospheric_Retrieval} and/or climate models yields constraints on the planet's atmospheric structure and composition. Such constraints yield valuable insights into a planet's habitability and could be used to infer the presence of a biosphere. In the following, we list observable signatures of habitability and biospheres in ascending order of difficulty to observe:
\begin{itemize}
    \setlength\itemsep{0em}
    \item \textbf{Planetary energy budget:} A planet's effective temperature and Bond albedo can be calculated from its thermal emission spectrum.
    \item  \textbf{Water and other molecules:} Important atmospheric species, such as water (\ce{H2O}), carbon dioxide (\ce{CO2}), or ozone (\ce{O3}), have strong spectral MIR features.
    \item \textbf{Atmospheric \pt{} structure:} The \pt{} structure can be constrained in retrievals and provides vital information about the atmospheric state. 
    \item \textbf{Surface conditions:} If not fully obscured by clouds, thermal emission spectra contain information about a planet's surface temperature and pressure.
    \item \textbf{Molecular biosignatures:} Important biogenic gases, such as methane (\ce{CH4}) or nitrous oxide (\ce{N2O}), have MIR features. The presence of a biosphere can be inferred if abiotic sources can be ruled out.
    \item \textbf{Atmospheric seasonality:} Seasonal periodicities in molecular abundances that are attributable to life are small for Earth \citep{mettler_2023} and thus challenging to detect in the MIR. However, they could be strong indicators for biological activity \citep{Olson2018}.
\end{itemize}
    For an in-depth review about evaluating planetary habitability and detectable signs of life, we refer to \citet{Schwieterman2018} and references therein.

\subsection{Context of and Goals for this Study}
\label{subsec:context_and_goals}

    In a previous study \citep{mettler_2020}, we analyzed 15 years of thermal emission Earth observation data for five spatially resolved locations. 
    We investigated flux levels and variations as a function of wavelength range and surface type (i.e., climate zone and surface thermal properties) and looked for periodic signals. From the spatially resolved single-surface-type measurements, we found that typically strong absorption bands from \ce{CO2} (\mic{15}) and \ce{O3} (\mic{9.65}) are significantly less pronounced and partially absent in polar regions. This implies that estimating correct abundance levels for these molecules might not be representative of the bulk abundances in these viewing geometries. Additionally, the time-resolved thermal emission spectrum provided insights into seasons/planetary obliquity, but its significance depended on viewing geometry and spectral band.
    
    In a follow-up study \citep{mettler_2023}, we expanded our analyses from spatially resolved locations to disk-integrated Earth views. We presented an exclusive dataset consisting of 2,690 disk-integrated mid-infrared (MIR) thermal emission spectra (\mic{3.75-15.4}, resolution $\R{}\approx1200$). The spectra were derived from remote sensing observations for four different viewing geometries at a high temporal resolution. Using this dataset, we investigated how Earth’s MIR spectral appearance changes as a function of viewing geometry, seasons, and phase angles and quantified the atmospheric seasonality of different bioindicators. We found that a representative, disk-integrated thermal emission spectrum of Earth does not exist. Instead, both the thermal emission spectrum and the strength of biosignature absorption features show seasonal variability and depend strongly on viewing geometry.

    In this paper, we treat Earth as a directly imaged exoplanet to assess the detectability of its characteristics from MIR observations with \life{}. For the first time, we perform a systematic retrieval analysis of real disk- and time-averaged Earth spectra. We investigate how the retrieval characterization depends on the viewing geometry and the season. Uniquely, in this study we do not only have access to the real Earth spectra, but also to ground truth data from remote sensing satellites. Hence, for the first time, we can compare retrieval results from real spectra to ground truth values. This allows us to evaluate the accuracy of the retrieved constraints and thereby validate our retrieval approach. Despite providing a unique opportunity to validate retrieval frameworks and their underlying assumptions, comparable retrieval studies on solar system observations are rare \citep[e.g.,][]{Tinetti_2006_1, Robinson_2023, Lustig-Yaeger_2023}. However, such studies are indispensable to obtain a correct characterization of terrestrial exoplanets in the future.

    {In Section~\ref{sec: datasets and Methodology}, we introduce the disk-integrated MIR thermal emission dataset and the level 3 satellite products used to derive the ground truths. We introduce our atmospheric retrieval routine and the used atmospheric model in Section~\ref{sec:retrievals}. In Sections~\ref{sec:results} and \ref{sec:discussion}, we present and discuss our retrieval results. We contextualize these results by discussing implications for characterizing terrestrial HZ exoplanets in Section~\ref{sec:implications_for_characterization}.
    Finally, in Section~\ref{sec:conclusion}, we summarize our findings and draw conclusions for future observations.} 

\begin{figure*}[ht!]
   \centering
    {\includegraphics[width= 0.9\hsize]{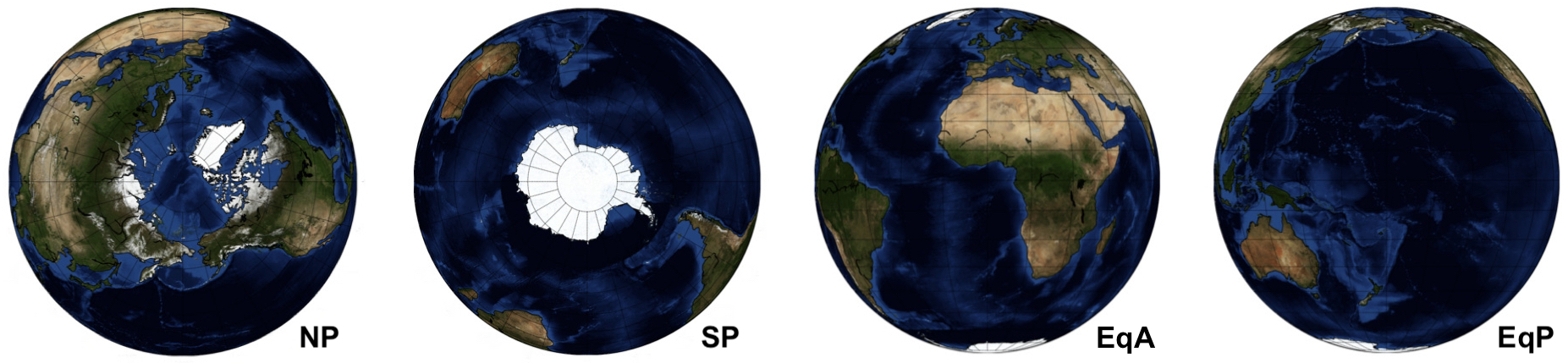}}
    \caption{The four observing geometries studied \citep[taken from][]{mettler_2023}. From left to right: North Pole (NP), South Pole (SP), Africa-centered equatorial view (EqA), and Pacific-centered equatorial view (EqP). Due to the continuously evolving view of low latitude viewing geometries as the planet rotates, the two equatorial views EqA \& EqP were combined to one observing geometry, EqC.}
         \label{fig:Earth viewing geometries}
    \end{figure*}

\section{Datasets and Methodology}
\label{sec: datasets and Methodology}

    In order to compile our ground truth and spectral radiance datasets, we make use of Earth remote sensing climate data obtained from NASA's Atmospheric Infrared Sounder \citep[AIRS;][]{AIRS} aboard the Aqua satellite. For comparison and validation, we have also analyzed data from the Infrared Atmospheric Sounding Interferometer \citep[IASI;][]{IASI} instrument aboard the MetOp satellite. The details of the datasets and the data reduction is discussed in Sections~\ref{SubSec: Spectra Dataset} and \ref{SubSec: GT Dataset}. Although we briefly cover the methodology behind our calculation of disk-averaged spectra and the dataset, we refer to \cite{mettler_2023} for a more comprehensive description.

\subsection{Using Earth Observation Data to Study Earth as an Exoplanet}
\label{subsec:Using Earth observation}

    While there are several methods to study Earth from afar, such as Earth-shine measurements or spacecraft flybys \citep[for a recent review see, e.g.,][and references therein]{Robinson2018}, we chose a remote sensing approach. This approach offers the extensive temporal, spatial, and spectral coverage needed to investigate the effect of observing geometries on disk-integrated thermal emission spectra and time-varying signals. However, for Earth-orbiting spacecrafts it is impossible to view the full disk of Earth and the spatially resolved satellite datasets have to be combined into a spatially resolved, global map of Earth, which can then be disk-integrated \citep[e.g.,][]{Tinetti_2006_1, Hearty2009, Gomez2012}. Furthermore, due to the swath geometry of satellites, daily remote sensing data contain gores, which are regions with no data points, between orbit passes near the equator. In the case of Aqua/AIRS these regions are filled within 48 hours as the satellite continues scanning Earth while orbiting it.

    For our analysis we defined four specific Earth observing geometries as shown in Figure~\ref{fig:Earth viewing geometries}: North (NP) and South Pole (SP), as well as Africa- (EqA) and Pacific-centered (EqP) equatorial views. For each viewing geometry, we mapped, calibrated, and geolocated radiances onto the globe and calculated the disk-integrated MIR thermal emission spectra. The spectra cover the \mic{3.75-15.4} wavelength range (with a gap between \mic{4.6-6.2}) at a nominal resolution of $\R{}\approx1200$ and comprise $2378$ spectral channels. The radiances originate from an AIRS Infrared (IR) level 1C product (V6.7) called AIRICRAD\footnote{\url{https://cmr.earthdata.nasa.gov/search/concepts/C1675477037-GES_DISC.html}} and are given in physical units of $\mathrm{W m^{-2} \mu m^{-1} sr^{-1}}$ \citep{AIRS_L1C}. The total dataset contains $2690$ disk-integrated thermal emission spectra for four consecutive years (2016-2019) at a high temporal resolution for the four full-disk observing geometries \citep[for an overview, see Table~1 in][]{mettler_2023}.

    The viewing geometries as portrayed in Figure~\ref{fig:Earth viewing geometries} evolve throughout the year for a distant observer due to Earth’s nonzero obliquity. Whereas the equatorial view blends seasons and has a diurnal cycle, the polar views show one season but blend day and night. Over the expected integration time of future direct imaging missions, the spectral appearance and characteristics of a planet change as it rotates around its spin axis and as spatial differences from clear and cloudy regions, contributions from different surface types as well as from different hemispheres vary with time. In accordance with the preliminary minimum \life{} requirements motivated in \citet{LIFE_III}, we adopt a typical integration time of 30 days, which is significantly longer than Earth's rotation period. Hence, we average over the EqA and EqP views and denote the resulting dataset EqC.

\begin{table}
\caption{Data and observation details, and spectral information. \label{Table: Study Overview}}
\begin{tabularx}{\linewidth}{l c c}
\toprule
\toprule
\textbf{Data and Observation Details} & \textbf{Value} & \textbf{Unit}\\
\midrule
Year of Data Origin & 2017 &  \\ 
Months of Observation & January \& July &  \\
Integration Time & 30 & days \\
Observed Viewing Geometries & NP, SP, EqC &  \\ 
&&\\
\midrule
\textbf{Spectral Information}&\textbf{Value} & \textbf{Unit}  \\
\midrule
Spectral Coverage & 3.75–15.4  & \mic{}  \\  
Nominal Resolution & 1200 & \\
Number of Spectral Channels & 2378 \\
\bottomrule
\end{tabularx}
\end{table}

    To capture the largest variability between observations, our analyses focus on observing Earth at its extremes in January and July. This choice is motivated by the measured relative flux change for these months at Earth's peaking wavelength in the disk-integrated thermal emission signal \citep[][]{mettler_2023}. Although a pacific-dominated view shows comparable variability to the South Pole view, Earth's rotation causes Africa and the Pacific to rotate in and out of the field of view. This rotation impacts the observed seasonal variability due to the different surface characteristics of EqA and EqP. The study details are summarized in Table~\ref{Table: Study Overview}.

\subsection{Compiling and Processing the MIR Spectra}
\label{SubSec: Spectra Dataset}
    
    The disk-integrated thermal emission spectra for this study are derived from our previously published dataset. Since Earth's MIR spectrum exhibited negligible differences between consecutive years for a fixed viewing geometry \citep[e.g.,][]{mettler_2020, mettler_2023}, we randomly chose the year 2017 and used the data of that year in order to calculate the monthly averages for January and July for the three viewing geometries: NP, SP and EqC. Blending day and night data to simulate the phase of Earth at its orbital position was unnecessary for polar views due to Earth's obliquity, so they naturally include data of both types. However, in the case of the EqC view, we blended day and night data to simulate a rotating Earth at quadrature. This orbital position is preferred for the direct imaging of exoplanets due to the large apparent angular separation between the exoplanet and its host star.

\begin{figure}
   \centering
    {\includegraphics[width= \hsize, trim=3cm 0.9cm 2.8cm 0.9cm, clip]{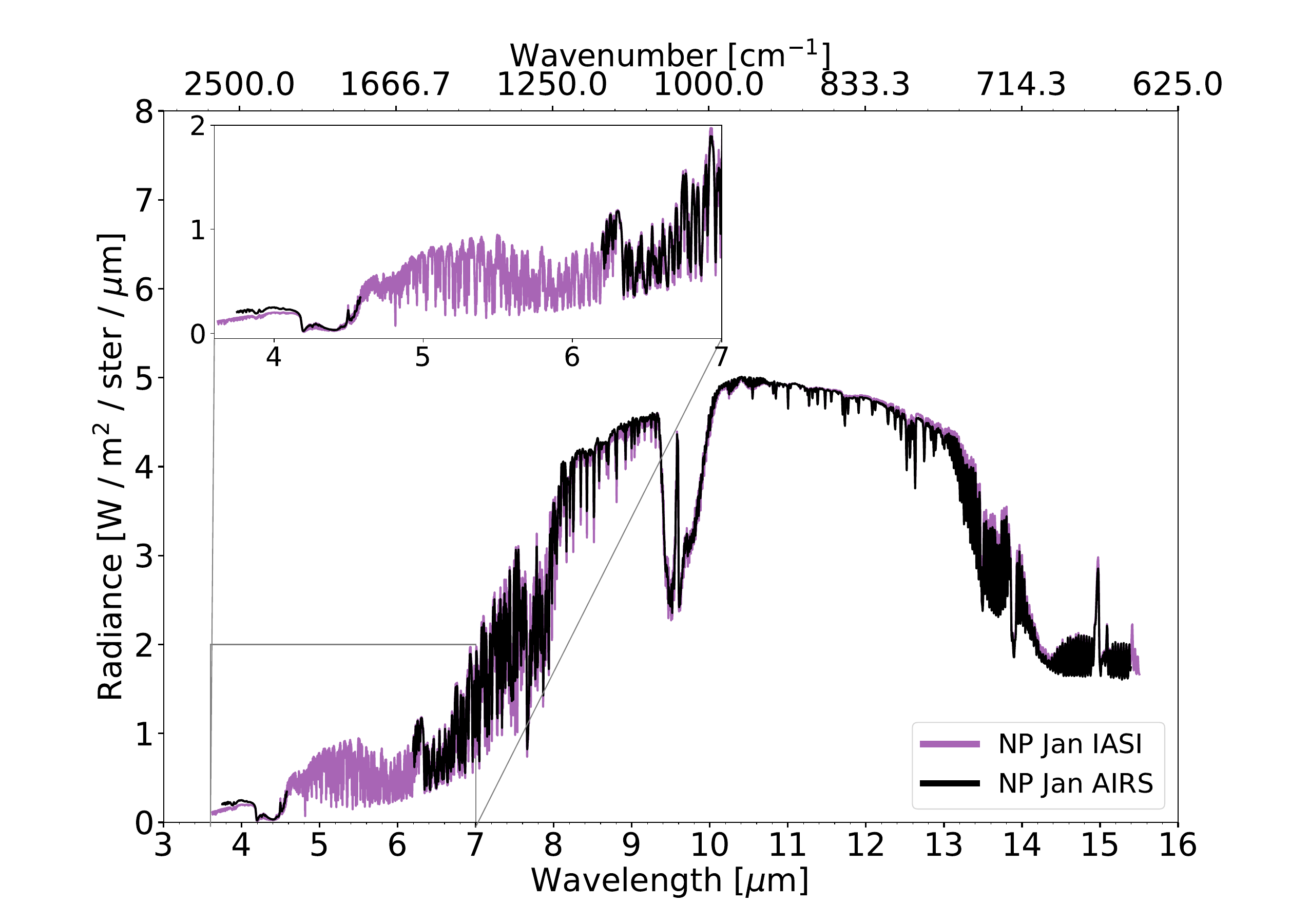}}
    \caption{Comparison between a single-day disk-integrated AIRS (black) and IASI (purple) spectrum for the NP view. The gap between \mic{4.6-6.2} is clearly visible in the AIRS spectrum. 
      }
         \label{fig:AIRSvsIASI}
\end{figure}

    AIRS spectra exhibit a gap between \mic{4.6-6.2} due to dead instrument channels. This gap lies in a \ce{H2O} absorption feature centered at \mic{6.2} \citep[e.g.,][]{Catling2018}. Due to concerns that the partially missing \ce{H2O} feature might deteriorate our retrieval results, we sourced level 1C data\footnote{IASI Level 1C - All Spectral Samples - Metop - Global, Collection ID: EO:EUM:DAT:METOP:IASIL1C-ALL, available at \hyperlink{https://data.eumetsat.int/product/EO:EUM:DAT:METOP:IASIL1C-ALL?query=iasi&s=extended}{EUMETSAT distribution center}}
    for the year 2017 from the IASI instrument aboard the MetOP satellite. We applied the same data reduction steps as for the AIRS dataset described in Section~\ref{subsec:Using Earth observation} and Section~2 of \citet{mettler_2023}. Covering the \mic{3.62-15.50} wavelength regime with $8461$ channels, IASI delivers a continuous spectrum comparable to that of AIRS, which makes it a suitable alternative instrument (see Figure~\ref{fig:AIRSvsIASI}). However, 
    test retrievals showed no significant discrepancies between the retrieval results obtained for the gapped AIRS and continuous IASI spectra. The lack of discrepancies can be attributed to \life{}'s noise level at these lower MIR wavelengths (e.g., Figure~\ref{fig:input_spectra}).
    Thus, since no significant differences were observed and the fact that our ground truth data introduced in Section~\ref{SubSec: GT Dataset} is based on Aqua/AIRS level 3 monthly standard physical retrievals, we opted to use AIRS spectra for this study for consistency.

\begin{figure*}[ht]
  \centering
  \begin{minipage}[c]{0.39\linewidth}
    \vfill
    \centering
    \includegraphics[width=\textwidth, trim=0cm 0cm 2.5cm 0cm, clip]{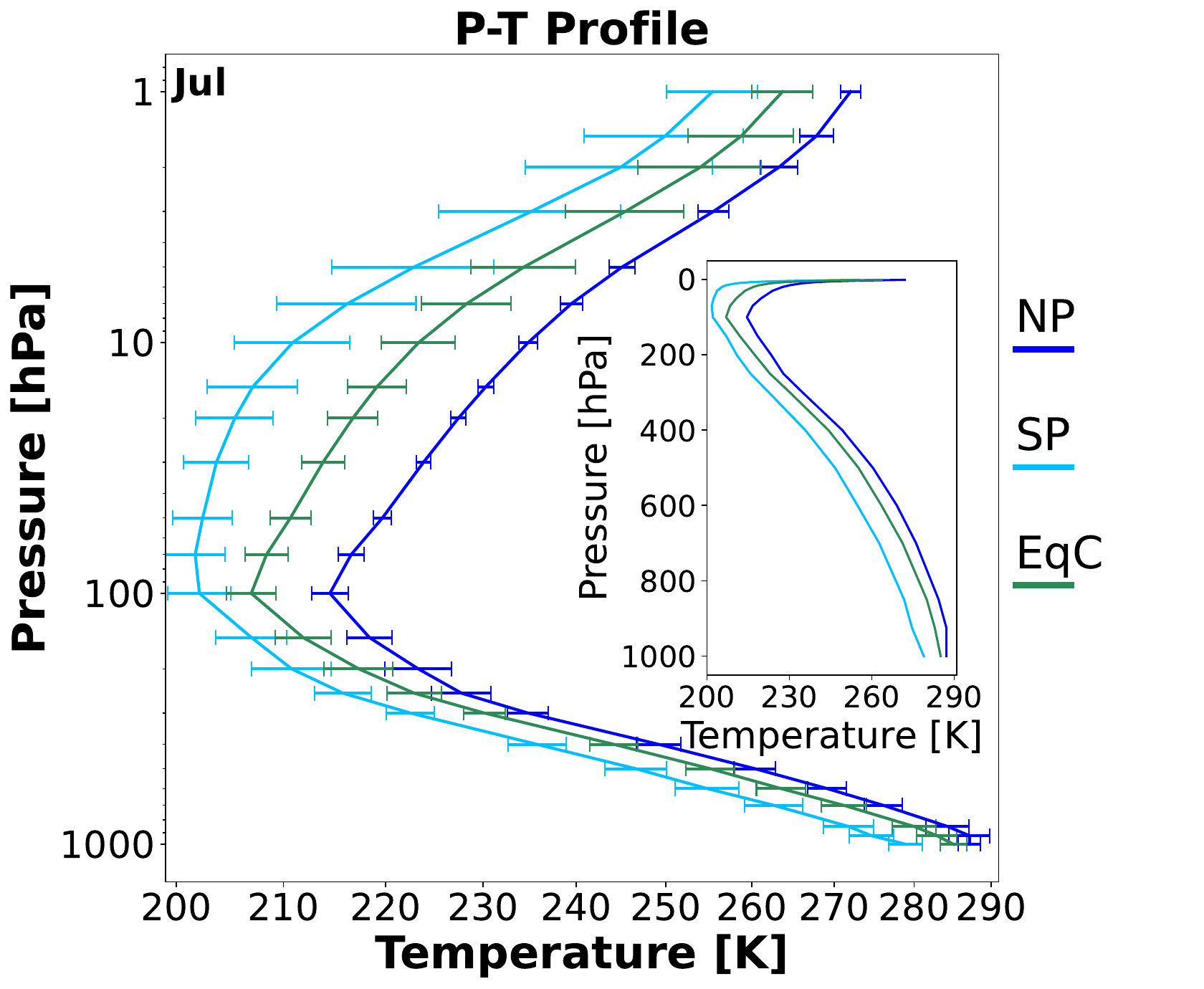} 

    \label{fig:subfigure_a}
    \vfill
  \end{minipage}
  \hfill
  \begin{minipage}[c]{0.6\linewidth}
    \centering
    \begin{minipage}[t]{0.49\textwidth}
      \centering
      \includegraphics[width=\textwidth, trim=0cm 0cm 2.5cm 0cm, clip]{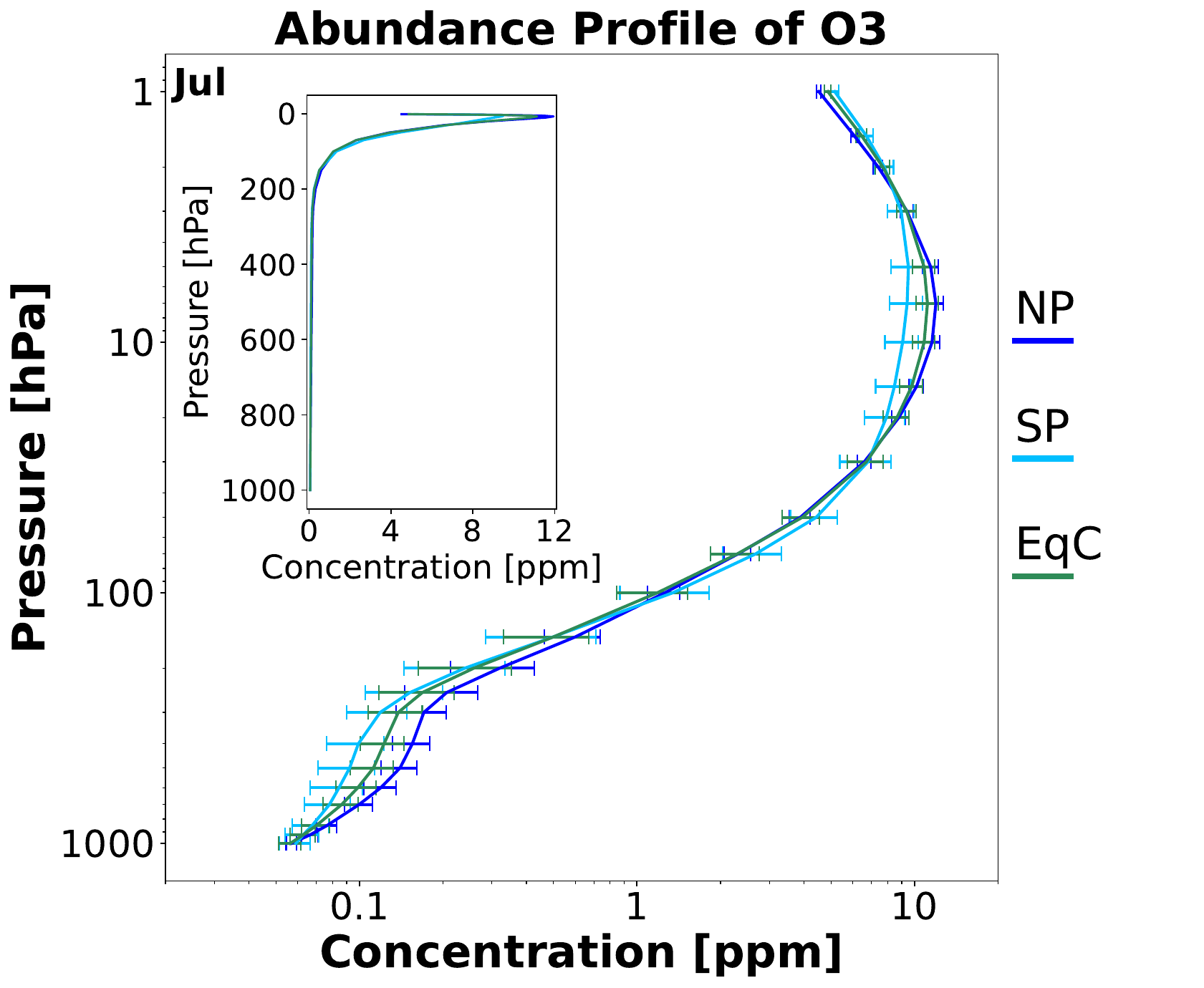}
      \label{fig:subfigure_b}
    \end{minipage}
    \hfill
    \begin{minipage}[t]{0.49\textwidth}
      \centering
      \includegraphics[width=\textwidth, trim=0cm 0cm 2.5cm 0cm, clip]{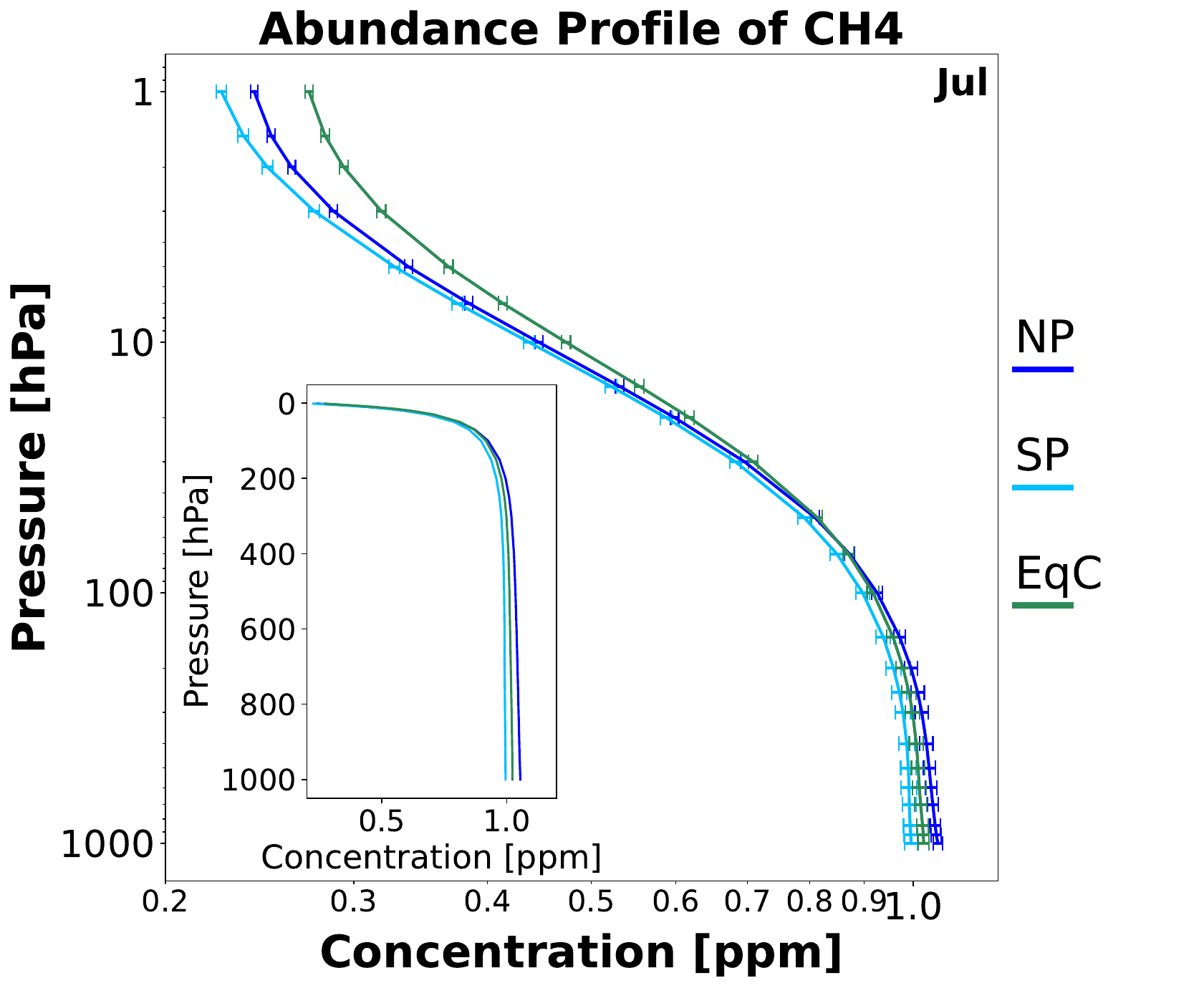}
      \label{fig:subfigure_c}
    \end{minipage}
    \\
    \begin{minipage}[t]{0.49\textwidth}
      \centering
      \includegraphics[width=\textwidth, trim=0cm 0cm 2.5cm 0cm, clip]{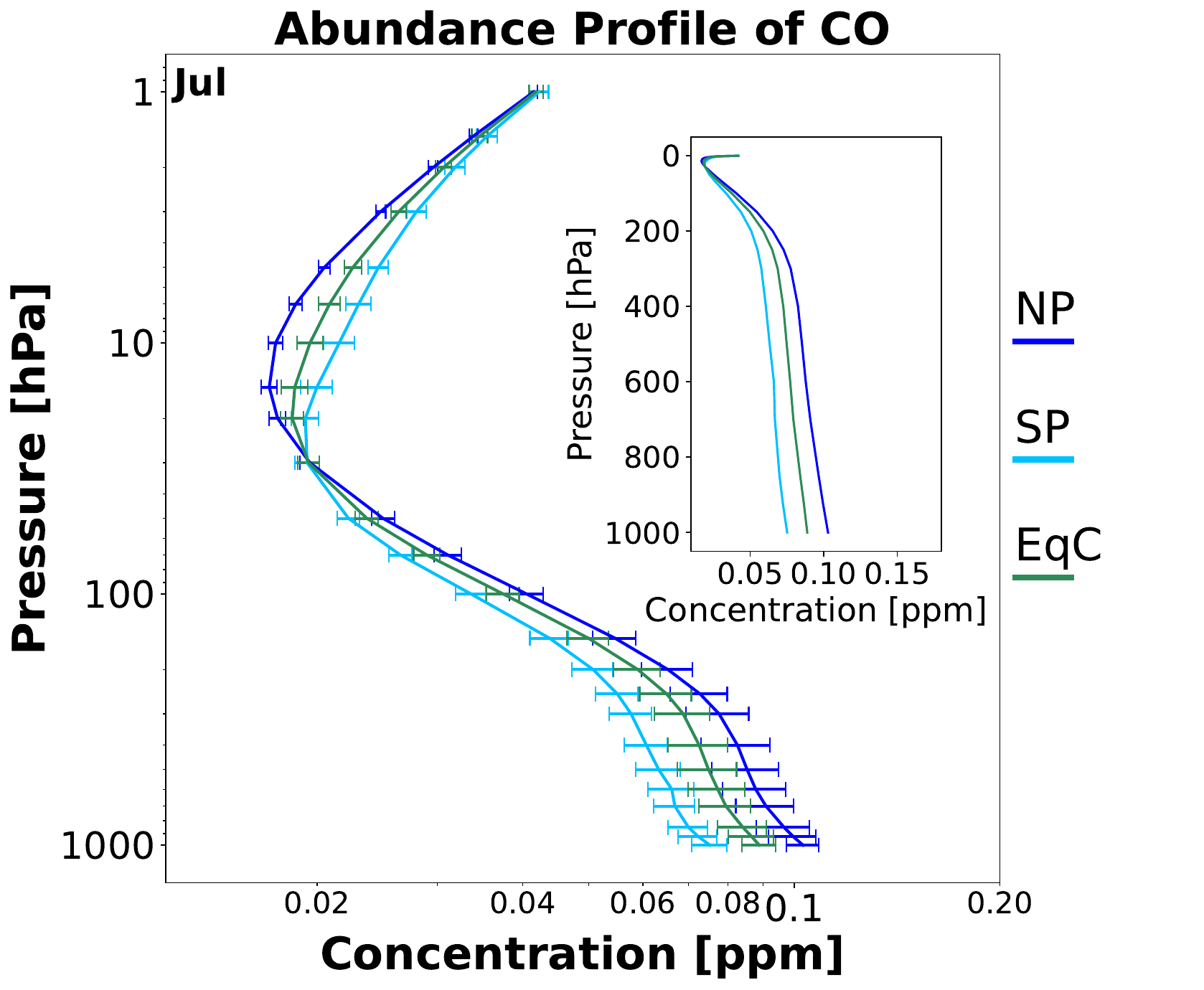}
      \label{fig:subfigure_d}
    \end{minipage}
    \hfill
    \begin{minipage}[t]{0.49\textwidth}
      \centering
      \includegraphics[width=\textwidth, trim=0cm 0cm 2.5cm 0cm, clip]{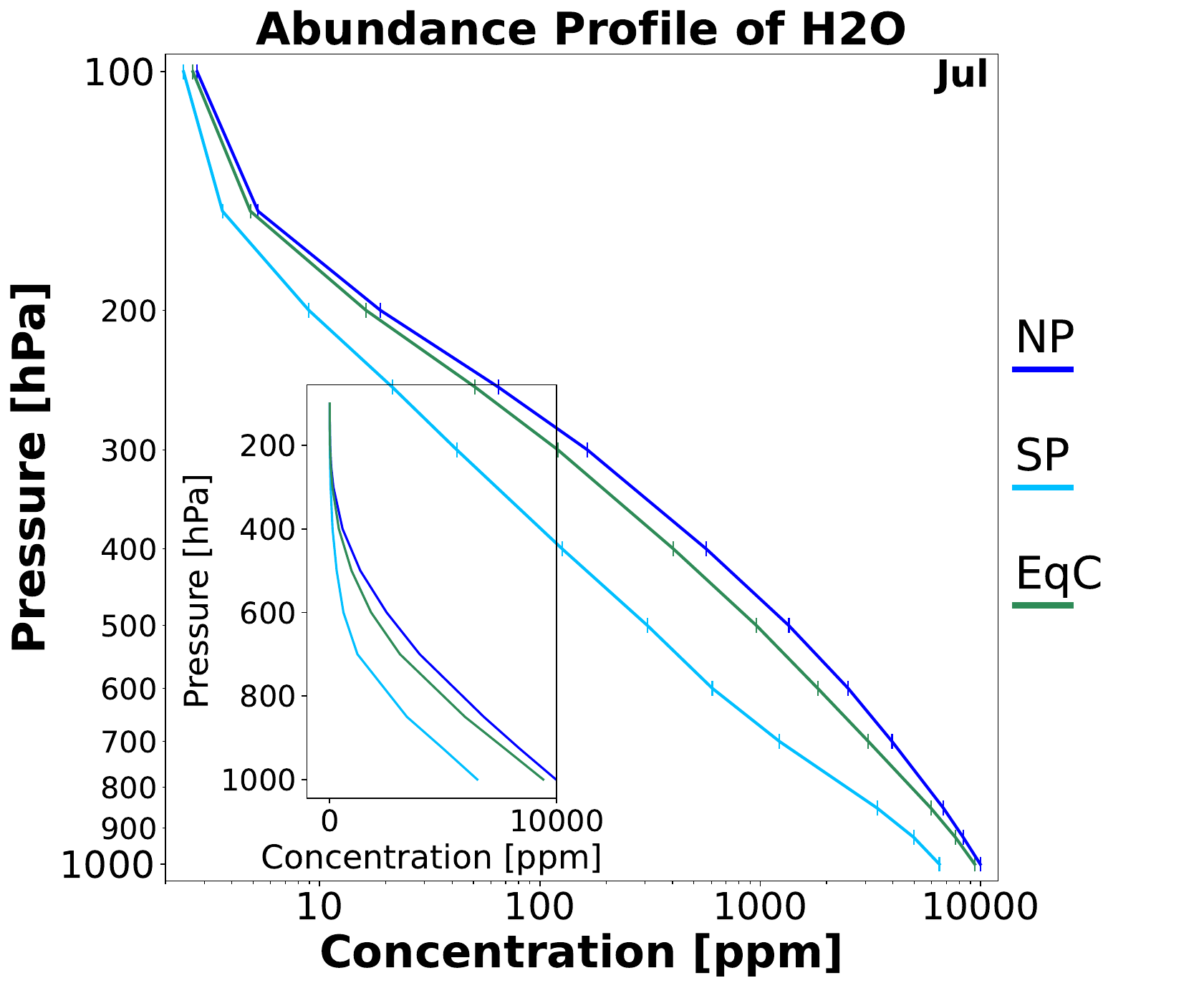}
      \label{fig:subfigure_e}
    \end{minipage}
  \end{minipage}
  \caption{Disk-integrated atmospheric profiles for July. From left to right: \pt{} profile followed by \ce{O3}, \ce{CH4}, \ce{CO}, and \ce{H2O} atmospheric profiles. The error bars are the error propagated uncertainties of the retrieved parameters from the AIRS L3 standard product. The different colors correspond to the viewing geometries: NP (blue), SP (turquoise), EqC (green). The insets display the profiles on a linear scale instead of a logarithmic one.}
  \label{fig:Profiles_Jul}
\end{figure*}
    
\subsection{Compiling and Processing the Ground Truths}
\label{SubSec: GT Dataset}
    
    In Section~\ref{sec:results}, we compare the retrieval outputs to a level 3 (L3) satellite product comprising the \pt{} profile and the trace-gas abundances. Specifically, we have used the Aqua/AIRS L3 Monthly Standard Physical Retrieval (AIRS-only) 1 degree x 1 degree V7.0 (AIRS3STM) product \citep{AIRS3STM_v07}, from which we extracted the surface temperature (land and sea surface) as well as the \pt{} profile. From the trace-gas parameters we extracted the total integrated column burdens and vertical profiles (mass mixing ratios) of \ce{H2O}, \ce{CO}, \ce{CH4}, and \ce{O3}. Both, the \pt{} profile and trace-gas abundances are reported on 24 standard pressure levels ranging from $1000$ to $1.0$~hPa, which are roughly matched to the instrument's vertical resolution \citep{AIRS_AIRS3STM_user_guide}. The \ce{H2O} profile is an exception, as it is only provided at twelve layers ranging from $1000$ to $100$~hPa, spanning from the surface to the tropopause. 

    Since the AIRS3STM product did not contain any \ce{CO2} abundances, we sourced the corresponding ground truth from a gridded monthly \ce{CO2} assimilated dataset\footnote{OCO-2 GEOS Level 3 monthly, 0.5x0.625 assimilated \ce{CO2} V10r (OCO2\_ GEOS\_L3CO2\_MONTH) at GES DISC \citep{OCO-2_v10}} based on observations from the Orbiting Carbon Observatory 2 (OCO-2). The OCO-2 mission provides the highest quality space-based XCO2 retrievals to date, where the level 3 data are produced by ingesting OCO-2 L2 retrievals every 6 hours with GEOS CoDAS, a modeling and data assimilation system maintained by NASA’s Global Modeling and Assimilation Office \citep[GMAO;][]{OCO-2_v10}. The data assimilation (or 'state estimation') technique is employed in order to estimate missing values based on the scientific understanding of Earth's carbon cycle and atmospheric transport. The missing values are mainly the result of the instrument's narrow $10$~km ground track and limited ability to penetrate through clouds and dense aerosols. 

    Following the data reduction of the radiances in Section~\ref{subsec:Using Earth observation}, the \pt{} profile and trace-gas abundances were mapped onto the globe for the different viewing geometries and then disk-integrated at each pressure level. For consistency, we also applied the empirical limb/weighting function to the ground truths. The uncertainties of the retrieved parameters from the AIRS L3 standard product were error propagated, and the resulting error bars are displayed for each data point. The results obtained for July and January are shown in Figure~\ref{fig:Profiles_Jul} and Appendix~\ref{app:Profiles_Jan}, respectively.

\section{Atmospheric Retrievals}
\label{sec:retrievals}

    First, we introduce the disk-integrated Earth spectra and the \lifesim{} noise model used as input for our retrievals (Section~\ref{subsec:input_spectra}). In Section~\ref{subsec:retrieval_routine}, we briefly describe our Bayesian atmospheric retrieval routine. Then, in Section~\ref{subsec:atmospheric_models}, we focus on the 1D plane-parallel atmosphere model used as retrieval forward model. Last, we motivate our choice of prior distributions (Section~\ref{subsec:priors}).

\subsection{Input Spectra for the Retrievals}\label{subsec:input_spectra}

    As input for our atmospheric retrievals, we use reduced-resolution versions of the disk-integrated ARIS spectra from Section~\ref{subsec:Using Earth observation} (NP, SP, and EqC viewing geometries for January and July). All spectra cover the \mic{3.8-15.3} wavelength range, with a gap between \mic{4.6} and \mic{6.2}.

\begin{figure}
  \centering
    \includegraphics[width=0.48\textwidth]{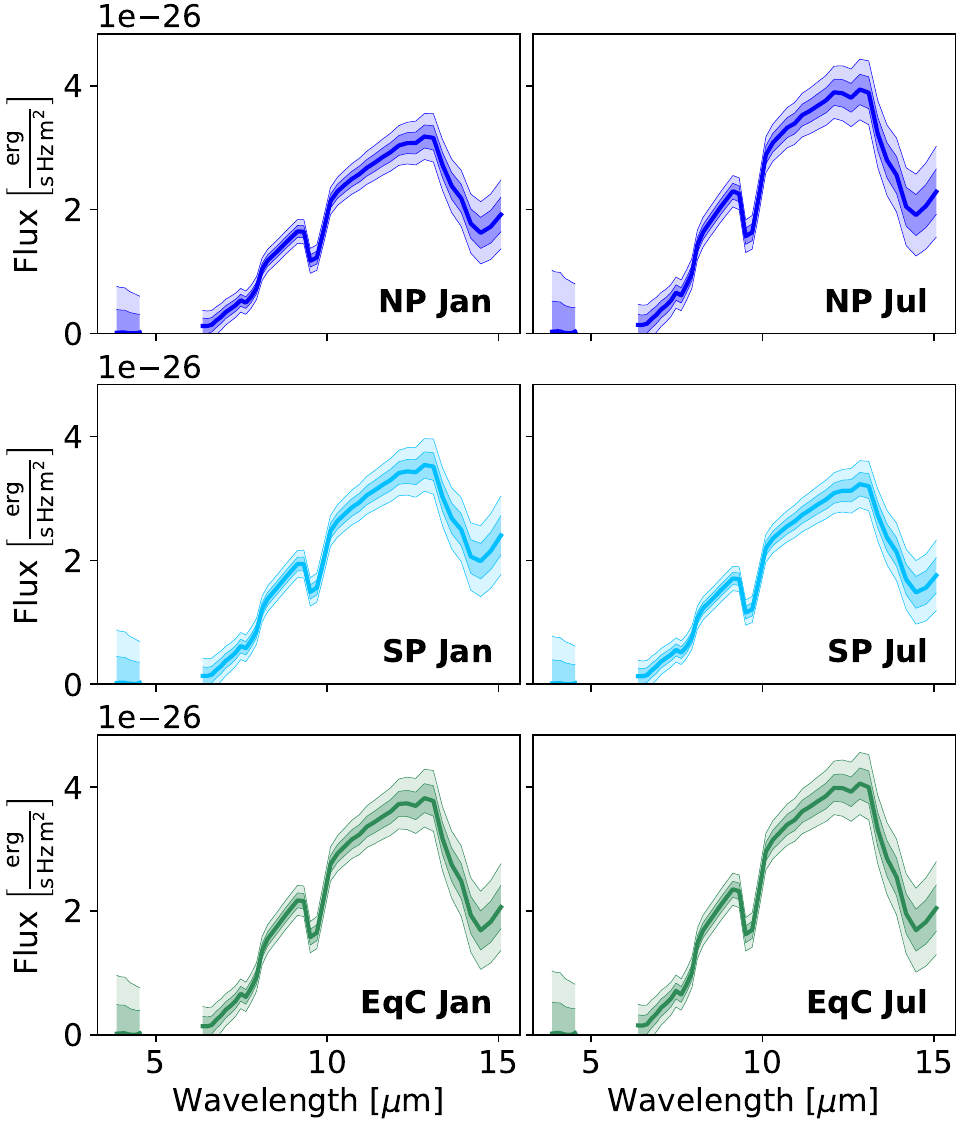}
  \caption{Disk-integrated \Rv{50} Earth spectra considered in our retrieval study. We indicate the \SNv{10} and \SNv{20} \lifesim{} noise levels as shaded areas. Spectra from the top left to the bottom right: NP Jan, NP Jul, SP Jan, SP Jul, EqC Jan, EqC Jul.}
  \label{fig:input_spectra}
\end{figure}

    Based on the preliminary minimal \life{} requirements presented in \citet{LIFE_III,LIFE_IX} and \citet{LIFE_V} (\Rv{50}, \SNv{10}), we consider two resolution cases (\Rv{50, 100}) and two signal-to-noise ratios (\SNv{10, 20}) for each of the six disk-integrated spectra. We define \R{} as $\lambda/\Delta\lambda$, with the width of a wavelength bin $\Delta\lambda$ and the wavelength at the bin center $\lambda$. Further, the \SN{} value corresponds to the \SN{} in the \mic{11.2} wavelength bin. We choose the \mic{11.2} bin because it does not coincide with any strong spectral features. In Figure~\ref{fig:input_spectra}, we show the six \Rv{50} input spectra together with the two different noise levels.

    We model the wavelength-dependent \SN{} expected for \life{} with \lifesim{} \citep{LIFE_II}, which accounts for astrophysical noise sources (photon noise of planet emission, stellar leakage, and local- as well as exozodiacal dust emission)\footnote{Thus, we implicitly assume that a large \life{}-like future space mission will not be dominated by instrumental noise terms (Dannert et al., in prep.).}. To estimate the \lifesim{} noise, we put Earth on a $1$~AU orbit around a G2V star located $10$~pc from the observer. The exozodiacal dust emission of the system was assumed to reach three times the local zodiacal level\footnote{This corresponds to the median level of exozodiacal dust emission found by the HOSTS survey for Sun-like stars \citep{ertel2020}.}.

    In our retrievals, we interpret the noise as uncertainty to the points of the disk-integrated spectra. Thus, the spectral points correspond to the true flux values and are not randomized according to the \lifesim{} \SN{}. While randomized spectra would provide a more accurate simulated observation, a retrieval study based on a single noise realization will yield biased parameter estimates. Ideally, we would run retrievals for multiple ($\gtrsim10$) noise realizations of each spectrum. However, the number of retrievals required make such a study computationally unfeasible. Yet, \citet{LIFE_III} motivate that results from retrievals on unrandomized spectra provide reliable estimates for the average expected retrieval performance on randomized spectra.

\subsection{Bayesian Retrieval Routine}\label{subsec:retrieval_routine}

    For this study, we utilized the Bayesian retrieval routine introduced in \citet{LIFE_III}. The initial routine was improved and modified in \citet{LIFE_V} and \citet{LIFE_IX}. We provide a brief summary of the routine here, and refer to the original publications for an in depth description.

    Our retrieval framework uses the radiative transfer code \texttt{petitRADTRANS} \citep{Molliere:petitRADTRANS, Molliere:petitRADTRANS2, LIFE_V} to calculate the theoretical emission spectrum of a 1D plane-parallel atmosphere model. \texttt{petitRADTRANS} assumes a black-body spectrum at the surface and models the interaction of each atmospheric layer with the radiation to calculate the spectrum at the top of the atmosphere. The model atmosphere is defined via a set of forward model parameters (see Section~\ref{subsec:atmospheric_models} for our forward model). In a retrieval, we search the space spanned by the prior probability distributions (or "priors") of the forward model parameters for the parameter combination that best reproduces the input spectrum. To efficiently search the prior volume, we use the \texttt{pyMultiNest} \citep{Buchner:PyMultinest} package, which uses the \texttt{MultiNest} \citep{Feroz:Multinest} implementation of the Nested Sampling algorithm \citep{Skilling:Nested_Sampling}. Here, we ran all retrievals using 700 live points and a sampling efficiency of 0.3\footnote{As suggested for evidence evaluation by the \texttt{MultiNest} documentation: \url{https://github.com/farhanferoz/MultiNest}}.

    The retrieval yields the posterior probability distribution (or "posterior") for the model parameters. The posterior estimates how likely a certain combination of model parameter values is given the observed spectrum. Further, our routine estimates the Bayesian evidence $\mathcal{Z}$, which is a measure for how well the used forward model fits the input spectrum and can be used for model comparison (see Appendix~\ref{app:model_selection}).

\subsection{Atmospheric Model in the Retrievals}\label{subsec:atmospheric_models}

    As in \citet{LIFE_III,LIFE_IX}, and \citet{LIFE_V}, we characterize each layer of the model atmosphere by its temperature, pressure, and the opacity sources present. We provide a list of all model parameters in Table~\ref{tab:model_parameters}. A comparison between different forward models to justify our choice is provided in Appendix~\ref{app:model_selection}. 

    In our forward model we parameterized the atmospheric \pt{} profile using a fourth order polynomial:
\begin{equation}\label{equ:3poly}
 T(P)=\sum_{i=0}^4a_iP^i.
\end{equation}
    Here, $P$ is the pressure, $T$ the corresponding temperature, and the $a_i$ are the parameters of the \pt{} model. As shown in \citet{LIFE_III}, a polynomial \pt{} model allows us to minimize the number of \pt{} parameters and thereby minimize the retrieval's computational complexity. Learning based \pt{} models require fewer parameters, but their accuracy for terrestrial planets is currently limited by the availability of sufficient training data \citep[e.g.,][]{TimmyPT2023}.

    We consider various opacities in our forward model. First, we account for the MIR absorption and emission by \ce{CO2}, \ce{H2O}, \ce{O3}, and \ce{CH4} (see Table~\ref{tab:opacities} for line lists, broadening coefficients, and cutoffs). We assume constant vertical abundance profiles for all molecules and discuss potential effects of this simplification in Section~\ref{sec:discussion}. Second, we model collision-induced absorption (CIA) and Rayleigh scattering features (CIA-pairs and Rayleigh-species are listed in Table~\ref{tab:opacities}).

    We neglect scattering and absorption by clouds. The patchy clouds in Earth's atmosphere partially block contributions from high-pressure atmosphere layers and thereby impede the characterization thereof. \citet{LIFE_IX} show that neglecting clouds in retrievals can lead to systematic errors in the retrieved surface temperature, surface pressure, and the planet radius. We provide a detailed discussion on potential effects of this simplification in Section~\ref{sec:discussion}.

\subsection{Prior Distributions}\label{subsec:priors}

    We list the priors assumed for all retrievals in Table~\ref{tab:model_parameters}. The priors on the \pt{} parameters $a_i$ and the surface pressure \Ps{} cover a wide range of atmospheric structures (from tenuous Mars-like to thick Venus-like atmospheres). For \ce{N2}, \ce{O2}, \ce{CO2}, \ce{H2O}, \ce{O3}, and \ce{CH4}, we select broad uniform priors that extend significantly below the minimal detectable abundances estimated in \citet{LIFE_III} ($\approx10^{-7}$ in mass fraction for our \R{} and \SN{} cases).

    As in \citet{LIFE_III,LIFE_IX} and \citet{LIFE_V}, we use Gaussian priors for the planet radius \Rpl{} and mass \Mpl{}. The \Rpl{} prior is based on \citet{LIFE_II}, who suggest that a planet detection with \life{} yields a constraint on \Rpl{}\footnote{For a HZ terrestrial planet, a radius estimate $R_\mathrm{est}$ for the true radius $R_\mathrm{true}$ with $R_\mathrm{est}/R_\mathrm{true}=0.97\pm0.18$ is predicted}. The statistical mass-radius relation \texttt{Forecaster}\footnote{\url{https://github.com/chenjj2/forecaster}} \citep{Kipping:Forecaster}, is then used to infer the prior on \lgrt{\Mpl{}} from the \Rpl{} prior. 

\begin{deluxetable}{lcc}
\tablecaption{Parameters of the retrieval forward model.}
\label{tab:model_parameters}
\tablehead{\colhead{Parameter} &\colhead{Description} &\colhead{Prior}}
\startdata 
$a_4$               &$\pt{}$ parameter (degree 4)                                 &$\mathcal{U}(0,10)$\\
$a_3$               &$\pt{}$ parameter (degree 3)                                 &$\mathcal{U}(0,100)$\\
$a_2$               &$\pt{}$ parameter (degree 2)                                 &$\mathcal{U}(0,500)$\\
$a_1$               &$\pt{}$ parameter (degree 1)                                 &$\mathcal{U}(0,500)$\\
$a_0$               &$\pt{}$ parameter (degree 0)                                 &$\mathcal{U}(0,1000)$\\
$\lgrt{\Ps{}}$      &$\lgrt{\textrm{Surface pressure }\left[\mathrm{bar}\right]}$    &$\mathcal{U}(-4,2)$\\
$\Rpl{}$            &Planet radius $\left[R_\oplus\right]$                      &$\mathcal{G}(1.0,0.2)$\\ 
$\lgrt{\Mpl{}}$     &$\lgrt{\textrm{Planet mass }\,\left[M_\oplus\right]}$    &$\mathcal{G}(0.0,0.4)$\\
$\lgrt{\ce{N2}}$    &$\lgrt{\textrm{\ce{N2} mass fraction}}$                &$\mathcal{U}(-10,0)$\\
$\lgrt{\ce{O2}}$    &$\lgrt{\textrm{\ce{O2} mass fraction}}$                &$\mathcal{U}(-10,0)$\\
$\lgrt{\ce{CO2}}$   &$\lgrt{\textrm{\ce{CO2} mass fraction}}$               &$\mathcal{U}(-10,0)$\\
$\lgrt{\ce{H2O}}$   &$\lgrt{\textrm{\ce{H2O} mass fraction}}$              &$\mathcal{U}(-10,0)$\\
$\lgrt{\ce{O3}}$    &$\lgrt{\textrm{\ce{O3} mass fraction}}$               &$\mathcal{U}(-10,0)$\\
$\lgrt{\ce{CH4}}$   &$\lgrt{\textrm{\ce{CH4} mass fraction}}$                &$\mathcal{U}(-10,0)$
\enddata
\tablecomments{The third column lists the priors assumed in the retrievals. We denote a boxcar prior with lower threshold $x$ and upper threshold $y$ as $\mathcal{U}(x,y)$; For a Gaussian prior with mean $\mu$ and standard deviation $\sigma$, we write $\mathcal{G}(\mu,\sigma)$.}
\end{deluxetable}
\begin{deluxetable*}{c|c|c|c|c|c|c|c}[ht!]
\tablecaption{Line and continuum opacities used in the retrievals.}
\label{tab:opacities}
\tablehead{\multicolumn{4}{c|}{Molecular Line Opacities}&\multicolumn{2}{c|}{CIA}&\multicolumn{2}{c}{Rayleigh Scattering}\\
\colhead{Molecule}\vline & \colhead{Line List}\vline& \colhead{Pressure-broadening}\vline& \colhead{Wing cutoff}\vline&\colhead{Pair}\vline&\colhead{Reference}\vline&\colhead{Molecule}\vline&\colhead{Reference}
}
\startdata
\ce{CO2}    &HN20     &$\gamma_{\mathrm{air}}$    &25 cm$^{-1}$ &\ce{N2}$-$\ce{N2}     &KA19           &\ce{N2}    &TH14, TH17\\
\ce{H2O}    &HN20     &$\gamma_{\mathrm{air}}$    &25 cm$^{-1}$ &\ce{N2}$-$\ce{O2}     &KA19           &\ce{O2}    &TH14, TH17\\
\ce{O3}     &HN20     &$\gamma_{\mathrm{air}}$    &25 cm$^{-1}$ &\ce{O2}$-$\ce{O2}     &KA19           &\ce{CO2}   &SU05\\
\ce{CH4}    &HN20     &$\gamma_{\mathrm{air}}$    &25 cm$^{-1}$ &\ce{CO2}$-$\ce{CO2}   &KA19           &\ce{CH4}   &SU05\\
\ce{CO}    &HN20     &$\gamma_{\mathrm{air}}$    &25 cm$^{-1}$&\ce{CH4}$-$\ce{CH4}   &KA19           &\ce{H2O}   &HA98\\
\ce{N2O}    &HN20     &$\gamma_{\mathrm{air}}$    &25 cm$^{-1}$ &\ce{H2O}$-$\ce{H2O}   &KA19           &           &\\
 &&&&\ce{H2O}$-$\ce{N2}    &KA19           &           &
\enddata
\tablecomments{The \ce{CO} and \ce{N2O} line opacities were used solely in the model selection retrievals presented in Appendix~\ref{app:model_selection}.}
\tablerefs{(HA98) \citet{Harvey1998}; (HN20) \citet{HN20}; (KA19) \citet{KARMAN2019160}; (SU05) \citet{2005JQSRT..92..293S}; (TH14) \citet{TH14}; (TH17) \citet{TH17}.}
\end{deluxetable*}

\section{Retrieval Results}
\label{sec:results}

\begin{figure*}
  \centering
    \includegraphics[width=\textwidth]{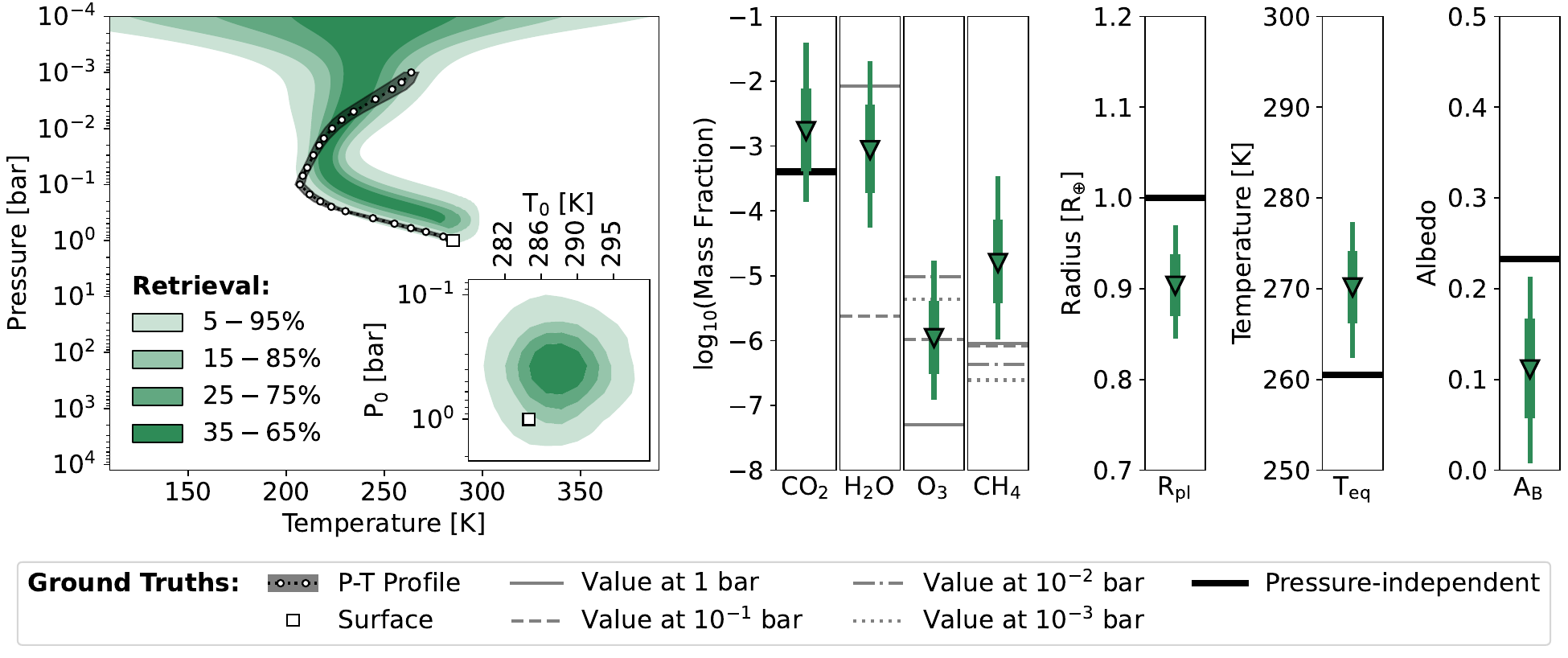}
  \caption{Retrieval results for the \Rv{100}, \SNv{20} EqC Jul Earth spectrum. The leftmost panel shows the retrieved \pt{} structure. Green-shaded areas indicate percentiles of the retrieved \pt{} profiles. The white square marks the true surface conditions (\Ps{}, \Ts{}). The white circles and the gray area show the true \pt{} structure and the uncertainty thereon. In the bottom right of the \pt{} panel, we show the retrieved constraints on the surface conditions. The remaining panels show the posteriors of the trace gas abundances and other parameters. Green lines indicate posterior percentiles (thick: $16\%-84\%$; thin: $2\%-98\%$). Thick black lines indicate pressure-independent ground truths. Thin gray lines show the true abundance at different atmospheric pressures (solid: 1~bar; dashed: $10^{-1}$~bar; dashed-dotted: $10^{-2}$~bar; dotted: $10^{-3}$~bar).}
  \label{fig:ret_summary_EqC_r100SN20}
\end{figure*}

    Here, we present the retrieval results obtained with the forward model from Section~\ref{subsec:atmospheric_models}. In Figure~\ref{fig:ret_summary_EqC_r100SN20}, we summarize the results from the retrieval on the \Rv{100}, \SNv{20} EqC Jul spectrum, which are representative of all retrieval results. We show the retrieved \pt{} structure, the posteriors of the atmospheric trace gases and radius \Rpl{}, and estimates for the equilibrium temperature \Teq{} and the Bond albedo \Ab{} (derived from the posteriors using the method outlined in Appendix~\ref{app:Bond_albedo}). The \ce{N2} and \ce{O2} posteriors are not shown since we did not constrain either abundance. We further plot the ground truths for all parameters. The true atmospheric abundances of \ce{H2O}, \ce{O3}, and \ce{CH4} depend on the atmospheric pressure (see Figure~\ref{fig:Profiles_Jul}). To indicate the range of these ground truth profiles, we plot the ground truths at four different pressures (1~bar, $10^{-1}$~bar, $10^{-2}$~bar, $10^{-3}$~bar).
    We provide the \pt{} profile results from all other retrievals in Appendix~\ref{app:additional_retrieval_data}. The posteriors for all retrievals (excluding the \pt{} parameters $a_i$) along with \Teq{} and \Ab{} estimates are shown in Figure~\ref{fig:PosteriorsM4}. We list the corresponding numeric values in Appendix~\ref{app:additional_retrieval_data}.

    From the results shown in Figure~\ref{fig:ret_summary_EqC_r100SN20}, we would rightly conclude that we are observing a potentially habitable planet. We find temperate surface conditions that would allow for liquid water to exist and easily detect the highly relevant atmospheric gases \ce{CO2}, \ce{H2O}, and \ce{O3}. Importantly, we also detect the potential biosignature \ce{CH4}. These findings hold for all considered viewing geometries, seasons, \R{}, and \SN{}. In the following, we address systematic differences between our retrieval results and the ground truths.
    
    From the retrieved \pt{} profiles in Figure~\ref{fig:ret_summary_EqC_r100SN20} and Appendix~\ref{app:additional_retrieval_data}, we see that our retrieved estimates for the surface conditions and the overall atmospheric \pt{} structure are inaccurate. While \Ts{} is well retrieved (roughly centered on the ground truth, uncertainty $\leq\pm10$~K), \Ps{} is underestimated by up to an order of magnitude (uncertainty $\leq\pm0.5$~dex). This observation does not only hold for the surface conditions but for the entire \pt{} profile. While the shape of the temperature structure is accurately retrieved, it is shifted relative to the ground truth to lower pressures. This effect is observable for all spectra, and becomes smaller for the higher \R{} and \SN{} cases. Further, constraints on the \pt{} structure in the upper atmosphere ($\lesssim10^{-3}$~bar) are weaker, which we expect due to negligible signatures from these layers in MIR emission spectra. The obtained constraints are due to extrapolation of the polynomial \pt{} model and thus not physical.

\begin{figure}[p]
  \centering
    \includegraphics[width=0.45\textwidth]{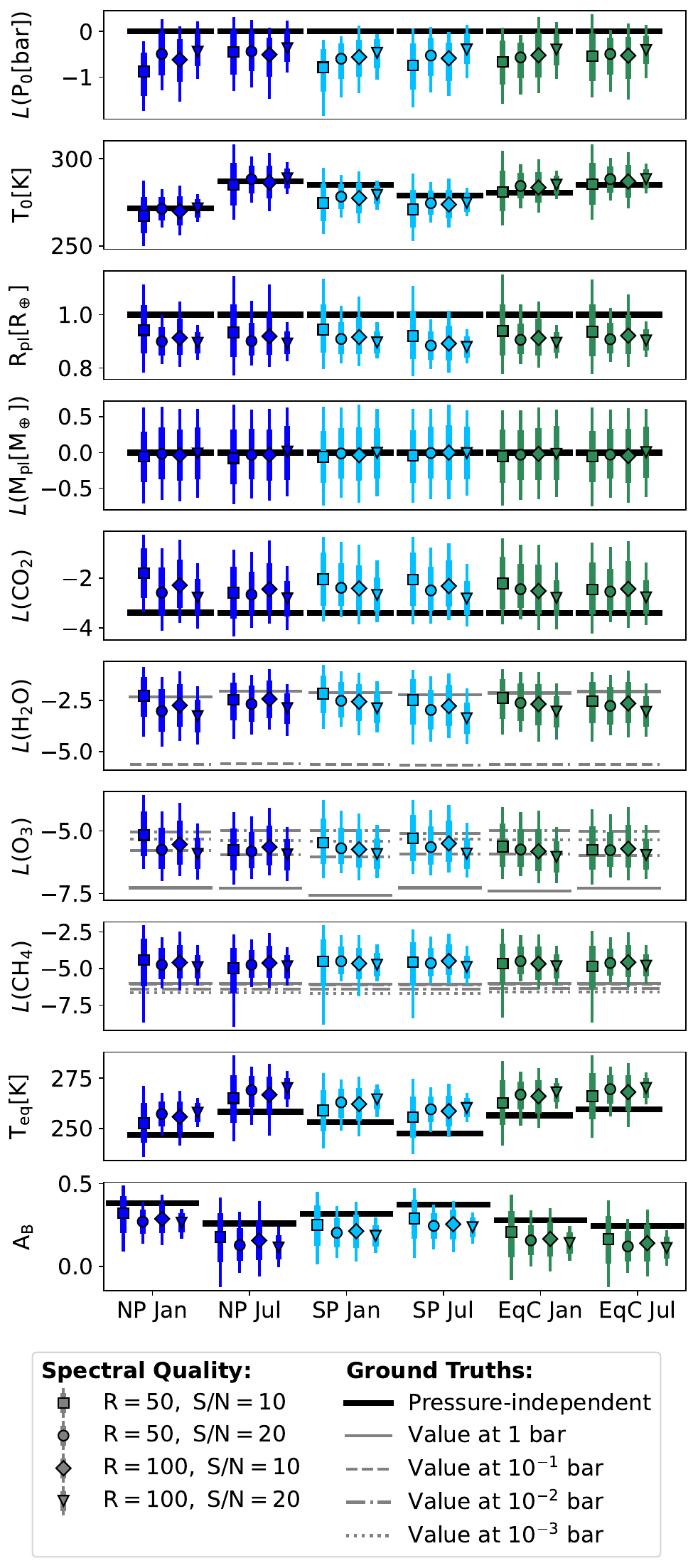}
  \caption{Posteriors for all combinations of \R{} (50, 100) and \SN{} (10, 20) for the six disk-integrated Earth spectra. Here, $L(\cdot)$ abbreviates $\lgrt{\cdot}$. The marker shape represents the spectral \R{} and \SN{}, the color shows the viewing geometry. Colored lines indicate posterior percentiles (thick: $16\%-84\%$; thin: $2\%-98\%$). Thick black lines indicate pressure-independent ground truths. Thin gray lines show the ground truth abundance at different atmospheric pressures (solid: 1~bar; dashed: $10^{-1}$~bar; dashed-dotted: $10^{-2}$~bar; dotted: $10^{-3}$~bar). Columns (left to right) show the results for: NP Jan, NP Jul, SP Jan, Sp Jul, EqC Jan, and EqC Jul.}
  \label{fig:PosteriorsM4}
\end{figure}

    Considering the parameter posteriors in Figure~\ref{fig:PosteriorsM4}, we observe that most parameters are well retrieved (i.e. at least one of the disk-integrated ground truths lies within the $16\%-84\%$ percentile of the posterior). Further, as expected, the constraints on the posteriors get stronger as we consider higher \R{} and \SN{} spectra, since these spectra contain more information and thus yield stronger constraints. {However, several parameter posteriors are biased relative to the ground truths.}

    {First, \Rpl{} is underestimated for all considered \R{} and \SN{} cases. This bias is strongest for the \SNv{20} results. The retrieved \Rpl{} biases are directly linked to the too low \Teq{} and \Ab{} estimates, since both parameters are derived from the \Rpl{} posterior (see Appendix~\ref{app:Bond_albedo}).}

    {Second, the aforementioned systematic underestimation of \Ps{} (and the \pt{} structure) is accompanied by a systematic overestimation of the trace-gas abundances.} This is most apparent for \ce{CO2} and \ce{CH4}, since their ground truths do not vary strongly throughout the atmosphere. The shifts in the retrieved\Ps{} and \pt{} structure translate to overestimated \ce{CO2} and \ce{CH4} abundances. This correlation is caused by a well-known degeneracy between the trace-gas abundances and the pressure-induced line-broadening by the bulk atmosphere \citep[see, e.g.,][]{Misra2014, Schwieterman2015}. This degeneracy also affects the \ce{H2O} and \ce{O3} posteriors. However, due to the strong dependence of the ground truth on the atmospheric pressure, biases are not directly visible (posteriors lie within the ground-truth range). Yet, lower retrieved \Ps{} lead to higher \ce{H2O} and \ce{O3} estimates, implying a degeneracy.

    {In Appendix~\ref{app:red_post_dist}, we provide a detailed analysis of the biases discussed above. We show that if correct estimates of \Ps{} or \Rpl{} are available, the retrieved biases on the remaining parameters can be largely eliminated.}

\subsection{{Reducing Abundance Biases by Considering Ratios}}\label{subsec:rel_abund_post}

    {As discussed above, our estimates for the trace-gas abundances are strongly affected by a degeneracy with \Ps{} and the \pt{} structure. Further, we expect the trace-gas posteriors to be impacted by a physical degeneracy with the planet's surface gravity $g_{pl}$ and thus \Mpl{} \citep[see,  e.g.,][]{Molliere:Gravity_Abundance_Degeneracy, 2018AJ....155..200F, Madhusudhan:Atmospheric_Retrieval, LIFE_III,LIFE_V,LIFE_IX}\footnote{The degeneracy with $g_{pl}$ (and $M_{\text{pl}}$) is caused by the dependence of the hydrostatic equilibrium on $g_{pl}$. In hydrostatic equilibrium, $g_{pl}$ is degenerate with the mean molecular weight of the atmosphere, which is directly linked to the trace-gas abundances.}. If the retrieved abundance posteriors of two different trace-gases are affected by these degeneracies in the same way, the biases in our retrieval results can be largely eliminated by considering their point-wise ratio (i.e., divide one posterior by another). Despite not providing information on the absolute trace-gas abundances, such ratios are of interest since they can help identify states of atmospheric chemical disequilibrium, which can indicate biological activity \citep[see Section~\ref{sec:Detect_Bioindicatiors} for an extended discussion;][]{Lovelock1965,Lovelock1975}.}

\begin{figure}
  \centering
    \includegraphics[width=0.45\textwidth]{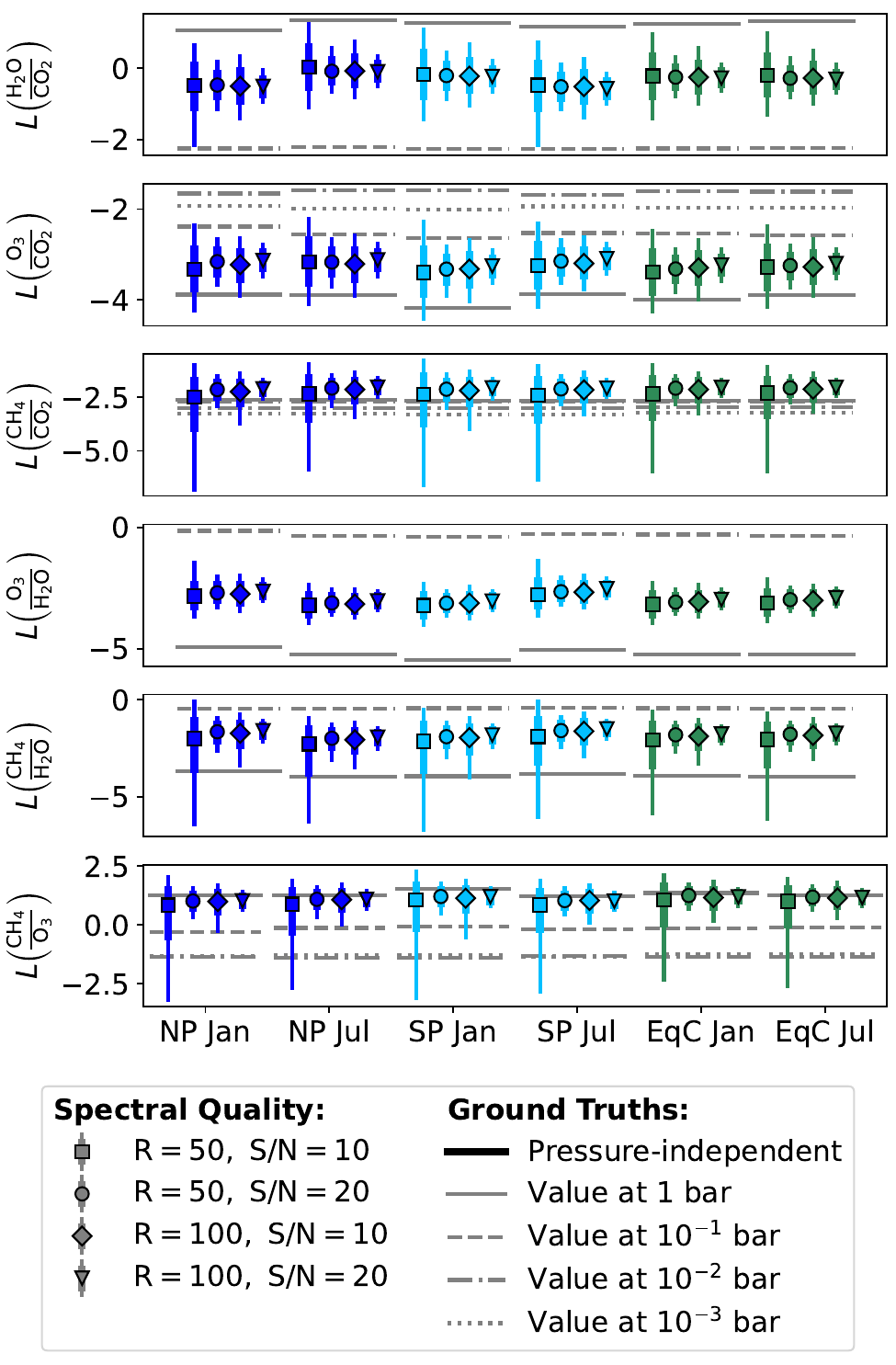}
  \caption{Posteriors of atmospheric trace-gas abundances relative to each other. The figure structure is equivalent to Figure~\ref{fig:PosteriorsM4}.}
  \label{fig:Posteriors_Rel_Abund}
\end{figure}

    We present the relative abundance posteriors for all trace-gas combinations in Figure~\ref{fig:Posteriors_Rel_Abund} (numerical values in Tables~\ref{tab:Rel_Abund_Values_NP} to \ref{tab:Rel_Abund_Values_EqC}). The uncertainties on the relative trace-gas abundances are significantly smaller than on the absolute abundances {due to the elimination of the aforementioned \Mpl{} degeneracy.} Further, in contrast to the absolute abundance posteriors (Figure~\ref{fig:PosteriorsM4}), all ratios lie within the range of the relative ground truths, indicating that the {biases invoked by the \Ps{} degeneracy are mostly eliminated.}

\section{{Discussion of Retrieval Results}}
\label{sec:discussion}

{The present study is, to our knowledge, the first to systematically run retrievals on real disk- and time-averaged MIR Earth spectra. By comparing our retrieval results to known ground truths, we can draw robust conclusions for the characterization performance of \life{} for Earth-like exoplanets. Further, by comparing our findings with other studies, we can find potential causes for the biases discussed in Section~\ref{sec:results}.}  

\subsection{{Comparing the LIFE Performance to Previous Studies}}\label{sec:comp_prev_stud}

    {Previous studies have evaluated how well \life{} could characterize terrestrial HZ exoplanets. \citet{LIFE_III} find preliminary estimates for \life{}'s minimal \R{} and \SN{} requirements by running retrievals on simulated Earth spectra. \citet{LIFE_V} run retrievals on simulated spectra from \citet{RugheimerETT}, which represent different stages in Earth's temporal evolution. \citet{LIFE_IX} investigate how well \life{} can characterize the atmosphere and clouds of Venus. All studies analyze spectra that were calculated with simplified, temporally constant, 1D atmosphere models. In contrast, we run retrievals on real disk- and time-averaged MIR Earth spectra.}

    {Despite large differences in the complexity of the considered spectra between our study and the previous ones, we confirm the previous findings for the detectability of different trace gases. Crucially, \ce{CH4}, the main driver for the minimum \life{} requirements from \citet{LIFE_III} (\Rv{50}, \SNv{10}), remains detectable here. Further,} the strength of the parameter constraints we retrieve here are equivalent to the prior studies, {which demonstrates their robustness. The retrieved $1~\sigma$ parameter uncertainties in all studies are} $<\pm0.5$~dex for pressures, $<\pm0.1\Rearth{}$ for radii, $<\pm20$~K for temperatures, and $<\pm1.0$~dex for trace-gas abundances.

\subsection{{Main Source for Radius Bias}}\label{sec:source_R_biases}

    {In Section~\ref{sec:results}, we state that our \Rpl{} estimates underestimate Earth's true radius. This leads to biased estimates for \Teq{} and \Ab{}, which are calculated from \Rpl{} (see Appendix~\ref{app:Bond_albedo}).}

    {We mainly attribute underestimation of \Rpl{} to neglecting Earth's patchy cloud coverage in our forward model (see Section~\ref{subsec:atmospheric_models}).} Clouds reduce the total MIR emission at the top of the atmosphere by partially absorbing the thermal emission from the warm, high-pressure atmosphere layers below them. {By using a first-order approximation (see Appendix~\ref{app:quant_radius_biases} for details), we can demonstrate that the magnitude of the bias on our \Rpl{} estimate can be fully attributed to the missing cloud treatment in our forward model.}
    
    {Further evidence for links between clouds and biased \Rpl{} estimates is provided by other thermal emission retrieval studies. \cite{LIFE_III}, who run retrievals on simulated cloud-free Earth spectra, retrieve bias-free \Rpl{} estimates. In contrast, \citet{LIFE_V} run cloud-free retrievals on simulated cloudy Earth spectra and also underestimate \Rpl{}.}

    

\subsection{{Main Source for Pressure Bias}}\label{sec:source_P_biases}
    
    {As stated in Section~\ref{sec:results}, the retrieved \Ps{} and \pt{} structure are offset to lower pressures relative to the ground truth. These biases are linked to offsets in the retrieved trace-gas abundances, which are degenerate with the pressure-induced line-broadening \citep[see, e.g.,][]{Misra2014,Schwieterman2015}.}

    {We attribute these biases to our assumption of vertically constant trace-gas abundances in our forward model (see Section~\ref{subsec:atmospheric_models}). This claim is motivated by comparison with previous \life{} retrieval studies. \citet{LIFE_III} assume constant abundance profiles both to generate their 1D Earth spectra and in their forward model, and retrieve unbiased \Ps{}, \pt{}, and abundance estimates. In contrast,} \citet{LIFE_V} assume constant abundance profiles to run retrievals on 1D Earth spectra from \citet{RugheimerETT}, which were generated using non-constant abundance profiles. {Their results show offsets in \Ps{}, the \pt{} structure, and the trace-gas abundances, which are comparable in magnitude to our offsets.}
    
    {Further, we argue that \ce{H2O} is the cause of the observed biases. First, in contrast to \ce{CO2}, \ce{O3}, and \ce{CH4}, \ce{H2O} has multiple strong absorption features in Earth's MIR spectrum \citep[see, e.g., Figure~3 in][]{LIFE_III}. Second, the ground truths in Figure~\ref{fig:Profiles_Jul} show that the variances for \ce{H2O} are more than two orders of magnitude greater than for the other species. Third, the main \ce{H2O} variance occurs in the lowermost atmosphere layers, where \ce{H2O} condensation occurs. These layers contribute most strongly to Earth's MIR thermal emission.}

\subsection{{Implications for Retrievals on Exoplanet Spectra}}

    {In the present study, ground truth measurements of Earth's atmosphere have allowed us to validate our results. We found important biases in the posteriors, which we attribute to simplifying assumptions made by our forward model. A proposed remedy is to derive quantities that are less affected, such as abundance ratios (see Section~\ref{subsec:rel_abund_post}). Also, in a future study, we aim to reduce biases by adding a parametrization for patchy clouds and a vertically non-constant \ce{H2O} profile (motivated by \ce{H2O} condensation) to our forward model.}
    
    {Independent of the success of this future effort,} intercomparison efforts \citep[e.g.,][]{BarstowRetComp} have shown that retrieval results also strongly depend on framework specificities (e.g., parameter estimation algorithms, radiative transfer implementations, and line-lists). {To ensure the correct characterization of exoplanets, robust and bias-free retrieval frameworks are required.} Thus, community efforts, such as the CUISINES Working Group\footnote{\url{https://nexss.info/cuisines/}}, that benchmark, compare, and validate different frameworks on real and simulated spectra with known ground truths are indispensable.

\section{{Implications For Characterizing Terrestrial HZ Exoplanets}}
\label{sec:implications_for_characterization}
\subsection{{Effects of Viewing Geometries and Seasons}}

    As described in Section~\ref{SubSec: Earth's Characteristics}, Earth exhibits an uneven distribution of land and ocean regions. Further, different surface types have different spectral and thermal characteristics \citep[e.g.,][]{Hearty2009, Gomez2012, Madden2020}. Also, the distribution of life on Earth is non-uniform with a measurable gradient in the abundance and diversity of life, both spatially (e.g., from deserts to rain forests) and temporally (e.g., from seasonal to geological timescales) \citep{Mendez_2021}. In \citet{mettler_2023}, we find that a representative, disk-integrated thermal emission spectrum of Earth does not exist. Instead, the MIR spectrum and the strength of the absorption features show seasonal variations and depend on the viewing geometry. For future observations of HZ terrestrial exoplanets, the viewing geometry will be unknown. Thus, we must understand how the viewing geometry impacts exoplanet characterization, observable habitability markers, and signatures of life.

    {As we see from Figure~\ref{fig:PosteriorsM4}, most parameter posteriors show no significant dependence on either the viewing geometry or the season (exceptions: \Ts{}, \Teq{}, and \Ab{}). For the \R{} and \SN{} levels studied here, both the retrieved \Rpl{} and the trace-gas abundance estimates show no measurable variations with the exoplanet's orientation relative to the observer. Thus, we conclude that their characterization depends on neither the viewing geometry nor the season for an Earth-like exoplanet.}
    
    {For \Ts{} and \Teq{}, the variations in the posteriors are largest for the NP view, where the differences between January and July are robustly detected in all \R{} and \SN{} scenarios. For the SP and EqC views, variations in \Ts{} and \Teq{} between January and July are much smaller and not confidently detected. This is in agreement with \citet{mettler_2023},} who find the seasonal disk-integrated thermal emission flux differences for the landmass dominated NP view to be $33\%$, as opposed to only $11\%$ for the ocean dominated SP and EqP (Pacific-centered equatorial) views. The increased \Ts{} variance observed for the NP view can be attributed to the large landmass fraction. Also, our results for \Ts{} and \Teq{} indicate that the NP Jul, SP, and EqC views cannot be differentiated from one another despite vastly different characteristics like climate zones and landmass fractions. This highlights the strong spectral degeneracy with respect to seasons and viewing geometries, and agrees with other studies \citep[e.g.,][]{Gomez2012,mettler_2023}.

    For \Ab{}, our retrieval results show small differences between the viewing angles and seasons. As for the temperatures, the variations are largest for the NP view (NP: $46\%$; SP: $17\%$; EqC: $16\%$). For the NP and SP views, the retrieved \Ab{} tends to be higher during winter, which agrees with the lower retrieved \Ts{} and \Teq{} values. {However, due to the uncertainties ($\pm0.05$ to $\pm0.10$) and biases on the posteriors, a confident detection of these \Ab{} differences is not possible. Thus, our \Ab{} characterization is independent of viewing geometry and season.}
    
    {However, as we discuss in Section~\ref{sec:results}, the accuracy and strength of our constraints for \Ts{}, \Teq{}, and \Ab{} are limited by our \Rpl{} estimates. As we demonstrate in Appendix~\ref{app:red_post_dist}, an accurate and strong \Rpl{} constraint 
    would yield detectable differences in \Ts{}, \Teq{}, and \Ab{}. In this case,} Earth-like seasonal \Ts{}, \Teq{}, and \Ab{} changes are easily detectable for the NP view with a \life{}-like observatory for all \R{} and \SN{} cases. Also for the SP view, detections of seasonal variations are possible (except for the \Rv{50}, \SNv{10} case). For the EqC view, which blends the two hemispheres, the seasonal variations remain undetected.

\subsection{{Detectability of Bioindicators}}\label{sec:Detect_Bioindicatiors}

    Earth's MIR spectrum contains features from numerous bioindicator gases. Examples are \ce{O3} (photochemical product of bioindicator \ce{O2}), \ce{CH4}, and \ce{N2O} \citep[see, e.g.,][for an extensive list]{Schwieterman2018}. While \ce{N2O} is not detectable at the \R{} and \SN{} considered, \ce{O3} and \ce{CH4} are (biases $\leq+1.0$~dex, uncertainties $\leq\pm1.0$~dex; see Appendix~\ref{app:model_selection}). {However, the sole detection of a bioindicator gases is not sufficient to infer the presence of life, since they can be produced abiotically \citep[see, e.g.,][]{Catling2018, Schwieterman2018, 2018haex.bookE..71H}. The simultaneous detection of multiple bioindicator gases provides a more robust marker for biological activity.}

    {One detectable multiple bioindicator is the 'triple fingerprint',} which is the simultaneous detection of atmospheric \ce{CO2}, \ce{H2O}, and \ce{O3} \citep[see, e.g.,][]{2002A&A...388..985S}. Depending on the viewing angle, the retrieved  $\lgrt{\ce{H2O}/\ce{CO2}}$ ranges from $-0.2$~dex to $-0.5$~dex and the $\lgrt{\ce{O3}/\ce{H2O}}$ from $-2.6$~dex to $-3.2$~dex (uncertainties $\leq\pm0.6$~dex). These values lie between the average $1$~bar and $0.1$~bar ground truths ($\lgrt{\ce{H2O}/\ce{CO2}}$: $1.2$~dex at $1$~bar, $-2.3$~dex at $0.1$~bar; $\lgrt{\ce{O3}/\ce{H2O}}$: $-0.3$~dex at $1$~bar, $-5.2$~dex at $0.1$~bar).
    
    Another promising multiple bioindicator is the simultaneous detection of reducing and oxidizing species in an atmosphere (i.e., a strong chemical disequilibrium). Since the two species will react rapidly with each other, simultaneous presence over large timescales is only possible if both are continually replenished at a high rate by life \citep{Lederberg1965}. {One example hereof that we confidently detect is the simultaneous presence of \ce{O2} (or its photochemical product \ce{O3})\footnote{{The atmospheric \ce{O2} abundance is not directly constrainable via MIR observations (see Section~\ref{sec:results}). However, as suggested by \citet{KozakisO3}, retrieved \ce{O3} abundance estimates can provide a first-order estimate for atmospheric \ce{O2} levels.}} and \ce{CH4} \citep{Lovelock1965,Lippincott1967}. For all but the \Rv{50}, \SNv{10} retrievals, we accurately constrain the $\lgrt{\ce{CH4}/\ce{O3}}$ abundance ratio to $1.1$~dex (uncertainty $\leq\pm0.5$~dex).} Especially, in the context of an Earth-like planet orbiting a Sun-like star, the detection of such an \ce{O2}/\ce{O3}-\ce{CH4} disequilibrium would represent a strong potential biosignature.

    \subsection{{Detectability of Seasonal Variations in Bioindicators}}\label{sec:TempVar_Bioindicatiors}

    Research on the detectability of exoplanet biosignatures has predominantly focused on static evidence for life (e.g., the coexistence of \ce{O2} and \ce{CH4}). However, the anticipated range of terrestrial planet atmospheres and the potential for both "false positives" and "false negatives" in conventional biosignatures \citep[e.g.,][]{Selsis2002, Meadows_2006, Reinhard_2017, Catling2018, Krissansen-Totton2022} underscore the necessity to explore additional life detection strategies. Time-varying signals, such as seasonal variations in atmospheric composition, have been proposed to be strong biosignatures \citep[e.g.,][]{Olson2018}, since they are biologically modulated phenomena that arise naturally on Earth and likely also occur on other non-zero obliquity and eccentricity planets. \citet{Olson2018} suggest, that atmospheric seasonality as a biosignature avoids many assumptions about specificities of metabolisms. Further, it offers a direct means to quantify biological fluxes, which would allow us to characterize, rather than simply identify, exoplanet biospheres.
    
    To assess the detectability of such time-dependent atmospheric modulations in exoplanets, we consider the retrieved abundance ratios in Figure~\ref{fig:Posteriors_Rel_Abund}. {Abundance ratios are less affected by parameter degeneracies and thus exhibit smaller uncertainties and biases ($\leq\pm0.4$~dex for the \Rv{50}, \SNv{20} and \Rv{100} cases). Independent of the viewing geometry, we see no significant differences between the trace-gas ratios retrieved for January and July. Since these months represent Earth's two extreme states, we do not expect differences in the trace-gas ratios to be observable for any two other months.} Consequentially, detecting the atmospheric seasonality of trace-gas abundances as a biosignature is not feasible for the studied \R{} and \SN{} cases.
    
    This agrees with our findings in \citet{mettler_2023}, where we studied disk-integrated Earth spectra and quantified the amplitudes of the seasonal variations in absorption strength by measuring the equivalent widths of the biosignature related absorption features. We detected small seasonal variations for \ce{O3}, \ce{CO2}, \ce{CH4}, and \ce{N2O}. For \ce{CO2} and \ce{CH4} the seasonal abundance variations of $1\%$ to $3\%$  are significantly smaller than the uncertainties on our retrieved abundance estimates ($\approx\pm0.5$~dex), which makes a detection unfeasible.

    
    {Significantly higher \R{} or \SN{} MIR spectra are required to be sensitive to the spectral variations evoked by Earth-like seasonal fluctuations in bioindicator gas abundances. Such observations require either a more sensitive instrument or an integration time greater than the assumed $30$~days (see Section~\ref{subsec:Using Earth observation}). However, while the magnitude of such spectral variations is unchanged for shorter integration times\footnote{{Local daily/weekly variations (e.g., in trace-gas abundances or the \pt{} structure) average out on the planetary scale and thus do not affect seasonality-related spectral features significantly. On Earth, seasonal changes in trace-gas abundances  occur on timescales of several months. Therefore, the magnitude of the associated spectral variations and thus their detectability remain unchanged for shorter observations.}} (e.g., $10$~days), it will decrease for extended integration times (e.g., $90$~days). For Earth, significant seasonal changes occur during such extended observations. However, the measured spectrum represents the average state of the observed atmosphere. Thus, the magnitude of the spectral variations evoked by seasonal fluctuations is diminished, which counteracts the sensitivity gain attained via an increase in observation time.}

    However, terrestrial exoplanets could display seasonality patterns that are very different from that of Earth or other Solar System planets. Given the extensive diversity among exoplanets (e.g., in terms of mass, size, host star type, and orbit), it is likely that some exhibit detectable seasonal variations. Seasonal signals could be amplified by several factors \citep[see, e.g., Section~4.3 in][]{mettler_2023} such as: shorter photochemical lifetimes and/or non-saturated spectral bands, increased orbital obliquity (leads to greater seasonal contrast due to varying ice and vegetation cover), biological activity promoted by moderately high obliquity (e.g. photosynthetic activity) consequently leading to heightened variations in biosignature gases, and the absence of competing effects from admixed hemispheres (particularly relevant for eccentric planets). The detectability of seasonality depends on both the magnitude of the biogenic signal and the degree to which the observation conditions mute that signal, and is likely maximized for an intermediate obliquity.

\section{Summary and Conclusion}
\label{sec:conclusion}

    In this study, we treated Earth as an exoplanet to examine how well it can be characterized from its MIR thermal emission spectrum. This is the first study that systematically ran atmospheric retrievals on simulated \life{} observations of real disk-integrated MIR Earth spectra for different viewing angles and seasons. By comparing the results to ground truths, we assessed the accuracy and robustness of the retrieved constraints and explored the applicability of simple 1D atmosphere models for characterizing the atmosphere of a real habitable planet with a global biosphere. 
    Further, we investigated whether the viewing geometry and season have a measurable impact on the characterization of an Earth-like exoplanet and searched for signs of atmospheric seasonality, indicative of a biosphere.
    
    {Our results at the minimal \life{} requirements (\Rv{50}, \SNv{10}) find Earth to be a temperate habitable planet with detectable levels of \ce{CO2}, \ce{H2O}, \ce{O3}, \ce{CH4}. We find that viewing geometry and the observed season do not affect the detectability of molecules, the retrieved relative abundances, and thus the characterization of Earth's atmospheric composition.} However, the seasonal flux difference of 33\% for the North Pole view causes variations in the retrieved surface temperature \Ts{}, equilibrium temperature \Teq{}, and Bond albedo \Ab{}, which are detectable with \life{} for all tested \R{} and \SN{} cases. If strong and unbiased estimates for the planet radius \Rpl{} are available, temporal variations in \Ts{}, \Teq{}, and \Ab{} are also observable for the South Pole and mixed equatorial views (for \Rv{50}, \SNv{20} and \Rv{100} retrievals). Finally, we find that Earth-like seasonal variations in biosignature gas abundances are not detectable with \life{} for all \R{} and \SN{} cases considered.

    In Summary, from the six MIR observables of habitable and inhabited worlds listed in Section~\ref{SubSec: Observable Indicators of Habitability in the MIR}, we are able to constrain four (planetary energy budget, the presence of water and other molecules, the \pt{} structure, and the molecular biosignatures). Regarding the surface conditions, we are able to accurately constrain \Ts{} despite Earth's patchy cloud nature. In contrast, all retrieved \Ps{} estimates are biased. In order to obtain a set of possible planetary surface condition solutions, climate models are required, which is beyond the scope of this work. Finally, we do not manage to detect atmospheric seasonality in biosignature gases, which is the last listed observable of habitable and inhabited worlds.
    
    {Further, by comparing our retrieval results for disk-integrated Earth spectra to the ground truths, we learn that biased parameter estimates will likely be obtained from retrievals on real exoplanet spectra. Importantly, we find that the commonly used simplifying assumptions of cloud-free atmospheres and vertically constant abundance profiles do bias retrieval results. Due to such biases, care needs to be taken when drawing conclusions from retrieval results. Derived quantities, such as abundance ratios, can be less affected by biases while retaining valuable information about the atmospheric state. However, community-wide efforts are required to develop robust and reliable frameworks for exoplanet characterization.}

    Nevertheless, from investigating Earth from afar, we learn that \life{} would correctly identify Earth as a planet where life could thrive, with detectable levels of bioindicators, a temperate climate, and surface conditions that allow for liquid surface water. The journey to characterize Earth-like planets and detect potentially habitable worlds has only started. Our work demonstrates that next generation, optimized space missions can assess whether nearby temperate terrestrial exoplanets are habitable or even inhabited. This provides a promising step forward in our quest to understand distant worlds.
   
\begin{acknowledgements}
{We thank an anonymous referee for the valuable comments.}
This work has been carried out within the framework of the National Center of Competence in Research PlanetS supported by the Swiss National Science Foundation under grants 51NF40\_182901 and 51NF40\_205606. J.N.M, S.P.Q., and R.H. acknowledge the financial support of the SNSF. B.S.K. acknowledges the support of an ETH Zurich Doc.Mobility Fellowship. Author contributions. J.N.M and B.S.K contributed equally to this work. Both carried out analyses, created figures, and wrote essential parts of the manuscript. S.P.Q. initiated the project. S.P.Q. and R.H. guided the project. All authors discussed the results and commented on the manuscript. 
\end{acknowledgements}

\bibliography{Lit.bib}
\bibliographystyle{aasjournal}

\appendix
\restartappendixnumbering
\section{Cloud Fractions for 2017}\label{app:patchy_clouds}

\begin{figure*}[htb!]
    \centering
    \includegraphics[width=0.64\textwidth, trim=1.5cm 2.2cm 0.2cm 3cm, clip]{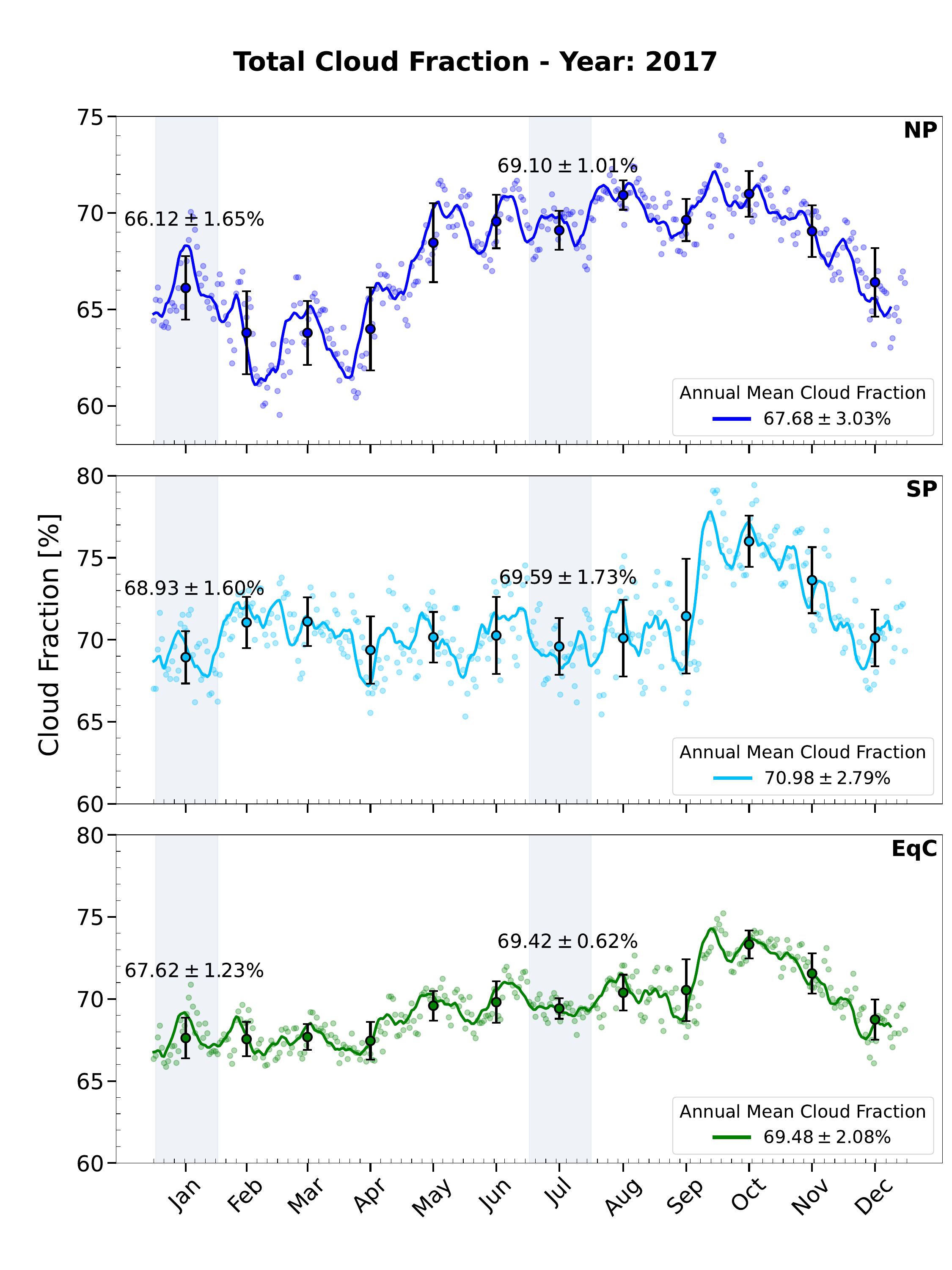}
    \caption{Total Cloud Fractions: This figure illustrates the total cloud fractions for the year 2017 across the investigated viewing geometries in this study. The data are derived from a level 3 satellite product \citep{SSF1DAY_L3}. The scattered points represent daily measurements, while the solid line depicts their rolling average with a window size of 8 days. The central points in the error bar scatter plot represent the monthly mean cloud coverage, while the accompanying error bars indicate the corresponding standard deviation. The shaded areas highlight the months January and July which were investigated in this study. The annotated cloud coverage values signify the monthly cloud fractions for these specific months.}
    \label{fig:Cloud Fraction 2017}
\end{figure*}

In order to compile Figure~\ref{fig:Cloud Fraction 2017}, we have sourced daily level 3 satellite data for the year 2017 from the CERES-Flight Model 3 (FM3) and FM4 instruments on the Aqua platform. Specifically, we have used the CERES Time-Interpolated TOA Fluxes, Clouds and Aerosols Daily Aqua Edition4A (\text{CER\_SSF1deg-Day\_Aqua-MODIS\_Edition4A}) data product \citep{SSF1DAY_L3}. The provided cloud properties are averaged for both day and night (24-hour) and day-only time periods. Furthermore, they are stratified into 4 atmospheric layers (surface-700 hPa, 700 hPa - 500 hPa, 500 hPa - 300 hPa, 300 hPa - 100 hPa) and a total of all layers. For our analysis we have used the latter, mapped the total cloud fractions onto the globe and calculated the disk-averaged value for each viewing geometry per day.

\clearpage
\FloatBarrier
\restartappendixnumbering
\section{Disk-Integrated Atmospheric Profiles Ground Truths for January}\label{app:Profiles_Jan}

\begin{figure*}[ht]
  \centering
  \begin{minipage}[c]{0.39\linewidth}
    \vfill
    \centering
    \includegraphics[width=\textwidth, trim=0cm 0cm 2.5cm 0cm, clip]{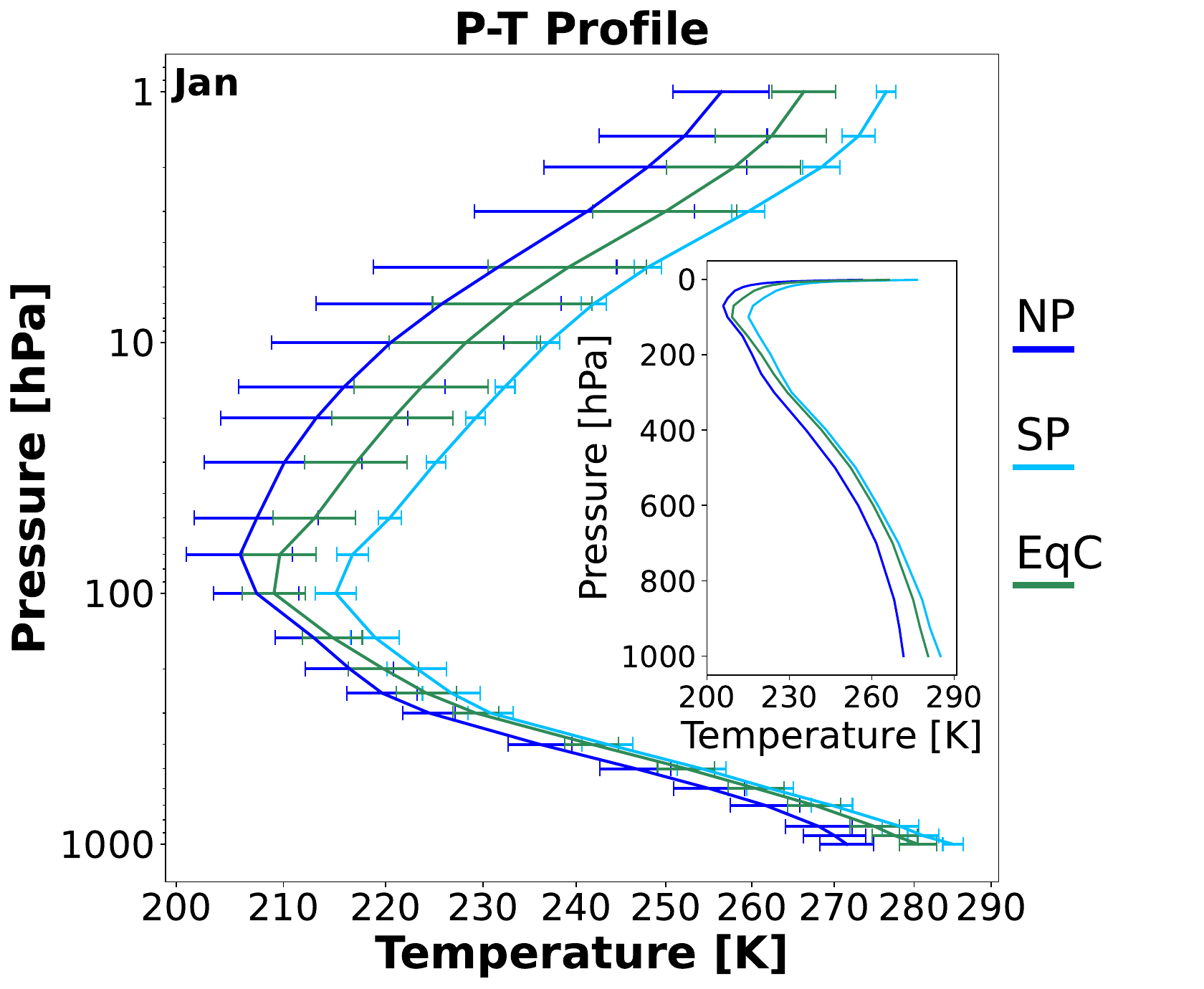}
    \vfill
  \end{minipage}
  \hfill
  \begin{minipage}[c]{0.6\linewidth}
    \centering
    \begin{minipage}[t]{0.48\textwidth}
      \centering
      \includegraphics[width=\textwidth, trim=0cm 0cm 2.5cm 0cm, clip]{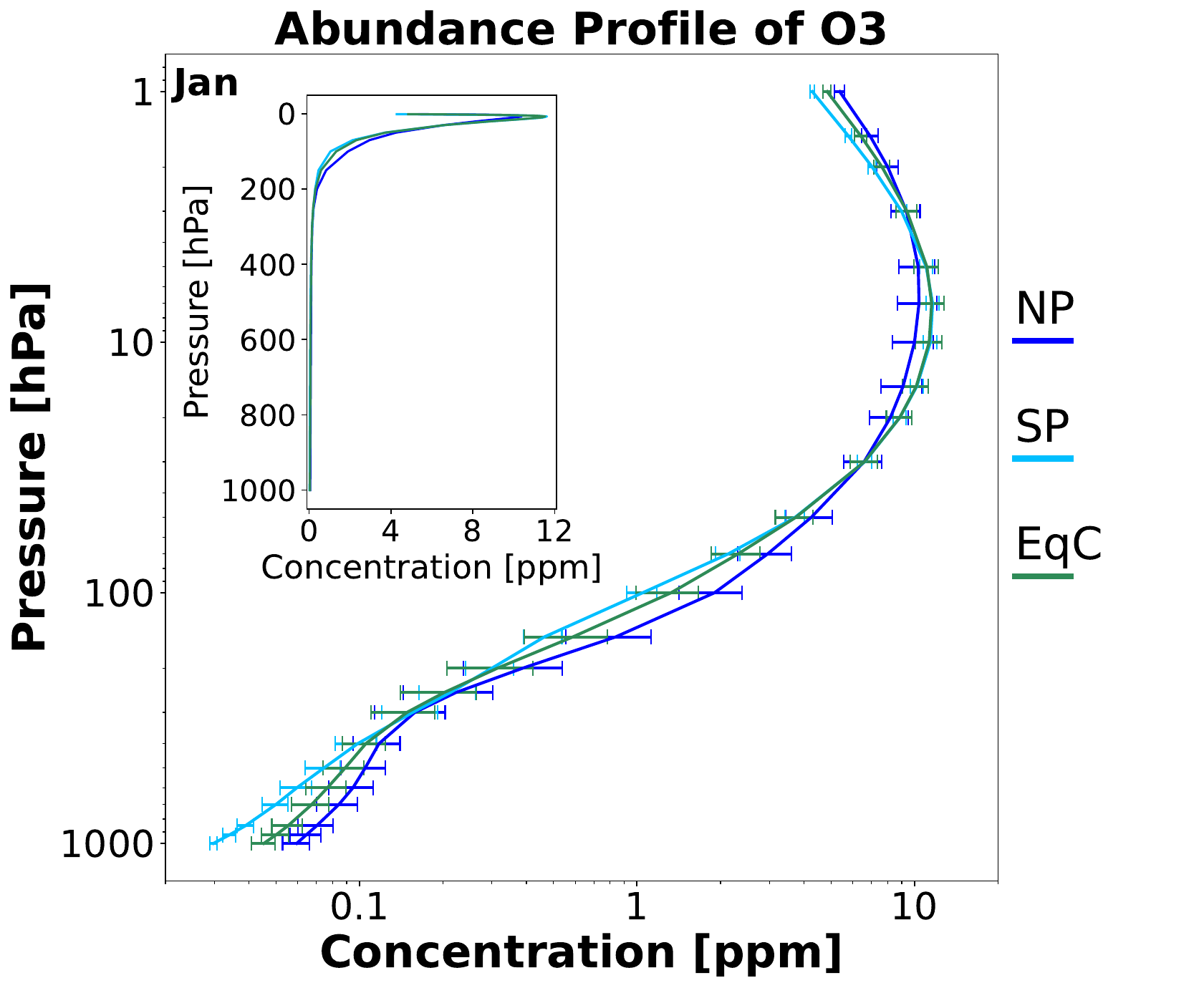}
    \end{minipage}
    \hfill
    \begin{minipage}[t]{0.48\textwidth}
      \centering
      \includegraphics[width=\textwidth, trim=0cm 0cm 2.5cm 0cm, clip]{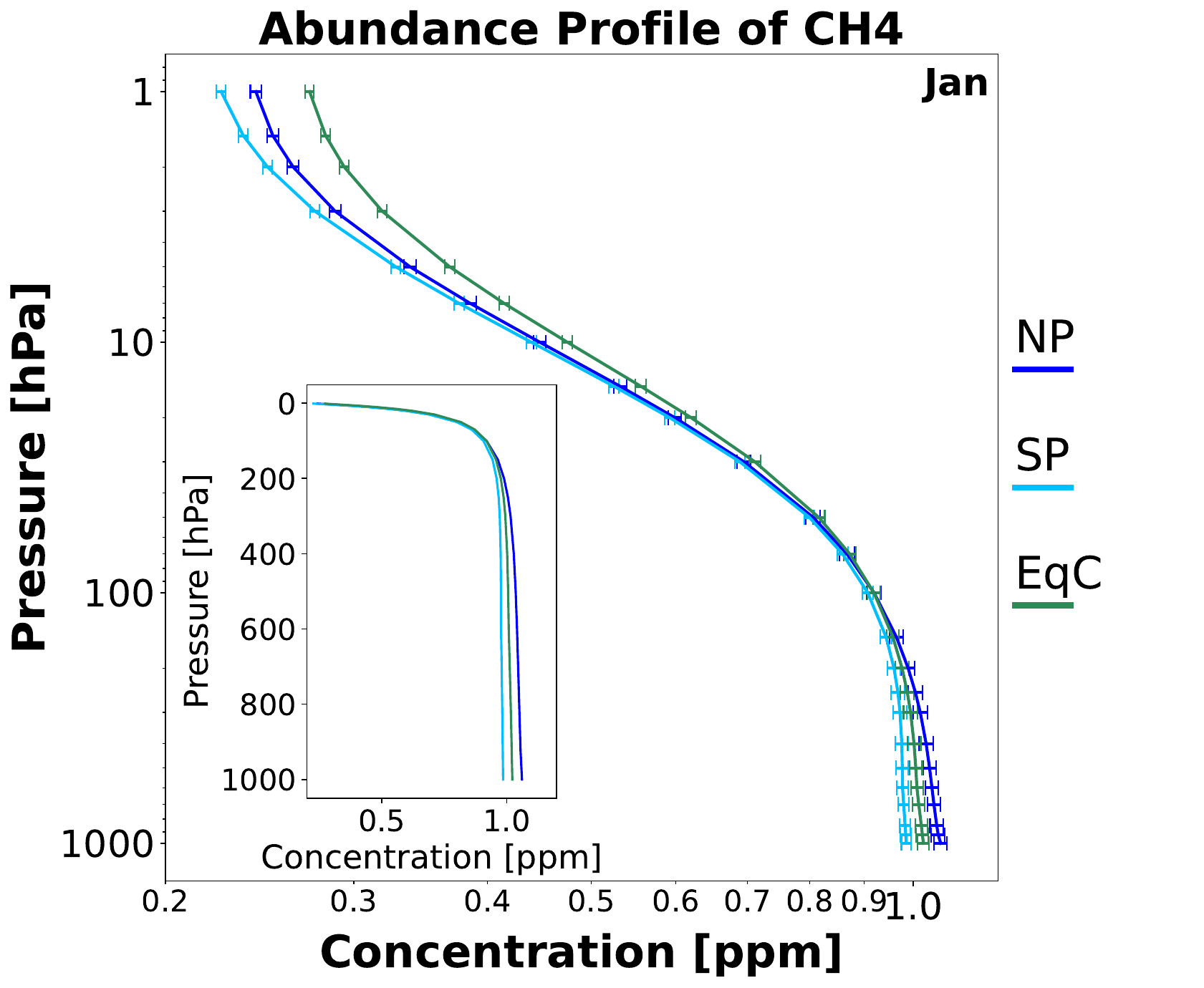}
    \end{minipage}
    \\
    \begin{minipage}[t]{0.48\textwidth}
      \centering
      \includegraphics[width=\textwidth, trim=0cm 0cm 2.5cm 0cm, clip]{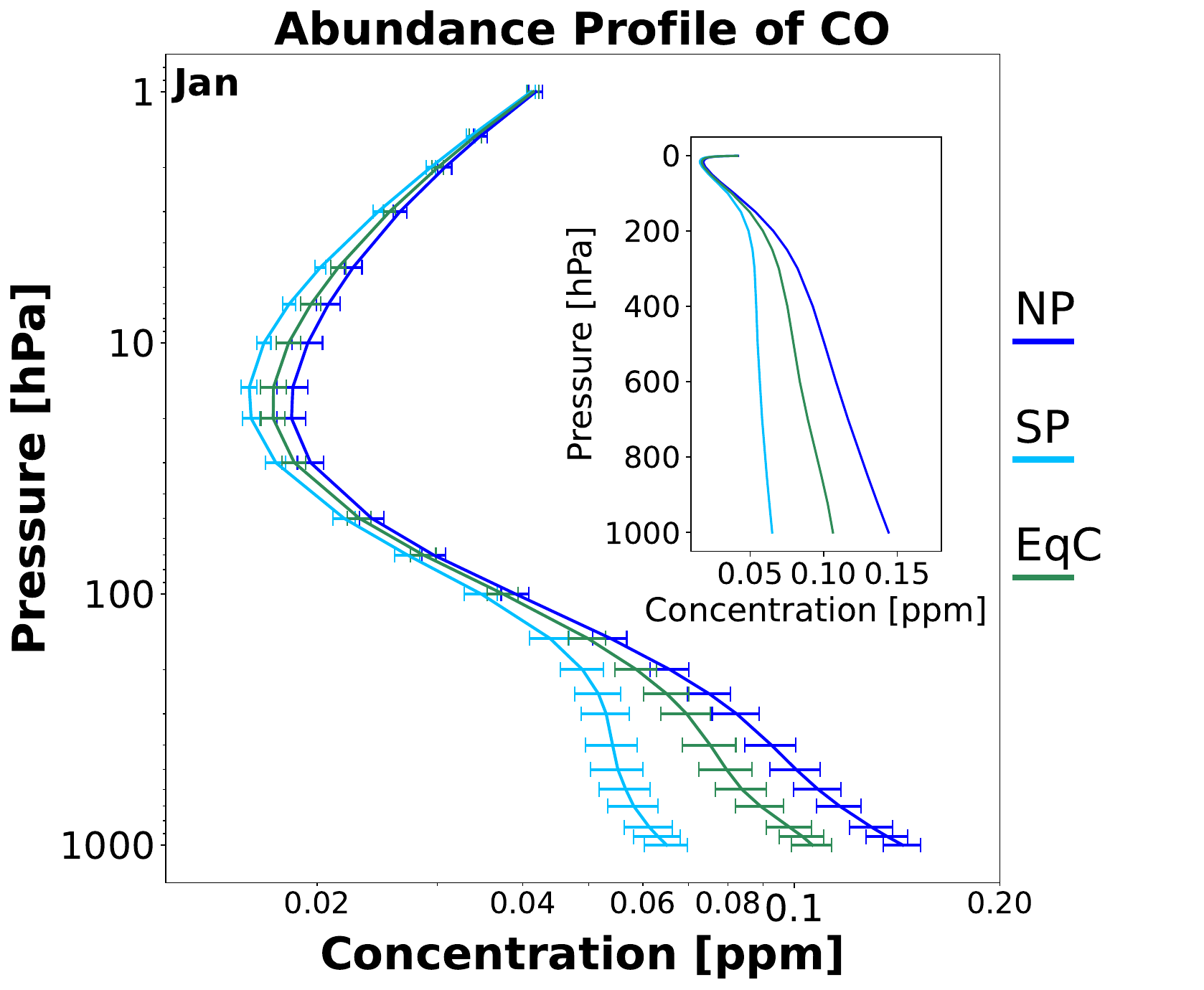}
    \end{minipage}
    \hfill
    \begin{minipage}[t]{0.48\textwidth}
      \centering
      \includegraphics[width=\textwidth, trim=0cm 0cm 2.5cm 0cm, clip]{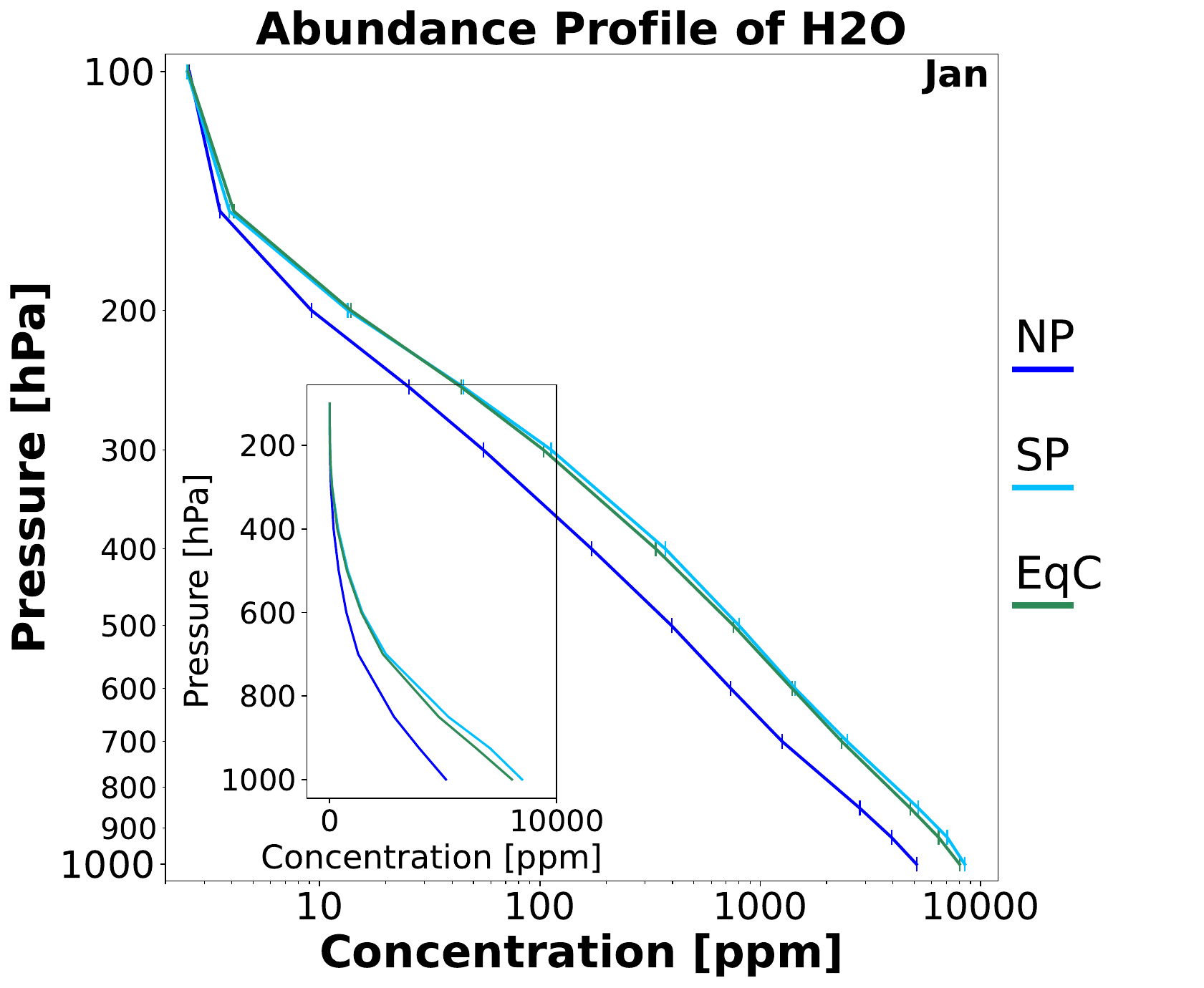}
    \end{minipage}
  \end{minipage}
  \caption{Disk-integrated atmospheric profiles for January: From left to right: \pt{} profile followed by \ce{O3}, \ce{CH4}, \ce{CO}, and \ce{H2O} atmospheric profiles. The different colors correspond to the viewing geometries: NP (blue), SP (turquoise), EqC (green)}
  \label{fig:Profiles_Jan}
\end{figure*}

In Figure \ref{fig:Profiles_Jan}, we show the disk-integrated ground truth profiles for Earth’s \pt{} structure and the abundances of trace gases for January.

\clearpage
\FloatBarrier


\restartappendixnumbering
\section{Retrieval Model Selection}\label{app:model_selection}

We performed a Bayesian model comparison to justify our choice of atmospheric forward model used for the retrieval analysis in this work (see Section~\ref{subsec:atmospheric_models} and Table~\ref{tab:model_parameters}). In our analysis, we ran atmospheric retrievals (using the routine introduced in Section~\ref{sec:retrievals}) assuming the following six atmospheric forward models $\mathcal{M}_\mathrm{i}$ of increasing complexity (see Table~\ref{tab:appendix_different_models} for the full parameter configuration of each model and the assumed priors):

\begin{enumerate}[start=1,label={$\mathcal{M}_\arabic*$:}]
\item (11 parameters) $-$ In addition to the five polynomial \pt{} parameters $a_i$ (see Eq.~\ref{equ:3poly} in Section~\ref{subsec:atmospheric_models}), we retrieve for the planet's radius \Rpl{}, mass \Mpl{}, and surface pressure \Ps{}. The model atmosphere only contains \ce{N2}, \ce{O2}, and \ce{CO2}.
\item (12 parameters) $-$ In addition to the $\mathcal{M}_1$ parameters, we add \ce{H2O} to the species present in the model atmosphere.
\item (13 parameters) $-$ In addition to the $\mathcal{M}_2$ parameters, we add \ce{O3} to the species present in the model atmosphere.
\item (14 parameters) $-$ In addition to the $\mathcal{M}_3$ parameters, we add \ce{CH4} to the species present in the model atmosphere.
\item (15 parameters) $-$ In addition to the $\mathcal{M}_4$ parameters, we add \ce{CO} to the species present in the model atmosphere.
\item (15 parameters) $-$ In addition to the $\mathcal{M}_4$ parameters, we add \ce{N2O} to the species present in the model atmosphere.
\end{enumerate}

Let us consider two retrievals assuming different atmospheric forward models $\mathcal{A}$ and $\mathcal{B}$ on the same disk-integrated Earth spectrum. Both results are characterized by their respective log-evidences $\ln\left(\mathcal{Z_A}\right)$ and $\ln\left(\mathcal{Z_B}\right)$. The Bayes' factor $K$ can be calculated from the evidences as follows:
\begin{equation}\label{eq:BayesFactor}
    \lgrt{K}=\frac{\ln\left(\mathcal{Z}_{\mathcal{A}}\right)-\ln\left(\mathcal{Z}_{\mathcal{B}}\right)}{\ln\left(10\right)}.
\end{equation}
The Bayes' factor $K$ provides a metric that quantifies which out of the two models $\mathcal{A}$ and $\mathcal{B}$ performs better for a given spectrum. The Jeffreys scale \citep[][Table \ref{Table:Jeffrey}]{Jeffreys:Theory_of_prob} provides a possible interpretation for the value of the Bayes factor $K$. A $\lgrt{K}$ value above zero marks a preference for model $\mathcal{A}$, whereas values below zero indicate preference for $\mathcal{B}$.

\begin{figure}
  \centering
    \includegraphics[width=0.99\textwidth]{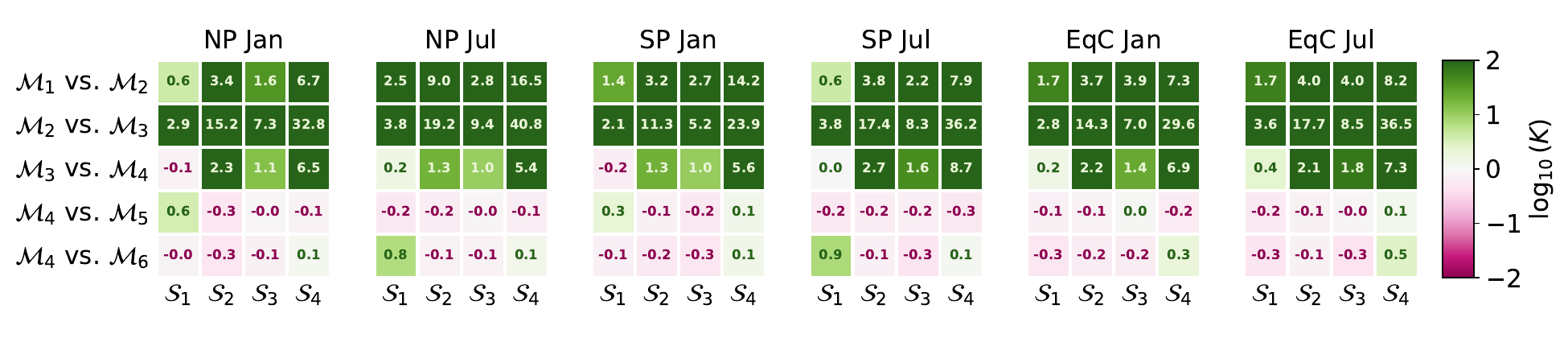}
  \caption{Bayes' factor \lgrt{K} for the comparison of the different models in Appendix~\ref{app:model_selection}. Positive values of \lgrt{K} (green) indicate preference of the model $\mathcal{M}_i$ with the higher $i$ value, while negative values (red) indicate the opposite. The color shading indicates the strength of the preference. The $\mathcal{S}_i$ represent different combinations of \R{} and \SN{} of the input spectra ($\mathcal{S}_1$: \Rv{50}, \SNv{10}; $\mathcal{S}_2$: \Rv{50}, \SNv{20}; $\mathcal{S}_3$: \Rv{100}, \SNv{10}; $\mathcal{S}_4$: \Rv{100}, \SNv{20}). Columns summarize the results obtained for the viewing geometries. From left to right: NP Jan, NP Jul, SP Jan, Sp Jul, EqC Jan, and EqC Jul.}
  \label{fig:Pmodel_comparison}
\end{figure}

Figure~\ref{fig:Pmodel_comparison} summarizes the results from our model comparison efforts for all considered disk-integrated Earth spectra (viewing geometries, \R{}, and \SN{}). The $\mathcal{M}_i$ correspond to the models listed above, while the $\mathcal{S}_i$ represent different combinations of \R{} and \SN{} of the input spectra ($\mathcal{S}_1$: \Rv{50}, \SNv{10}; $\mathcal{S}_2$: \Rv{50}, \SNv{20}; $\mathcal{S}_3$: \Rv{100}, \SNv{10}; $\mathcal{S}_4$: \Rv{100}, \SNv{20}). Green squares indicate positive \lgrt{K} values and preference of the $\mathcal{M}_i$ with the high $i$, while red squares represent negative \lgrt{K} values and preference of the low $i$ $\mathcal{M}_i$. The color shading indicates the strength of the preference.

We observe that $\mathcal{M}_3$ is generally preferred over $\mathcal{M}_2$ and $\mathcal{M}_1$ for all considered spectra. Thus, \ce{H2O} and \ce{O3} are confidently detectable with \life{}. Further, $\mathcal{M}_4$ is preferred over $\mathcal{M}_3$ for all but the \Rv{50}, \SNv{10} cases, suggesting that also \ce{CH4} is detectable. In contrast, the \lgrt{K} value of roughly $0$ indicates that models $\mathcal{M}_5$ and $\mathcal{M}_6$ perform similarly well as model $\mathcal{M}_4$. However, since $\mathcal{M}_5$ and $\mathcal{M}_6$ each require one additional parameter (abundance of \ce{CO} or \ce{N2O}, respectively), we prefer model $\mathcal{M}_4$. This indicates that neither \ce{CO} nor \ce{N2O} are detectable in Earth's atmosphere at the \R{} and \SN{} considered here, which is in agreement with the findings in \cite{LIFE_III}. In conclusion, $\mathcal{M}_4$ shows the best performance of all models considered. Therefore, we used $\mathcal{M}_4$ as forward model in the retrieval analyses presented in main part of this manuscript.

\begin{deluxetable}{lcccccccc}[ht!]
\tablecaption{Parameter configurations of the nine tested retrieval forward models.}
\label{tab:appendix_different_models} 
\tablehead{
\colhead{\multirow{2}{*}{Parameter}} &\colhead{\multirow{2}{*}{Description}} &\colhead{\multirow{2}{*}{Prior}} &\multicolumn{6}{c}{Parameter Configuration}
\\
\cline{4-9}
&&&\colhead{$\mathcal{M}_1$} &\colhead{$\mathcal{M}_2$} &\colhead{$\mathcal{M}_3$} &\colhead{$\mathcal{M}_4$} &\colhead{$\mathcal{M}_5$} &\colhead{$\mathcal{M}_6$}
}
\startdata 
 $a_4$               &\pt{} parameter (degree 4)                            &$\mathcal{U}(0,10)$ &\modelparam{} &\modelparam{} &\modelparam{} &\modelparam{} &\modelparam{} &\modelparam{}\\
 $a_3$               &\pt{} parameter (degree 3)                            &$\mathcal{U}(0,100)$    
&\modelparam{} &\modelparam{} &\modelparam{} &\modelparam{} &\modelparam{} &\modelparam{}\\
 $a_2$               &\pt{} parameter (degree 2)                            &$\mathcal{U}(0,500)$
        &\modelparam{} &\modelparam{} &\modelparam{} &\modelparam{} &\modelparam{} &\modelparam{}\\
 $a_1$               &\pt{} parameter (degree 1)                            &$\mathcal{U}(0,500)$    
        &\modelparam{} &\modelparam{} &\modelparam{} &\modelparam{} &\modelparam{} &\modelparam{}\\
 $a_0$               &\pt{} parameter (degree 0)                            &$\mathcal{U}(0,1000)$  
        &\modelparam{} &\modelparam{} &\modelparam{} &\modelparam{} &\modelparam{} &\modelparam{}\\
 $\lgrt{\Ps{}}$      &$\lgrt{\textrm{Surface pressure }\left[\mathrm{bar}\right]}$          &$\mathcal{U}(-4,2)$    
        &\modelparam{} &\modelparam{} &\modelparam{} &\modelparam{} &\modelparam{} &\modelparam{}\\
 $\Rpl{}$            &Planet radius $\left[R_\oplus\right]$                 &$\mathcal{G}(1.0,0.2)$ 
        &\modelparam{} &\modelparam{} &\modelparam{} &\modelparam{} &\modelparam{} &\modelparam{}\\
 $\lgrt{\Mpl{}}$     &$\lgrt{\textrm{Planet mass } \left[M_\oplus\right]}$                   &$\mathcal{G}(0.0,0.4)$ 
        &\modelparam{} &\modelparam{} &\modelparam{} &\modelparam{} &\modelparam{} &\modelparam{} \\
 $\lgrt{\ce{N2}}$    &$\lgrt{\textrm{\ce{N2} mass fraction}}$             &$\mathcal{U}(-10,0)$
        &\modelparam{} &\modelparam{} &\modelparam{} &\modelparam{} &\modelparam{} &\modelparam{}\\
 $\lgrt{\ce{O2}}$    &$\lgrt{\textrm{\ce{O2} mass fraction}}$             &$\mathcal{U}(-10,0)$
        &\modelparam{} &\modelparam{} &\modelparam{} &\modelparam{} &\modelparam{} &\modelparam{}\\
 $\lgrt{\ce{CO2}}$   &$\lgrt{\textrm{\ce{CO2} mass fraction}}$            &$\mathcal{U}(-10,0)$
        &\modelparam{} &\modelparam{} &\modelparam{} &\modelparam{} &\modelparam{} &\modelparam{}\\
 $\lgrt{\ce{H2O}}$   &$\lgrt{\textrm{\ce{H2O} mass fraction}}$             &$\mathcal{U}(-10,0)$
        &\notparam{}   &\modelparam{} &\modelparam{} &\modelparam{} &\modelparam{} &\modelparam{}\\
 $\lgrt{\ce{O3}}$    &$\lgrt{\textrm{\ce{O3} mass fraction}}$            &$\mathcal{U}(-10,0)$
        &\notparam{} &\notparam{} &\modelparam{} &\modelparam{} &\modelparam{} &\modelparam{}\\
 $\lgrt{\ce{CH4}}$   &$\lgrt{\textrm{\ce{CH4} mass fraction}}$             &$\mathcal{U}(-10,0)$
        &\notparam{} &\notparam{} &\notparam{} &\modelparam{} &\modelparam{} &\modelparam{}\\
 $\lgrt{\ce{CO}}$    &$\lgrt{\textrm{\ce{CO} mass fraction}}$          &$\mathcal{U}(-10,0)$
        &\notparam{} &\notparam{} &\notparam{} &\notparam{} &\modelparam{} &\notparam{}\\
 $\lgrt{\ce{N2O}}$   &$\lgrt{\textrm{\ce{N2O} mass fraction}}$             &$\mathcal{U}(-10,0)$
        &\notparam{} &\notparam{} &\notparam{} &\notparam{} &\notparam{} &\modelparam{}\\
\enddata
\tablecomments{In the third column we specify the priors assumed in the retrievals. We denote a boxcar prior with lower threshold $x$ and upper threshold $y$ as $\mathcal{U}(x,y)$; For a Gaussian prior with mean $\mu$ and standard deviation $\sigma$, we write $\mathcal{G}(\mu,\sigma)$. The last nine columns summarize the model parameters used by each of the different forward models tested in the retrievals (\modelparam{}~$=$~used, \notparam{}~$=$~unused).}
\end{deluxetable}
\begin{table}[ht!]
\caption{Jeffreys scale \citep{Jeffreys:Theory_of_prob}.}
\label{Table:Jeffrey}      
\centering                          
\begin{tabular}{l c c}        
\hline\hline                 
$\lgrt{K}$ &Probability &Strength of Evidence\\    
\hline 
   $<0$     &$<0.5$         &Support for $\mathcal{B}$\\
   $0-0.5$  &$0.5-0.75$     &Very weak support for $\mathcal{A}$\\
   $0.5-1$  &$0.75-0.91$    &Substantial support for $\mathcal{A}$\\
   $1-2$    &$0.91-0.99$    &Strong support for $\mathcal{A}$\\
   $>2$     &$>0.99$        &Decisive support for $\mathcal{A}$\\ 
\hline 
\end{tabular}
\tablecomments{Scale for interpretation of the Bayes' factor $K$ for two models $\mathcal{A}$ and $\mathcal{B}$. The scale is symmetrical, i.e., negative values of $\lgrt{K}$ correspond to very weak, substantial, strong, or decisive support for model $\mathcal{B}$.}
\end{table}

\clearpage
\FloatBarrier
\restartappendixnumbering
\section{Calculation of the Equilibrium Temperature and Bond albedo}\label{app:Bond_albedo}

The equilibrium temperature \Teq{} and the Bond albedo \Ab{} are not directly determined in our atmospheric retrievals. However, both parameters provide important information about the energy budget of Earth. In the following, we summarize how we derive estimates for \Teq{} and \Ab{} from the retrieved parameter posteriors.

To determine \Teq{}, we first calculate the MIR spectra corresponding to the retrieved parameter posteriors over a wide wavelength range. For each spectrum, we then integrate the flux to estimate the total emitted flux and use the Stefan-Boltzmann law to compute the effective temperature \Teff{} of a black-body with the same flux, which corresponds to the \Teq{} of the planet.  From the resulting \Teq{} distribution, we can deduce the planetary \Ab{} distribution using:
\begin{equation}\label{eq:bond_albedo}
    \Ab = 1 - 16\,\pi\sigma \frac{a^2_P\Teq^4}{L_*}.
\end{equation}
Here, $\sigma$ is the Stefan–Boltzmann constant, $a_P$ is the semi-major axis of the planet orbit around its star, and $L_*$ is the luminosity of the star. To calculate \Ab{}, we assume that $a_P$ and $L_*$ to be known with an accuracy of $\pm$1\% (i.e., for an exo-Earth, $a_P=1.00\pm0.01$~AU, $L_*=1.00\pm0.01\,\,L_\odot$ with the solar luminosity $L_\odot$). For each value in the \Teq{} distribution, we randomly draw an $a_P$ and $L_*$ value from two uncorrelated normal distributions and calculate the corresponding \Ab{} value. This yields the distribution for the planetary Bond albedo \Ab{}.

\clearpage
\FloatBarrier
\restartappendixnumbering
\section{Supplementary Data from the Retrievals}\label{app:additional_retrieval_data}

\begin{figure*}
  \centering
    \includegraphics[width=0.32\textwidth]{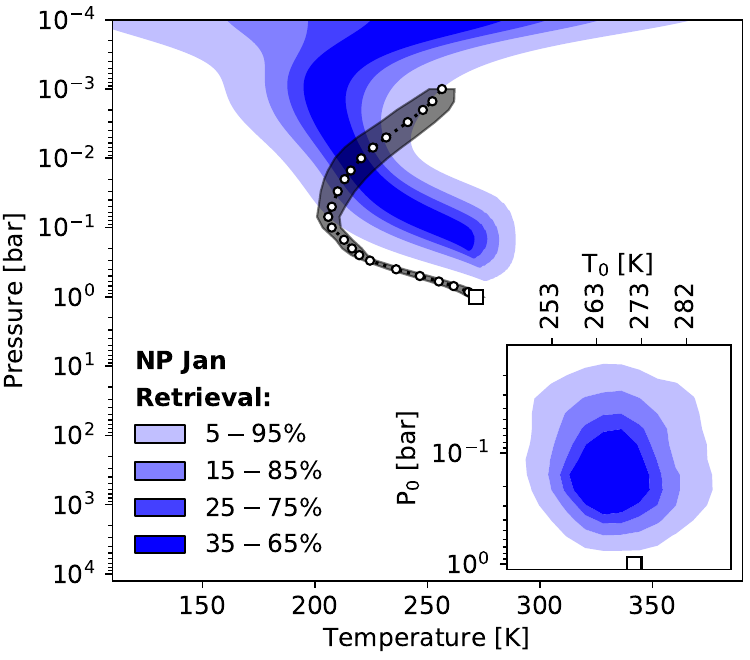}\quad
    \includegraphics[width=0.32\textwidth]{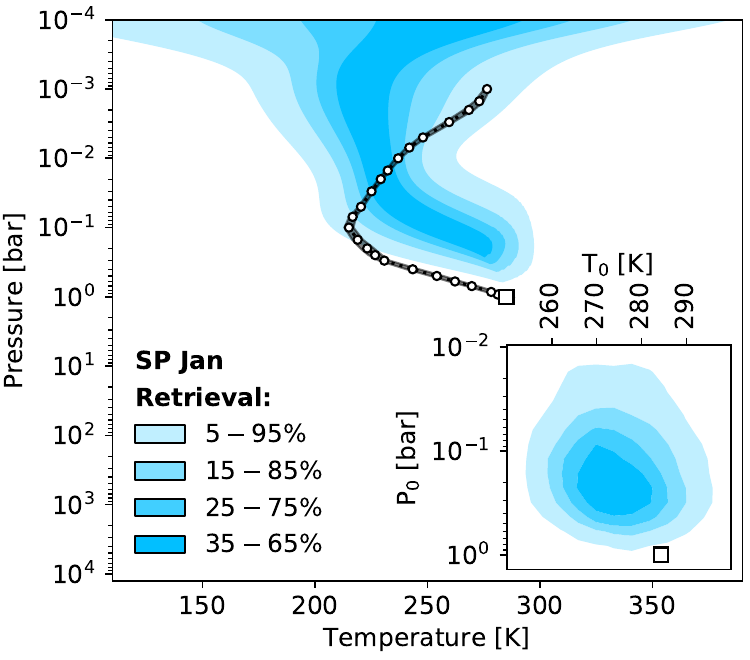}\quad
    \includegraphics[width=0.32\textwidth]{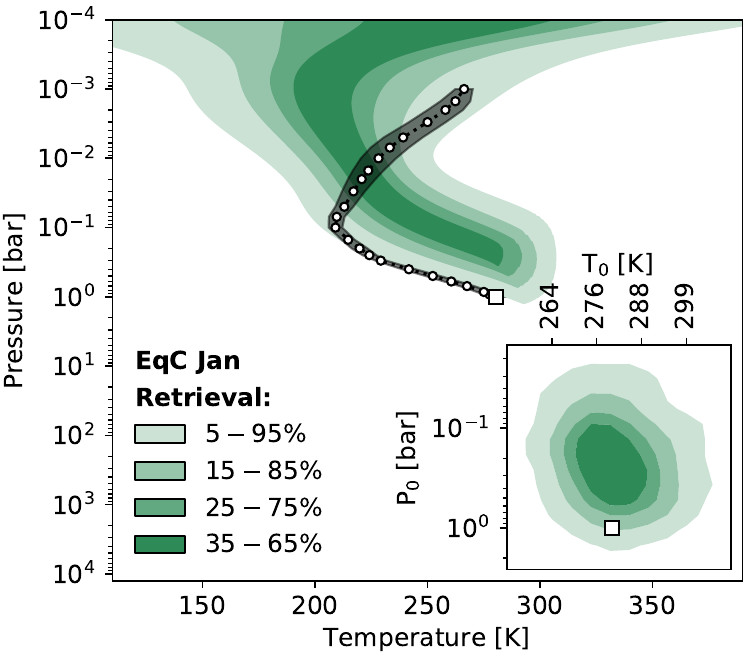}\\
    \includegraphics[width=0.32\textwidth]{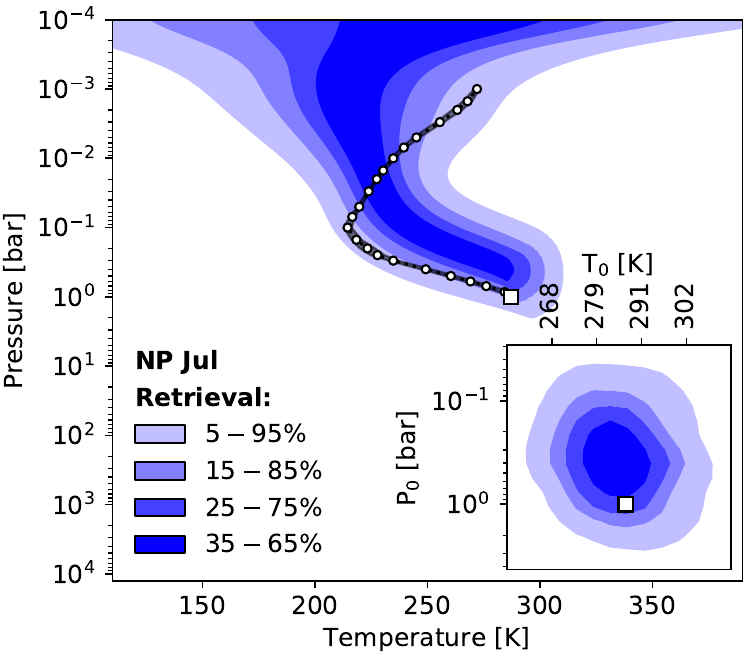}\quad
    \includegraphics[width=0.32\textwidth]{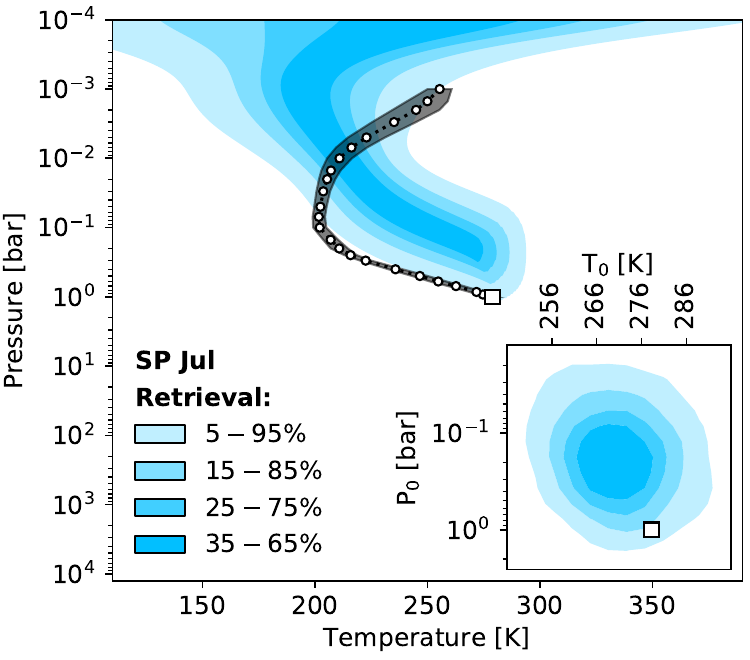}\quad
    \includegraphics[width=0.32\textwidth]{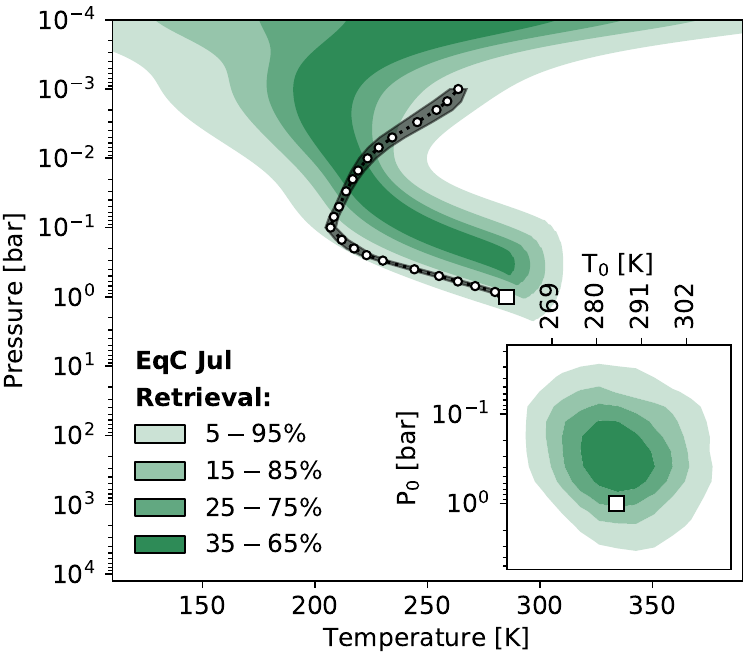} 
  \caption{\pt{} profiles retrieved for the six disk-integrated, \Rv{50} and \lifesim{} \SNv{10} Earth spectra. The color-shaded areas indicate percentiles of the retrieved \pt{} profiles. The white square marker shows the true surface pressure \Ps{} and temperature \Ts{}, the white circular markers show the true \pt{} structure, and the gray shaded area indicates the uncertainty thereon. In the bottom right of each panel, we plot the 2D \Ps{}-\Ts{} posterior, to visualize the constraints on the retrieved surface conditions. Each panel shows the result for one viewing angle. From top-left to bottom-right: NP Jan, SP Jan, EqC Jan, NP Jul, SP Jul, and EqC Jul.}
  \label{figapp:ret_PT_profiles_r50SN10}
\end{figure*}

\begin{figure}
  \centering
    \includegraphics[width=0.32\textwidth]{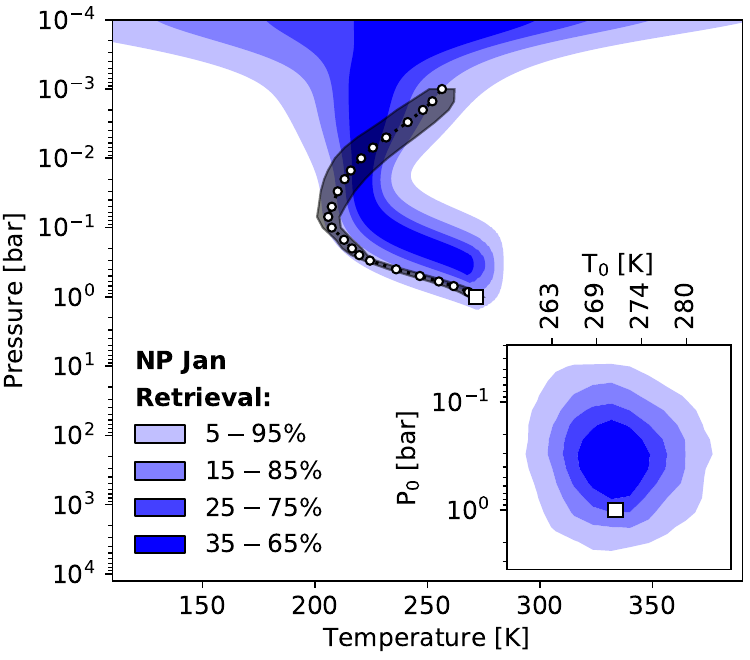}\quad
    \includegraphics[width=0.32\textwidth]{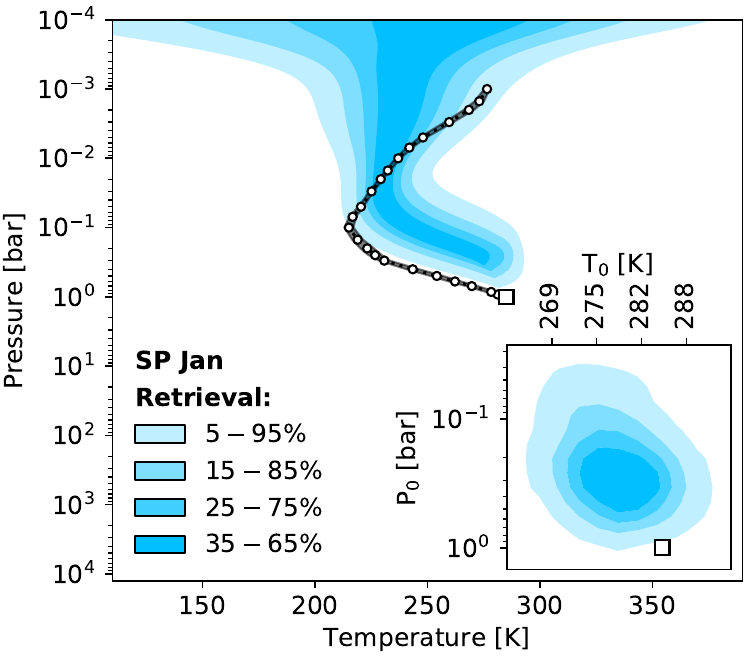}\quad
    \includegraphics[width=0.32\textwidth]{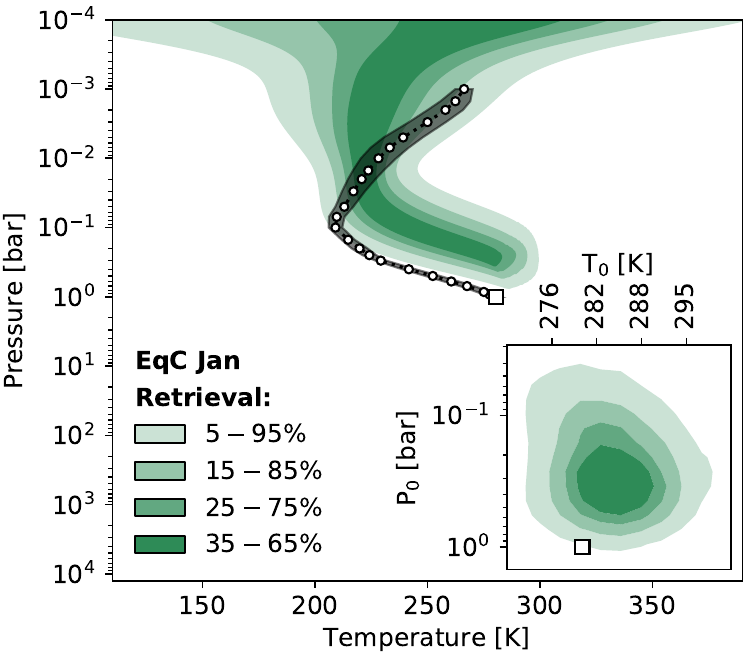}\\
    \includegraphics[width=0.32\textwidth]{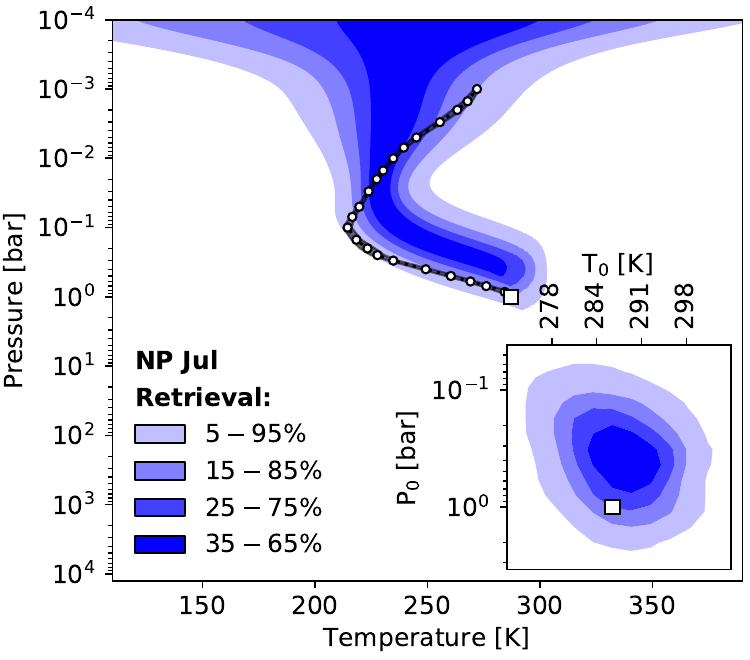}\quad
    \includegraphics[width=0.32\textwidth]{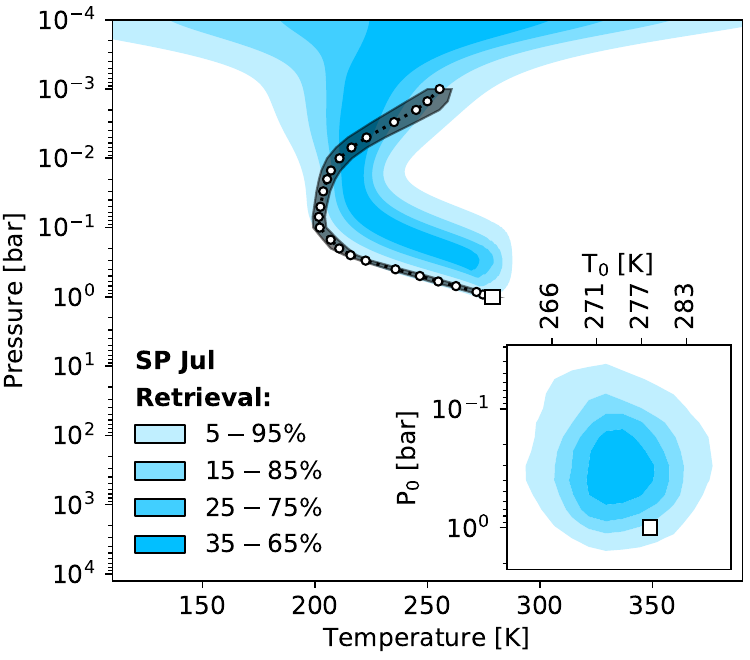}\quad
    \includegraphics[width=0.32\textwidth]{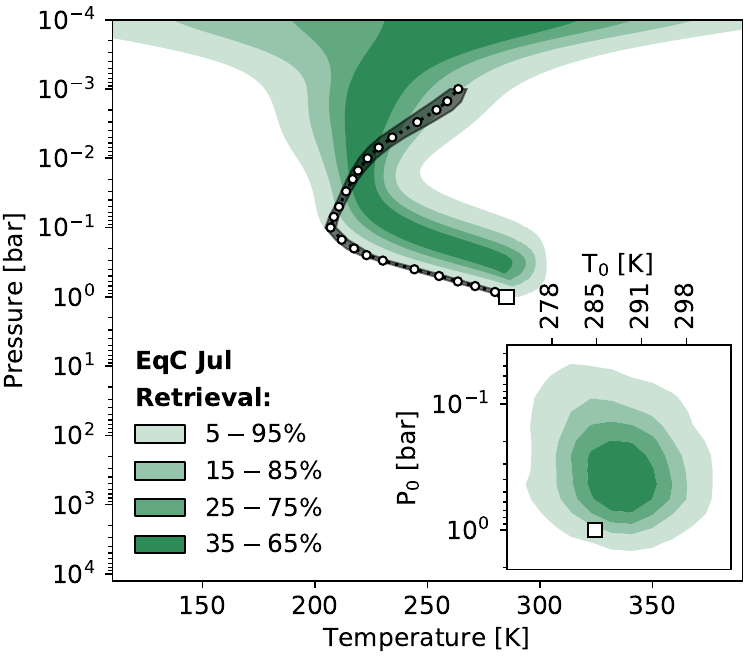} 
  \caption{As for Figure \ref{figapp:ret_PT_profiles_r50SN10}, but for the \Rv{50} and \lifesim{} \SNv{20} Earth spectra.}
  \label{figapp:ret_PT_profiles_r50SN20}
\end{figure}

\begin{figure}
  \centering
    \includegraphics[width=0.32\textwidth]{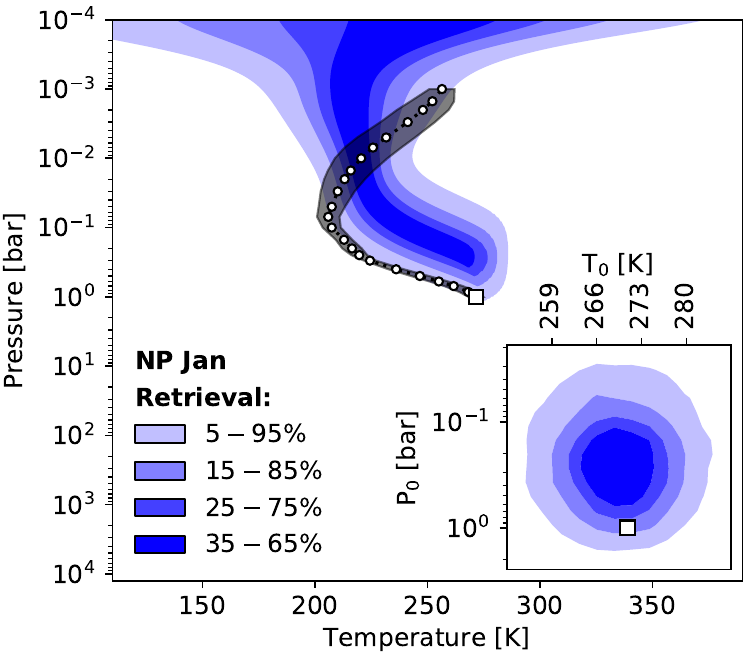}\quad
    \includegraphics[width=0.32\textwidth]{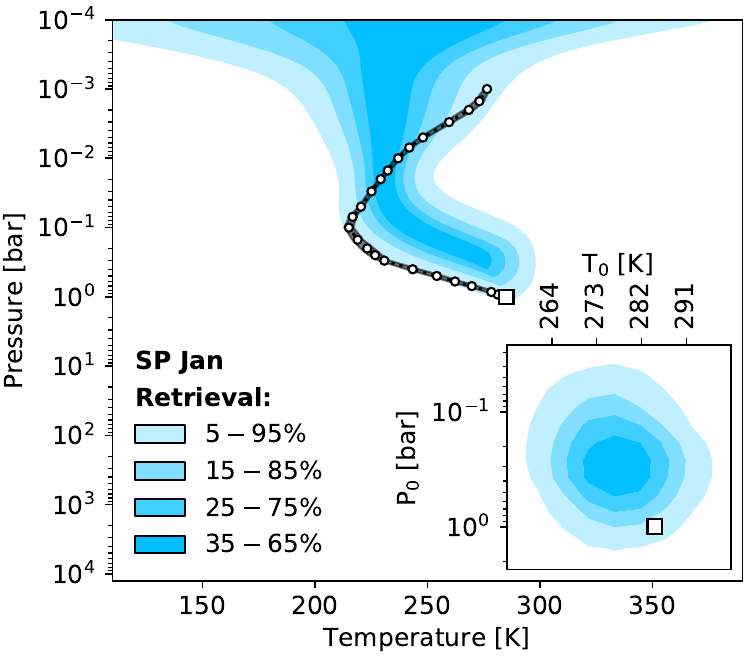}\quad
    \includegraphics[width=0.32\textwidth]{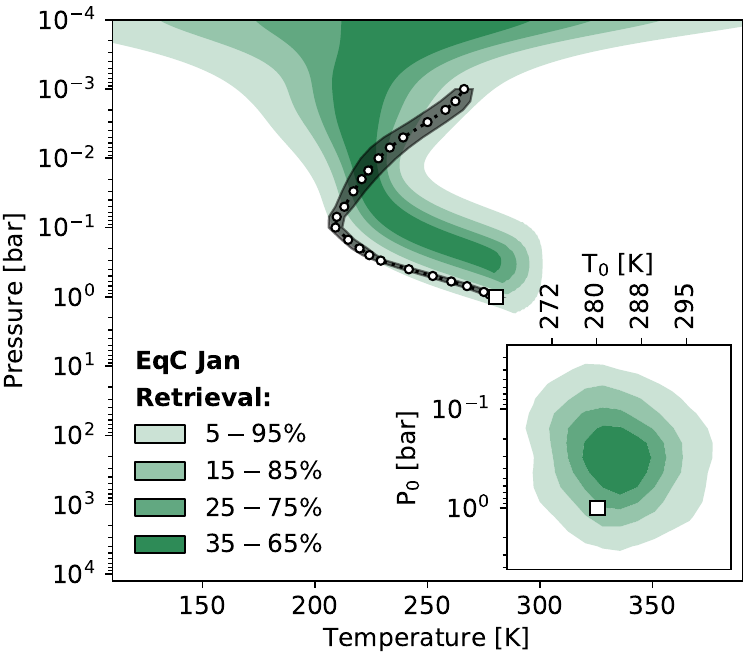}\\
    \includegraphics[width=0.32\textwidth]{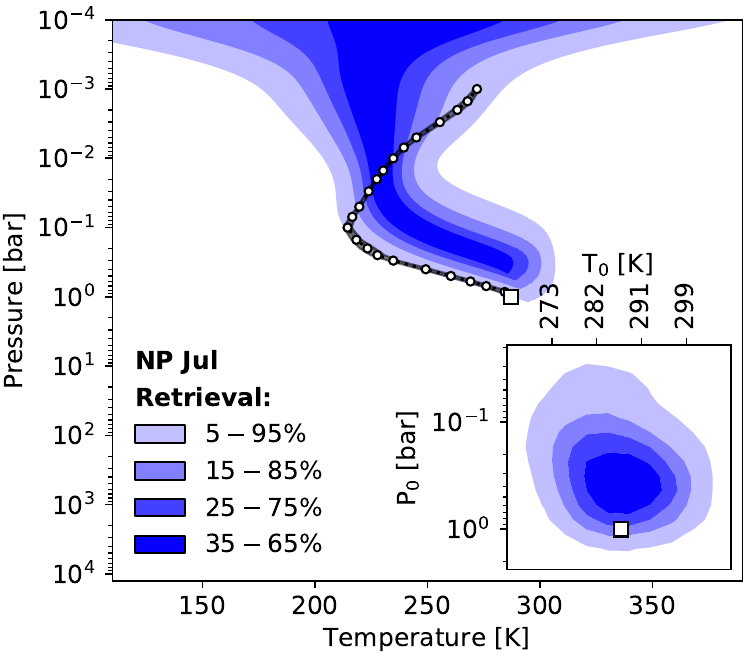}\quad
    \includegraphics[width=0.32\textwidth]{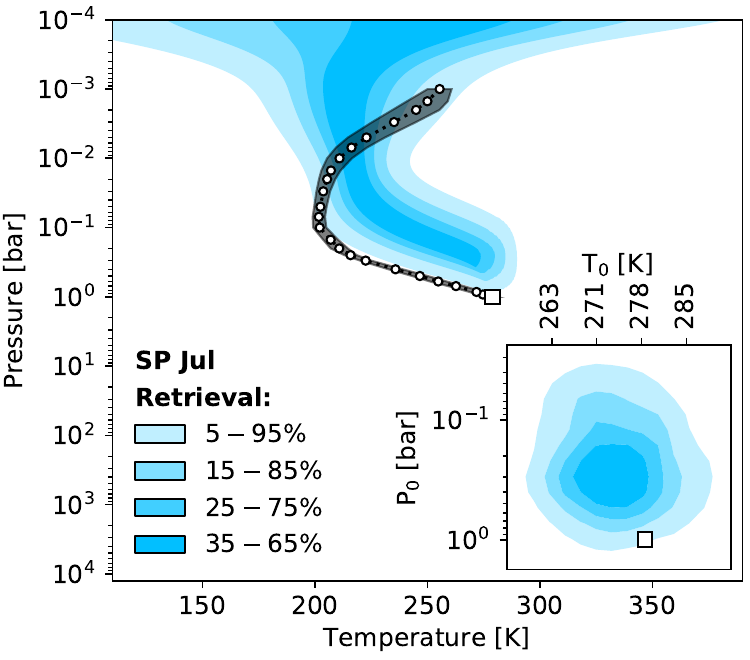}\quad
    \includegraphics[width=0.32\textwidth]{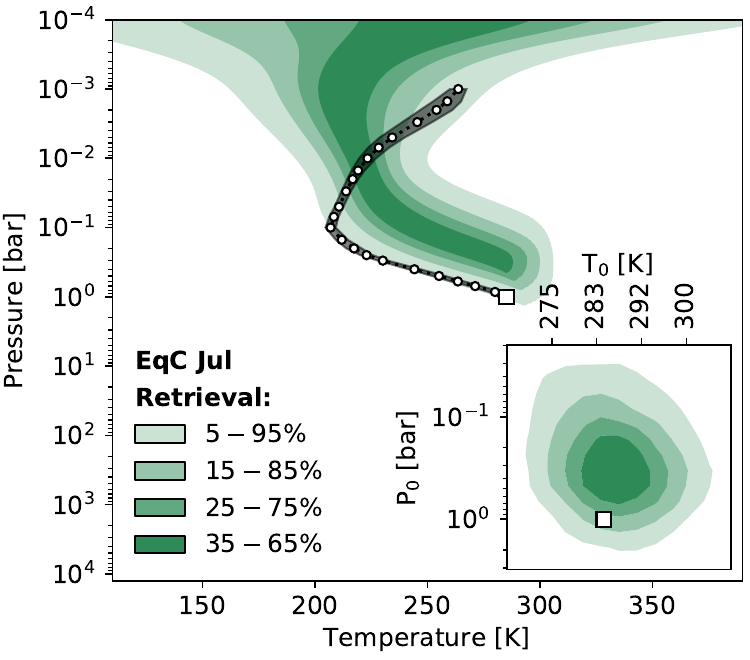} 
  \caption{As for Figure \ref{figapp:ret_PT_profiles_r50SN10}, but for the \Rv{100} and \lifesim{} \SNv{10} Earth spectra.}
  \label{figapp:ret_PT_profiles_r100SN10}
\end{figure}

\begin{figure}
  \centering
    \includegraphics[width=0.32\textwidth]{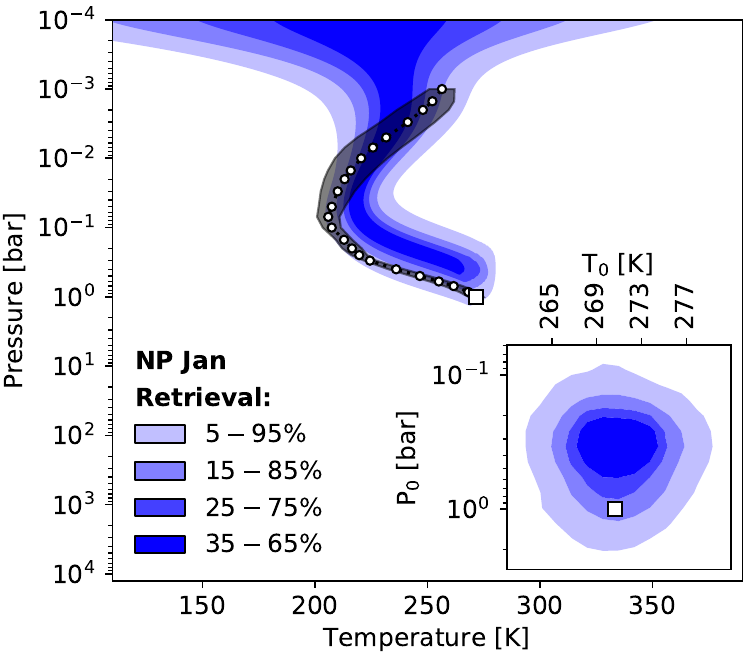}\quad
    \includegraphics[width=0.32\textwidth]{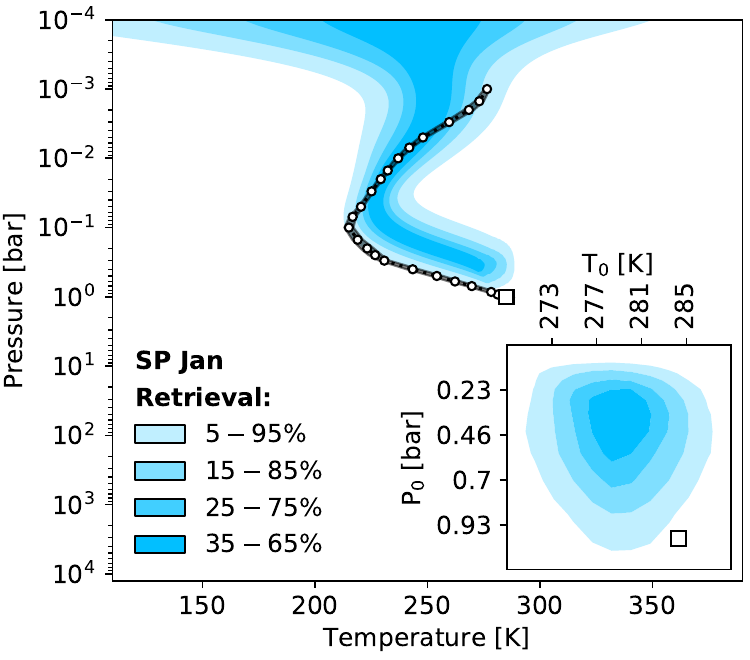}\quad
    \includegraphics[width=0.32\textwidth]{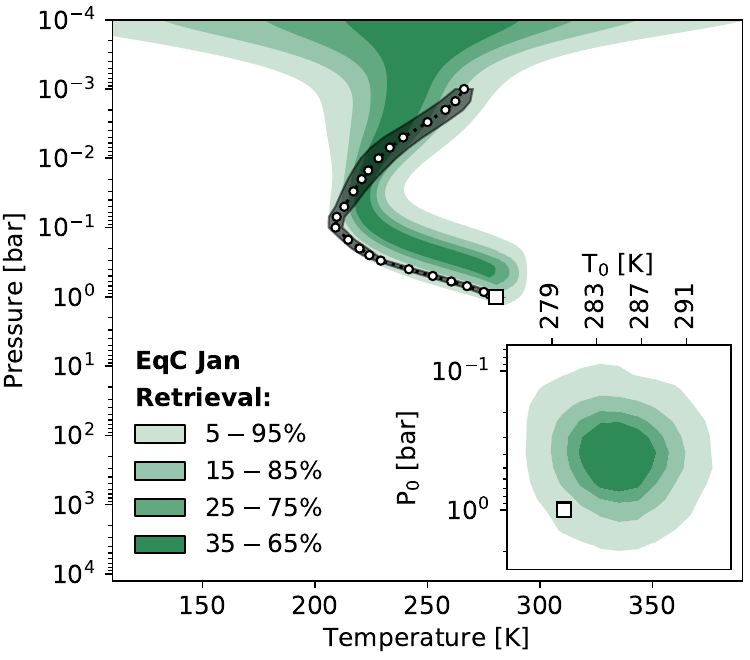}\\
    \includegraphics[width=0.32\textwidth]{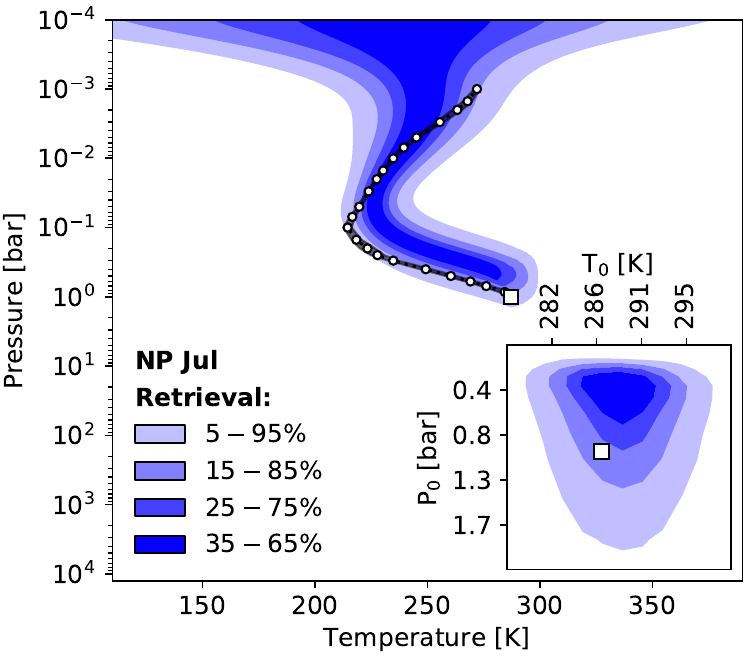}\quad
    \includegraphics[width=0.32\textwidth]{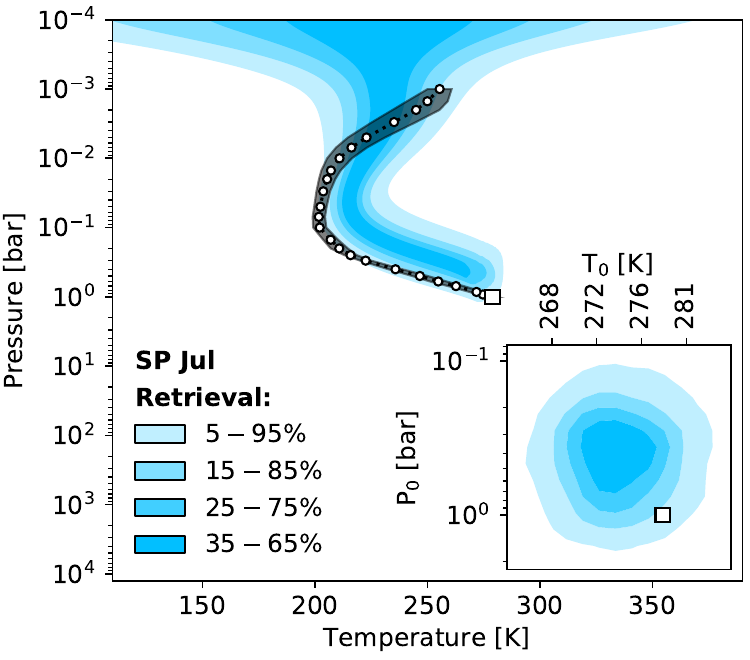}\quad
    \includegraphics[width=0.32\textwidth]{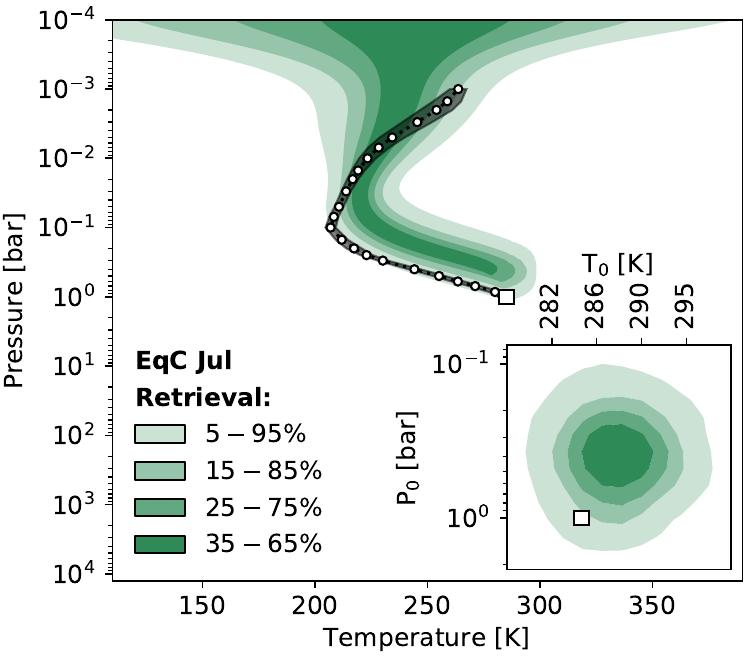} 
  \caption{As for Figure \ref{figapp:ret_PT_profiles_r50SN10}, but for the \Rv{100} and \lifesim{} \SNv{20} Earth spectra.}
  \label{figapp:ret_PT_profiles_r100SN20}
\end{figure}

In Figures~\ref{figapp:ret_PT_profiles_r50SN10} to \ref{figapp:ret_PT_profiles_r100SN20}, we provide the retrieved \pt{} structures for all viewing geometries, seasons, \R{}, and \SN{}:
\begin{itemize}
    \item Figure \ref{figapp:ret_PT_profiles_r50SN10} $-$ Retrieved \pt{} structures for \Rv{50}, \SNv{10},
    \item Figure \ref{figapp:ret_PT_profiles_r50SN20} $-$ Retrieved \pt{} structures for \Rv{50}, \SNv{20},
    \item Figure \ref{figapp:ret_PT_profiles_r100SN10} $-$ Retrieved \pt{} structures for \Rv{100}, \SNv{10},
    \item Figure \ref{figapp:ret_PT_profiles_r100SN20} $-$ Retrieved \pt{} structures for \Rv{100}, \SNv{20}.
\end{itemize}

In Tables~\ref{tab:Rel_Abund_Values_NP} to \ref{tab:Rel_Abund_Values_EqC}, we provide the numerical values corresponding to Figures \ref{fig:PosteriorsM4}, \ref{fig:Posteriors_Rel_Abund}, and \ref{fig:Reduced Posteriors} for the different viewing angles:
\begin{itemize}
    \item Table \ref{tab:Rel_Abund_Values_NP} $-$ NP viewing angle posteriors,
    \item Table \ref{tab:Rel_Abund_Values_SP} $-$ SP viewing angle posteriors,
    \item Table \ref{tab:Rel_Abund_Values_EqC} $-$ EqC viewing angle posteriors.
\end{itemize}

\movetabledown=50mm
\begin{rotatetable}
\begin{deluxetable}{cc|cccc|c|c|c|c|cccc|c|c|c|c|}
\tablecaption{Molecular line opacities and continuum opacities used in the retrievals.}
\tablecaption{Numeric values corresponding to the retrieved parameter posteriors in Figures~\ref{fig:PosteriorsM4}, \ref{fig:Reduced Posteriors}, and \ref{fig:Posteriors_Rel_Abund} for the NP viewing angle.}
\tablehead{
&&\multicolumn{4}{c}{NP Jan $-$ Ground Truths}\vline&\multicolumn{4}{c}{NP Jan $-$ Posteriors}\vline&\multicolumn{4}{c}{NP Jul $-$ Ground Truths}\vline&\multicolumn{4}{c}{NP Jul $-$ Posteriors}\vline\\
&&\multicolumn{4}{c}{Pressure Levels [bar]}\vline&\multicolumn{2}{c}{\Rv{50}} &\multicolumn{2}{c}{\Rv{100}} \vline&\multicolumn{4}{c}{Pressure Levels [bar]}\vline &\multicolumn{2}{c}{\Rv{50}} &\multicolumn{2}{c}{\Rv{100}}\vline\\
&\colhead{Posterior}\vline&\colhead{1}&\colhead{$10^{-1}$}&\colhead{$10^{-2}$}&\colhead{$10^{-3}$}\vline&\colhead{\SNv{10}}   &\colhead{\SNv{20}}\vline   &\colhead{\SNv{10}}   &\colhead{\SNv{20}}\vline&\colhead{1}&\colhead{$10^{-1}$}&\colhead{$10^{-2}$}&\colhead{$10^{-3}$}\vline &\colhead{\SNv{10}}   &\colhead{\SNv{20}}\vline   &\colhead{\SNv{10}}   &\colhead{\SNv{20}} \vline
}
\startdata
\multirow{10}{*}{\rotatebox[origin=c]{90}{Figure~\ref{fig:PosteriorsM4}}}&$L($$\mathrm{P_{0} [bar]}$$)$ &\multicolumn{4}{c|}{$0.0$}   &$-0.9^{+0.3}_{-0.5}$  &$-0.5^{+0.4}_{-0.4}$  &$-0.6^{+0.4}_{-0.4}$  &$-0.4^{+0.3}_{-0.3}$ &\multicolumn{4}{c|}{$0.0$} &$-0.4^{+0.4}_{-0.4}$  &$-0.4^{+0.4}_{-0.4}$  &$-0.5^{+0.3}_{-0.4}$  &$-0.4^{+0.3}_{-0.2}$  \\
&$\mathrm{T_{0} [K]}$   &\multicolumn{4}{c|}{$271.5$} &$267.3^{+9.2}_{-8.4}$  &$271.3^{+5.2}_{-4.8}$  &$270.0^{+6.6}_{-6.4}$  &$271.6^{+3.5}_{-3.5}$ &\multicolumn{4}{c|}{$287.2$ } &$285.1^{+10.9}_{-10.2}$  &$288.3^{+6.0}_{-6.3}$  &$286.5^{+7.8}_{-7.6}$  &$288.9^{+4.0}_{-4.3}$  \\
&$\mathrm{R_{pl} [R_\oplus]}$  &\multicolumn{4}{c|}{$1.0$}  &$0.9^{+0.1}_{-0.1}$  &$0.9^{+0.0}_{-0.0}$  &$0.9^{+0.1}_{-0.1}$  &$0.9^{+0.0}_{-0.0}$ &\multicolumn{4}{c|}{$1.0$} &$0.9^{+0.1}_{-0.1}$  &$0.9^{+0.1}_{-0.0}$  &$0.9^{+0.1}_{-0.1}$  &$0.9^{+0.0}_{-0.0}$  \\
&$L($$\mathrm{M_{pl} [M_\oplus]}$$)$  &\multicolumn{4}{c|}{$0.0$}  &$-0.1^{+0.3}_{-0.3}$  &$-0.0^{+0.3}_{-0.3}$  &$-0.0^{+0.3}_{-0.3}$  &$-0.0^{+0.3}_{-0.3}$ &\multicolumn{4}{c|}{$0.0$} &$-0.1^{+0.4}_{-0.3}$  &$-0.0^{+0.3}_{-0.3}$  &$-0.0^{+0.3}_{-0.3}$  &$0.0^{+0.3}_{-0.4}$  \\
&$L($$\mathrm{CO_2}$$)$  &\multicolumn{4}{c|}{$-3.4$}  &$-1.8^{+0.9}_{-0.9}$  &$-2.6^{+0.9}_{-0.8}$  &$-2.3^{+0.9}_{-0.8}$  &$-2.8^{+0.6}_{-0.6}$ &\multicolumn{4}{c|}{$-3.4$} &$-2.6^{+0.9}_{-0.9}$  &$-2.7^{+0.8}_{-0.7}$  &$-2.4^{+0.9}_{-0.7}$  &$-2.8^{+0.6}_{-0.6}$  \\
&$L($$\mathrm{H_2O}$$)$ &$-2.3$ &$-5.6$ &$-$ &$-$   &$-2.3^{+0.8}_{-0.9}$  &$-3.0^{+0.9}_{-0.9}$  &$-2.7^{+0.9}_{-0.9}$  &$-3.3^{+0.6}_{-0.6}$ &$-2.1$ &$-5.6$ &$-$ &$-$ &$-2.5^{+0.7}_{-0.9}$  &$-2.7^{+0.7}_{-0.7}$  &$-2.4^{+0.8}_{-0.8}$  &$-2.9^{+0.6}_{-0.6}$  \\
&$L($$\mathrm{O_3}$$)$ &$-7.3$ &$-5.8$ &$-5.0$ &$-5.3$   &$-5.2^{+0.8}_{-0.7}$  &$-5.8^{+0.8}_{-0.6}$  &$-5.5^{+0.8}_{-0.7}$  &$-5.9^{+0.5}_{-0.5}$ &$-7.3$ &$-6.0$ &$-5.0$ &$-5.4$ &$-5.8^{+0.8}_{-0.7}$  &$-5.8^{+0.7}_{-0.6}$  &$-5.6^{+0.8}_{-0.6}$  &$-5.9^{+0.5}_{-0.5}$  \\
&$L($$\mathrm{CH_4}$$)$  &$-6.0$ &$-6.1$ &$-6.4$ &$-6.7$  &$-4.4^{+1.2}_{-1.6}$  &$-4.7^{+0.9}_{-0.8}$  &$-4.6^{+1.0}_{-0.9}$  &$-4.9^{+0.6}_{-0.6}$ &$-6.0$ &$-6.1$ &$-6.4$ &$-6.7$  &$-5.0^{+1.2}_{-1.5}$  &$-4.7^{+0.8}_{-0.7}$  &$-4.6^{+1.0}_{-0.8}$  &$-4.8^{+0.6}_{-0.6}$  \\
&$\mathrm{T_{eq} [K]}$  &\multicolumn{4}{c|}{$246.8$}  &$252.6^{+8.5}_{-8.2}$  &$257.2^{+4.7}_{-4.6}$   &$255.8^{+6.1}_{-6.5}$  &$257.9^{+3.2}_{-3.2}$ &\multicolumn{4}{c|}{$258.3$} &$265.1^{+10.0}_{-10.7}$  &$269.0^{+5.5}_{-6.8}$  &$266.8^{+7.4}_{-8.3}$  &$270.1^{+3.6}_{-4.1}$  \\
&$\mathrm{A_B}$   &\multicolumn{4}{c|}{$0.38$} &$0.32^{+0.09}_{-0.10}$  &$0.27^{+0.05}_{-0.06}$  &$0.29^{+0.07}_{-0.07}$  &$0.26^{+0.04}_{-0.04}$ &\multicolumn{4}{c|}{$0.26$} &$0.18^{+0.13}_{-0.13}$  &$0.13^{+0.09}_{-0.08}$  &$0.16^{+0.10}_{-0.10}$  &$0.11^{+0.05}_{-0.05}$ \\\hline
\multirow{4}{*}{\rotatebox[origin=c]{90}{Figure~\ref{fig:Reduced Posteriors}}}&$L($$\mathrm{CO_2}$$)$ &\multicolumn{4}{c|}{$-3.4$}   &$-3.3^{+0.5}_{-0.5}$  &$-3.5^{+0.4}_{-0.4}$  &$-3.4^{+0.4}_{-0.4}$  &$-3.5^{+0.3}_{-0.3}$ &\multicolumn{4}{c|}{$-3.4$} &$-3.4^{+0.6}_{-0.5}$  &$-3.4^{+0.4}_{-0.4}$  &$-3.4^{+0.4}_{-0.4}$  &$-3.4^{+0.4}_{-0.4}$  \\
&$L($$\mathrm{H_2O}$$)$  &$-2.3$ &$-5.6$ &$-$ &$-$  &$-3.5^{+0.5}_{-0.6}$  &$-4.0^{+0.4}_{-0.4}$  &$-3.9^{+0.5}_{-0.5}$  &$-4.1^{+0.4}_{-0.4}$ &$-2.1$ &$-5.6$ &$-$ &$-$  &$-3.2^{+0.5}_{-0.6}$  &$-3.4^{+0.4}_{-0.4}$  &$-3.4^{+0.4}_{-0.4}$  &$-3.5^{+0.3}_{-0.4}$  \\
&$L($$\mathrm{O_3}$$)$   &$-7.3$ &$-5.8$ &$-5.0$ &$-5.3$ &$-6.6^{+0.4}_{-0.4}$  &$-6.5^{+0.4}_{-0.4}$  &$-6.5^{+0.4}_{-0.4}$  &$-6.5^{+0.3}_{-0.3}$  &$-7.3$ &$-6.0$ &$-5.0$ &$-5.4$  &$-6.4^{+0.5}_{-0.4}$  &$-6.4^{+0.3}_{-0.3}$  &$-6.4^{+0.4}_{-0.4}$  &$-6.4^{+0.3}_{-0.4}$  \\
&$L($$\mathrm{CH_4}$$)$  &$-6.0$ &$-6.1$ &$-6.4$ &$-6.7$ &$-6.0^{+0.8}_{-1.5}$  &$-5.6^{+0.4}_{-0.5}$  &$-5.7^{+0.5}_{-0.6}$  &$-5.6^{+0.4}_{-0.4}$ &$-6.0$ &$-6.1$ &$-6.4$ &$-6.7$ &$-5.9^{+0.7}_{-1.2}$  &$-5.5^{+0.4}_{-0.4}$  &$-5.6^{+0.5}_{-0.5}$  &$-5.5^{+0.4}_{-0.4}$  \\\hline
\multirow{4}{*}{\rotatebox[origin=c]{90}{Figure~\ref{fig:Reduced Posteriors}}}&$L($$\mathrm{P_{0} [bar]}$$)$  &\multicolumn{4}{c|}{$0.0$}  &$-0.9^{+0.3}_{-0.5}$  &$-0.7^{+0.4}_{-0.4}$  &$-0.7^{+0.4}_{-0.4}$  &$-0.5^{+0.3}_{-0.2}$  &\multicolumn{4}{c|}{$0.0$} &$-0.5^{+0.4}_{-0.4}$  &$-0.7^{+0.3}_{-0.3}$  &$-0.7^{+0.3}_{-0.4}$  &$-0.6^{+0.3}_{-0.2}$  \\
&$\mathrm{T_{0} [K]}$  &\multicolumn{4}{c|}{$271.5$}  &$261.3^{+2.9}_{-2.0}$  &$260.1^{+1.3}_{-1.0}$  &$260.7^{+1.9}_{-1.3}$  &$259.1^{+0.7}_{-0.6}$ &\multicolumn{4}{c|}{$287.2$} &$278.4^{+4.1}_{-3.0}$  &$278.2^{+2.2}_{-1.6}$  &$278.9^{+3.3}_{-2.3}$  &$275.8^{+1.1}_{-0.9}$  \\
&$\mathrm{T_{eq} [K]}$  &\multicolumn{4}{c|}{$246.8$}  &$247.1^{+1.9}_{-1.8}$  &$246.3^{+0.9}_{-0.9}$  &$246.8^{+1.2}_{-1.3}$  &$246.4^{+0.6}_{-0.6}$ &\multicolumn{4}{c|}{$258.3$} &$258.2^{+2.1}_{-2.0}$  &$257.4^{+1.0}_{-1.0}$  &$258.0^{+1.5}_{-1.4}$  &$257.2^{+0.7}_{-0.7}$  \\
&$\mathrm{A_B}$  &\multicolumn{4}{c|}{$0.38$}  &$0.38^{+0.03}_{-0.03}$  &$0.39^{+0.02}_{-0.02}$  &$0.38^{+0.02}_{-0.02}$  &$0.39^{+0.02}_{-0.02}$ &\multicolumn{4}{c|}{$0.26$} &$0.26^{+0.03}_{-0.04}$  &$0.27^{+0.02}_{-0.03}$  &$0.26^{+0.03}_{-0.03}$  &$0.28^{+0.02}_{-0.02}$  \\\hline
\multirow{6}{*}{\rotatebox[origin=c]{90}{Figure~\ref{fig:Posteriors_Rel_Abund}}}&$L\left(\frac{\mathrm{H_2O}}{\mathrm{CO_2}}\right)$ &$1.1$ &$-2.2$ &$-$ &$-$  &$-0.5^{+0.6}_{-0.6}$  &$-0.5^{+0.3}_{-0.3}$  &$-0.5^{+0.4}_{-0.4}$  &$-0.5^{+0.2}_{-0.2}$ &$1.3$ &$-2.2$ &$-$ &$-$ &$0.0^{+0.6}_{-0.6}$  &$-0.1^{+0.3}_{-0.3}$  &$-0.1^{+0.4}_{-0.4}$  &$-0.1^{+0.2}_{-0.2}$\\  
&$L\left(\frac{\mathrm{O_3}}{\mathrm{CO_2}}\right)$  &$-3.9$ &$-2.4$ &$-1.7$ &$-1.9$  &$-3.3^{+0.5}_{-0.4}$  &$-3.2^{+0.3}_{-0.3}$  &$-3.2^{+0.3}_{-0.3}$  &$-3.1^{+0.2}_{-0.2}$ &$-3.9$ &$-2.6$ &$-1.6$ &$-2.0$ &$-3.2^{+0.5}_{-0.5}$  &$-3.2^{+0.3}_{-0.3}$  &$-3.2^{+0.3}_{-0.3}$  &$-3.1^{+0.2}_{-0.2}$  \\
&$L\left(\frac{\mathrm{CH_4}}{\mathrm{CO_2}}\right)$  &$-2.6$ &$-2.7$ &$-3.0$ &$-3.3$   &$-2.5^{+0.8}_{-1.5}$  &$-2.1^{+0.3}_{-0.4}$  &$-2.2^{+0.4}_{-0.6}$  &$-2.1^{+0.2}_{-0.2}$ &$-2.6$ &$-2.7$ &$-3.0$ &$-3.3$ &$-2.3^{+0.7}_{-1.2}$  &$-2.1^{+0.3}_{-0.4}$  &$-2.1^{+0.4}_{-0.5}$  &$-2.0^{+0.2}_{-0.2}$  \\
&$L\left(\frac{\mathrm{O_3}}{\mathrm{H_2O}}\right)$  &$-4.9$ &$-0.1$ &$-$ &$-$  &$-2.8^{+0.5}_{-0.4}$  &$-2.7^{+0.3}_{-0.3}$  &$-2.7^{+0.4}_{-0.4}$  &$-2.6^{+0.2}_{-0.2}$ &$-5.2$ &$-0.4$ &$-$ &$-$ &$-3.2^{+0.4}_{-0.4}$  &$-3.1^{+0.3}_{-0.2}$  &$-3.2^{+0.3}_{-0.3}$  &$-3.0^{+0.2}_{-0.2}$  \\
&$L\left(\frac{\mathrm{CH_4}}{\mathrm{H_2O}}\right)$   &$-3.7$ &$-0.4$ &$-$ &$-$ &$-2.0^{+0.9}_{-1.6}$  &$-1.7^{+0.4}_{-0.4}$  &$-1.7^{+0.5}_{-0.6}$  &$-1.6^{+0.3}_{-0.3}$ &$-4.0$ &$-0.5$ &$-$ &$-$  &$-2.3^{+0.7}_{-1.5}$  &$-2.0^{+0.4}_{-0.4}$  &$-2.1^{+0.5}_{-0.6}$  &$-1.9^{+0.2}_{-0.3}$  \\
&$L\left(\frac{\mathrm{CH_4}}{\mathrm{O_3}}\right)$  &$1.3$ &$-0.3$ &$-1.3$ &$-1.3$  &$0.8^{+0.6}_{-1.3}$  &$1.0^{+0.3}_{-0.3}$  &$1.0^{+0.4}_{-0.4}$  &$1.0^{+0.2}_{-0.2}$ &$1.3$ &$-0.1$ &$-1.4$ &$-1.3$  &$0.9^{+0.6}_{-1.2}$  &$1.1^{+0.3}_{-0.3}$  &$1.1^{+0.3}_{-0.4}$  &$1.1^{+0.2}_{-0.2}$  
\enddata
\addtolength{\leftskip} {-0cm}
\addtolength{\rightskip}{-4.8cm}
\tablecomments{Retrieved model parameter posteriors for the all combinations of spectral resolutions (\Rv{50, 100}) and noise levels (\SNv{10, 20}) of disk-integrated NP Earth spectra. Here, $L(\cdot)$ abbreviates \lgrt{\cdot}. We provide the median of the retrieved posterior and indicate the $16\% - 84\%$ range via $+/-$ indices. We Further provide the ground truth values. If independent of the atmospheric pressure, we provide a single value. Otherwise, we provide the ground truths at pressures $1$~bar, $10^-1$~bar, $10^-2$~bar, and $10^-3$~bar if available.}
\label{tab:Rel_Abund_Values_NP}
\end{deluxetable}
\end{rotatetable}

\movetabledown=50mm
\begin{rotatetable}
\begin{deluxetable}{cc|cccc|c|c|c|c|cccc|c|c|c|c|}
\tablecaption{Numeric values corresponding to the retrieved parameter posteriors in Figures~\ref{fig:PosteriorsM4}, \ref{fig:Reduced Posteriors}, and \ref{fig:Posteriors_Rel_Abund} for the SP viewing angle.}
\tablehead{
&&\multicolumn{4}{c}{SP Jan $-$ Ground Truths}\vline&\multicolumn{4}{c}{SP Jan $-$ Posteriors}\vline&\multicolumn{4}{c}{SP Jul $-$ Ground Truths}\vline&\multicolumn{4}{c}{SP Jul $-$ Posteriors}\vline\\
&&\multicolumn{4}{c}{Pressure Levels [bar]}\vline&\multicolumn{2}{c}{\Rv{50}} &\multicolumn{2}{c}{\Rv{100}} \vline&\multicolumn{4}{c}{Pressure Levels [bar]}\vline &\multicolumn{2}{c}{\Rv{50}} &\multicolumn{2}{c}{\Rv{100}}\vline\\
&\colhead{Posterior}\vline&\colhead{1}&\colhead{$10^{-1}$}&\colhead{$10^{-2}$}&\colhead{$10^{-3}$}\vline&\colhead{\SNv{10}}   &\colhead{\SNv{20}}\vline   &\colhead{\SNv{10}}   &\colhead{\SNv{20}}\vline&\colhead{1}&\colhead{$10^{-1}$}&\colhead{$10^{-2}$}&\colhead{$10^{-3}$}\vline &\colhead{\SNv{10}}   &\colhead{\SNv{20}}\vline   &\colhead{\SNv{10}}   &\colhead{\SNv{20}} \vline
}
\startdata
\multirow{10}{*}{\rotatebox[origin=c]{90}{Figure~\ref{fig:PosteriorsM4}}} &$L($$\mathrm{P_{0} [bar]}$$)$ &\multicolumn{4}{c|}{$0.0$}   &$-0.8^{+0.3}_{-0.5}$  &$-0.6^{+0.3}_{-0.4}$  &$-0.6^{+0.3}_{-0.4}$  &$-0.5^{+0.2}_{-0.3}$ &\multicolumn{4}{c|}{$0.0$} &$-0.7^{+0.4}_{-0.5}$  &$-0.5^{+0.3}_{-0.4}$  &$-0.6^{+0.3}_{-0.4}$  &$-0.4^{+0.3}_{-0.3}$  \\
&$\mathrm{T_{0} [K]}$  &\multicolumn{4}{c|}{$285.0$}  &$274.7^{+9.6}_{-8.9}$  &$278.3^{+5.6}_{-5.6}$  &$277.5^{+7.0}_{-6.9}$  &$279.3^{+3.6}_{-3.7}$ &\multicolumn{4}{c|}{$278.9$} &$271.1^{+9.4}_{-8.9}$  &$274.6^{+5.4}_{-5.0}$  &$273.8^{+6.7}_{-6.4}$  &$274.8^{+3.8}_{-3.5}$  \\
&$\mathrm{R_{pl} [R_\oplus]}$  &\multicolumn{4}{c|}{$1.0$}  &$0.9^{+0.1}_{-0.1}$  &$0.9^{+0.1}_{-0.0}$  &$0.9^{+0.1}_{-0.1}$  &$0.9^{+0.0}_{-0.0}$ &\multicolumn{4}{c|}{$1.0$} &$0.9^{+0.1}_{-0.1}$  &$0.9^{+0.0}_{-0.0}$  &$0.9^{+0.1}_{-0.1}$  &$0.9^{+0.0}_{-0.0}$  \\
&$L($$\mathrm{M_{pl} [M_\oplus]}$$)$  &\multicolumn{4}{c|}{$0.0$}  &$-0.1^{+0.3}_{-0.3}$  &$-0.0^{+0.3}_{-0.3}$  &$-0.0^{+0.3}_{-0.3}$  &$-0.0^{+0.3}_{-0.3}$ &\multicolumn{4}{c|}{$0.0$} &$-0.0^{+0.3}_{-0.3}$  &$-0.0^{+0.3}_{-0.3}$  &$-0.0^{+0.3}_{-0.3}$  &$-0.0^{+0.3}_{-0.3}$  \\
&$L($$\mathrm{CO_2}$$)$  &\multicolumn{4}{c|}{$-3.4$}  &$-2.0^{+0.9}_{-0.9}$  &$-2.4^{+0.9}_{-0.6}$  &$-2.4^{+0.8}_{-0.8}$  &$-2.7^{+0.6}_{-0.6}$ &\multicolumn{4}{c|}{$-3.4$} &$-2.1^{+1.0}_{-0.9}$  &$-2.5^{+0.8}_{-0.7}$  &$-2.3^{+0.9}_{-0.7}$  &$-2.8^{+0.6}_{-0.6}$  \\
&$L($$\mathrm{H_2O}$$)$  &$-2.1$ &$-5.7$ &$-$ &$-$ &$-2.2^{+0.7}_{-0.8}$  &$-2.5^{+0.8}_{-0.7}$  &$-2.6^{+0.7}_{-0.8}$  &$-2.9^{+0.6}_{-0.6}$ &$-2.2$ &$-5.7$ &$-$ &$-$  &$-2.5^{+0.9}_{-1.1}$  &$-3.0^{+0.8}_{-0.8}$  &$-2.8^{+0.9}_{-0.8}$  &$-3.4^{+0.6}_{-0.6}$  \\
&$L($$\mathrm{O_3}$$)$ &$-7.6$ &$-6.0$ &$-5.0$ &$-5.4$  &$-5.5^{+0.9}_{-0.7}$  &$-5.7^{+0.7}_{-0.5}$  &$-5.7^{+0.7}_{-0.6}$  &$-5.9^{+0.5}_{-0.5}$ &$-7.3$ &$-5.9$ &$-5.1$ &$-5.3$ &$-5.3^{+0.8}_{-0.8}$  &$-5.6^{+0.7}_{-0.6}$  &$-5.5^{+0.8}_{-0.6}$  &$-5.9^{+0.5}_{-0.5}$  \\
&$L($$\mathrm{CH_4}$$)$  &$-6.1$ &$-6.1$ &$-6.4$ &$-6.7$  &$-4.5^{+1.3}_{-1.5}$  &$-4.5^{+0.8}_{-0.7}$  &$-4.7^{+0.9}_{-0.9}$  &$-4.7^{+0.6}_{-0.6}$ &$-6.1$ &$-6.1$ &$-6.4$ &$-6.7$ &$-4.6^{+1.1}_{-1.4}$  &$-4.6^{+0.9}_{-0.7}$  &$-4.5^{+0.9}_{-0.8}$  &$-4.9^{+0.6}_{-0.6}$  \\
&$\mathrm{T_{eq} [K]}$  &\multicolumn{4}{c|}{$253.1$}  &$259.0^{+8.7}_{-8.9}$  &$263.0^{+5.3}_{-6.1}$  &$262.1^{+6.6}_{-6.8}$  &$264.5^{+3.5}_{-3.6}$ &\multicolumn{4}{c|}{$247.6$} &$255.6^{+8.7}_{-8.9}$  &$259.5^{+4.9}_{-4.8}$  &$258.7^{+6.2}_{-6.1}$  &$260.2^{+3.4}_{-3.2}$  \\
&$\mathrm{A_B}$  &\multicolumn{4}{c|}{$0.32$}  &$0.25^{+0.10}_{-0.11}$  &$0.2^{+0.07}_{-0.07}$  &$0.21^{+0.08}_{-0.08}$  &$0.18^{+0.05}_{-0.05}$ &\multicolumn{4}{c|}{$0.37$} &$0.29^{+0.10}_{-0.10}$  &$0.24^{+0.06}_{-0.06}$  &$0.25^{+0.07}_{-0.08}$  &$0.24^{+0.04}_{-0.04}$\\\hline
\multirow{4}{*}{\rotatebox[origin=c]{90}{Figure~\ref{fig:Reduced Posteriors}}}&$L($$\mathrm{CO_2}$$)$  &\multicolumn{4}{c|}{$-3.4$}  &$-3.5^{+0.5}_{-0.5}$  &$-3.6^{+0.4}_{-0.4}$  &$-3.4^{+0.5}_{-0.4}$  &$-3.6^{+0.3}_{-0.3}$  &\multicolumn{4}{c|}{$-3.4$} &$-3.4^{+0.5}_{-0.5}$  &$-3.5^{+0.4}_{-0.4}$  &$-3.4^{+0.5}_{-0.4}$  &$-3.5^{+0.3}_{-0.3}$  \\
&$L($$\mathrm{H_2O}$$)$  &$-2.1$ &$-5.7$ &$-$ &$-$  &$-3.4^{+0.4}_{-0.6}$  &$-3.6^{+0.3}_{-0.4}$  &$-3.5^{+0.5}_{-0.5}$  &$-3.8^{+0.3}_{-0.4}$ &$-2.2$ &$-5.7$ &$-$ &$-$  &$-3.6^{+0.5}_{-0.7}$  &$-4.0^{+0.4}_{-0.4}$  &$-3.9^{+0.5}_{-0.5}$  &$-4.1^{+0.4}_{-0.4}$  \\
&$L($$\mathrm{O_3}$$)$ &$-7.6$ &$-6.0$ &$-5.0$ &$-5.4$   &$-6.8^{+0.4}_{-0.4}$  &$-6.6^{+0.3}_{-0.4}$  &$-6.6^{+0.4}_{-0.4}$  &$-6.7^{+0.3}_{-0.3}$ &$-7.3$ &$-5.9$ &$-5.1$ &$-5.3$  &$-6.4^{+0.4}_{-0.4}$  &$-6.4^{+0.3}_{-0.4}$  &$-6.5^{+0.4}_{-0.4}$  &$-6.5^{+0.4}_{-0.3}$  \\
&$L($$\mathrm{CH_4}$$)$ &$-6.1$ &$-6.1$ &$-6.4$ &$-6.7$   &$-6.2^{+0.8}_{-1.3}$  &$-5.6^{+0.4}_{-0.4}$  &$-5.7^{+0.6}_{-0.6}$  &$-5.6^{+0.4}_{-0.4}$ &$-6.1$ &$-6.1$ &$-6.4$ &$-6.7$  &$-5.7^{+0.7}_{-1.1}$  &$-5.6^{+0.4}_{-0.4}$  &$-5.6^{+0.5}_{-0.5}$  &$-5.6^{+0.4}_{-0.4}$  \\\hline
\multirow{4}{*}{\rotatebox[origin=c]{90}{Figure~\ref{fig:Reduced Posteriors}}}&$L($$\mathrm{P_{0} [bar]}$$)$  &\multicolumn{4}{c|}{$0.0$}  &$-0.8^{+0.3}_{-0.5}$  &$-0.9^{+0.3}_{-0.3}$  &$-0.7^{+0.3}_{-0.3}$  &$-0.7^{+0.2}_{-0.2}$  &\multicolumn{4}{c|}{$0.0$} &$-0.8^{+0.4}_{-0.5}$  &$-0.7^{+0.3}_{-0.4}$  &$-0.7^{+0.3}_{-0.4}$  &$-0.5^{+0.3}_{-0.2}$  \\
&$\mathrm{T_{0} [K]}$  &\multicolumn{4}{c|}{$285.0$}  &$269.1^{+3.6}_{-2.3}$  &$269.0^{+2.0}_{-1.4}$  &$268.7^{+2.3}_{-1.6}$  &$267.3^{+1.0}_{-0.7}$ &\multicolumn{4}{c|}{$278.9$} &$262.5^{+2.9}_{-1.9}$  &$260.9^{+1.4}_{-0.9}$  &$261.4^{+1.8}_{-1.3}$  &$259.9^{+0.6}_{-0.6}$  \\
&$\mathrm{T_{eq} [K]}$  &\multicolumn{4}{c|}{$253.1$}  &$253.5^{+1.8}_{-1.9}$  &$252.7^{+0.9}_{-1.0}$  &$253.1^{+1.2}_{-1.3}$  &$252.6^{+0.6}_{-0.6}$ &\multicolumn{4}{c|}{$247.6$} &$247.5^{+1.8}_{-2.0}$  &$246.5^{+0.9}_{-0.9}$  &$246.9^{+1.2}_{-1.3}$  &$246.7^{+0.6}_{-0.6}$  \\
&$\mathrm{A_B}$ &\multicolumn{4}{c|}{$0.32$}   &$0.31^{+0.03}_{-0.03}$  &$0.32^{+0.02}_{-0.02}$  &$0.32^{+0.02}_{-0.03}$  &$0.33^{+0.02}_{-0.02}$ &\multicolumn{4}{c|}{$0.37$} &$0.37^{+0.03}_{-0.03}$  &$0.39^{+0.02}_{-0.02}$  &$0.39^{+0.02}_{-0.02}$  &$0.39^{+0.02}_{-0.02}$  \\\hline
\multirow{6}{*}{\rotatebox[origin=c]{90}{Figure~\ref{fig:Posteriors_Rel_Abund}}}&$L\left(\frac{\mathrm{H_2O}}{\mathrm{CO_2}}\right)$  &$1.3$ &$-2.3$ &$-$ &$-$  &$-0.2^{+0.6}_{-0.6}$  &$-0.2^{+0.3}_{-0.3}$  &$-0.2^{+0.4}_{-0.4}$  &$-0.2^{+0.2}_{-0.2}$ &$1.2$ &$-2.3$ &$-$ &$-$ &$-0.5^{+0.6}_{-0.6}$  &$-0.5^{+0.3}_{-0.3}$  &$-0.5^{+0.4}_{-0.4}$  &$-0.6^{+0.2}_{-0.2}$  \\
&$L\left(\frac{\mathrm{O_3}}{\mathrm{CO_2}}\right)$  &$-4.2$ &$-2.6$ &$-1.6$ &$-2.0$  &$-3.4^{+0.5}_{-0.5}$  &$-3.3^{+0.3}_{-0.3}$  &$-3.3^{+0.3}_{-0.3}$  &$-3.3^{+0.2}_{-0.2}$ &$-3.9$ &$-2.5$ &$-1.7$ &$-1.9$  &$-3.2^{+0.5}_{-0.4}$  &$-3.1^{+0.2}_{-0.2}$  &$-3.2^{+0.3}_{-0.3}$  &$-3.1^{+0.2}_{-0.2}$  \\
&$L\left(\frac{\mathrm{CH_4}}{\mathrm{CO_2}}\right)$  &$-2.7$ &$-2.7$ &$-3.0$ &$-3.3$  &$-2.4^{+0.8}_{-1.4}$  &$-2.1^{+0.3}_{-0.4}$  &$-2.2^{+0.5}_{-0.6}$  &$-2.1^{+0.2}_{-0.2}$ &$-2.7$ &$-2.7$ &$-3.0$ &$-3.3$ &$-2.4^{+0.7}_{-1.2}$  &$-2.1^{+0.3}_{-0.3}$  &$-2.2^{+0.4}_{-0.5}$  &$-2.1^{+0.2}_{-0.2}$  \\
&$L\left(\frac{\mathrm{O_3}}{\mathrm{H_2O}}\right)$  &$-5.5$ &$-0.4$ &$-$ &$-$  &$-3.2^{+0.4}_{-0.4}$  &$-3.1^{+0.3}_{-0.3}$  &$-3.1^{+0.3}_{-0.3}$  &$-3.0^{+0.2}_{-0.2}$ &$-5.0$ &$-0.3$ &$-$ &$-$  &$-2.8^{+0.5}_{-0.5}$  &$-2.7^{+0.3}_{-0.3}$  &$-2.7^{+0.3}_{-0.3}$  &$-2.5^{+0.2}_{-0.2}$  \\
&$L\left(\frac{\mathrm{CH_4}}{\mathrm{H_2O}}\right)$  &$-3.9$ &$-0.4$ &$-$ &$-$  &$-2.1^{+0.9}_{-1.5}$  &$-1.9^{+0.4}_{-0.5}$  &$-2.0^{+0.5}_{-0.7}$  &$-1.8^{+0.2}_{-0.3}$ &$-3.8$ &$-0.4$ &$-$ &$-$ &$-1.9^{+0.8}_{-1.3}$  &$-1.6^{+0.4}_{-0.4}$  &$-1.6^{+0.5}_{-0.5}$  &$-1.5^{+0.2}_{-0.3}$  \\
&$L\left(\frac{\mathrm{CH_4}}{\mathrm{O_3}}\right)$ &$1.5$ &$-0.1$ &$-1.4$ &$-1.3$    &$1.1^{+0.6}_{-1.2}$  &$1.2^{+0.3}_{-0.3}$  &$1.1^{+0.4}_{-0.5}$  &$1.2^{+0.2}_{-0.2}$ &$1.2$ &$-0.2$ &$-1.3$ &$-1.4$  &$0.8^{+0.6}_{-1.0}$  &$1.0^{+0.3}_{-0.3}$  &$1.0^{+0.3}_{-0.4}$  &$1.0^{+0.2}_{-0.2}$ 
\enddata
\addtolength{\leftskip} {-0cm}
\addtolength{\rightskip}{-4.8cm}
\tablecomments{Retrieved model parameter posteriors for the all combinations of spectral resolutions (\Rv{50, 100}) and noise levels (\SNv{10, 20}) of disk-integrated SP Earth spectra. Here, $L(\cdot)$ abbreviates \lgrt{\cdot}. We provide the median of the retrieved posterior and indicate the $16\% - 84\%$ range via $+/-$ indices. We Further provide the ground truth values. If independent of the atmospheric pressure, we provide a single value. Otherwise, we provide the ground truths at pressures $1$~bar, $10^-1$~bar, $10^-2$~bar, and $10^-3$~bar if available.}
\label{tab:Rel_Abund_Values_SP}
\end{deluxetable}
\end{rotatetable}

\movetabledown=50mm
\begin{rotatetable}
\begin{deluxetable}{cc|cccc|c|c|c|c|cccc|c|c|c|c|}
\tablecaption{Numeric values corresponding to the retrieved parameter posteriors in Figures~\ref{fig:PosteriorsM4}, \ref{fig:Reduced Posteriors}, and \ref{fig:Posteriors_Rel_Abund} for the EqC viewing angle.}
\tablehead{
&&\multicolumn{4}{c}{EqC Jan $-$ Ground Truths}\vline&\multicolumn{4}{c}{EqC Jan $-$ Posteriors}\vline&\multicolumn{4}{c}{EqC Jul $-$ Ground Truths}\vline&\multicolumn{4}{c}{EqC Jul $-$ Posteriors}\vline\\
&&\multicolumn{4}{c}{Pressure Levels [bar]}\vline&\multicolumn{2}{c}{\Rv{50}} &\multicolumn{2}{c}{\Rv{100}} \vline&\multicolumn{4}{c}{Pressure Levels [bar]}\vline &\multicolumn{2}{c}{\Rv{50}} &\multicolumn{2}{c}{\Rv{100}}\vline\\
&\colhead{Posterior}\vline&\colhead{1}&\colhead{$10^{-1}$}&\colhead{$10^{-2}$}&\colhead{$10^{-3}$}\vline&\colhead{\SNv{10}}   &\colhead{\SNv{20}}\vline   &\colhead{\SNv{10}}   &\colhead{\SNv{20}}\vline&\colhead{1}&\colhead{$10^{-1}$}&\colhead{$10^{-2}$}&\colhead{$10^{-3}$}\vline &\colhead{\SNv{10}}   &\colhead{\SNv{20}}\vline   &\colhead{\SNv{10}}   &\colhead{\SNv{20}} \vline
}
\startdata
\multirow{10}{*}{\rotatebox[origin=c]{90}{Figure~\ref{fig:PosteriorsM4}}}&$L($$\mathrm{P_{0} [bar]}$$)$  &\multicolumn{4}{c|}{0.0}  &$-0.7^{+0.4}_{-0.4}$  &$-0.6^{+0.3}_{-0.4}$  &$-0.5^{+0.4}_{-0.4}$  &$-0.4^{+0.3}_{-0.3}$&\multicolumn{4}{c|}{$0.0$}  &$-0.5^{+0.4}_{-0.5}$  &$-0.5^{+0.3}_{-0.4}$  &$-0.5^{+0.3}_{-0.4}$  &$-0.4^{+0.3}_{-0.2}$  \\  
&$\mathrm{T_{0} [K]}$ &\multicolumn{4}{c|}{$280.5$}   &$281.0^{+10.6}_{-9.4}$  &$284.4^{+5.7}_{-5.5}$  &$283.5^{+7.2}_{-6.9}$  &$285.1^{+3.7}_{-3.5}$ &\multicolumn{4}{c|}{$285.1$} &$285.5^{+10.6}_{-10.3}$  &$288.3^{+5.8}_{-5.9}$  &$287.0^{+7.8}_{-7.4}$  &$288.5^{+3.8}_{-3.7}$  \\  
&$\mathrm{R_{pl} [R_\oplus]}$ &\multicolumn{4}{c|}{$1.0$}   &$0.9^{+0.1}_{-0.1}$  &$0.9^{+0.0}_{-0.0}$  &$0.9^{+0.1}_{-0.1}$  &$0.9^{+0.0}_{-0.0}$ &\multicolumn{4}{c|}{$1.0$} &$0.9^{+0.1}_{-0.1}$  &$0.9^{+0.0}_{-0.0}$  &$0.9^{+0.1}_{-0.1}$  &$0.9^{+0.0}_{-0.0}$  \\  
&$L($$\mathrm{M_{pl} [M_\oplus]}$$)$  &\multicolumn{4}{c|}{$0.0$}  &$-0.1^{+0.3}_{-0.3}$  &$-0.0^{+0.3}_{-0.3}$  &$-0.0^{+0.3}_{-0.3}$  &$-0.0^{+0.3}_{-0.3}$ &\multicolumn{4}{c|}{$0.0$} &$-0.1^{+0.3}_{-0.3}$  &$-0.0^{+0.3}_{-0.3}$  &$-0.0^{+0.3}_{-0.3}$  &$0.0^{+0.3}_{-0.3}$  \\  
&$L($$\mathrm{CO_2}$$)$  &\multicolumn{4}{c|}{$-3.4$}  &$-2.2^{+1.0}_{-0.9}$  &$-2.4^{+0.9}_{-0.7}$  &$-2.5^{+0.9}_{-0.8}$  &$-2.8^{+0.6}_{-0.6}$ &\multicolumn{4}{c|}{$-3.4$} &$-2.5^{+1.0}_{-0.9}$  &$-2.5^{+0.8}_{-0.7}$  &$-2.4^{+1.0}_{-0.8}$  &$-2.8^{+0.6}_{-0.5}$  \\  
&$L($$\mathrm{H_2O}$$)$  &$-2.1$ &$-5.6$ &$-$ &$-$  &$-2.4^{+0.8}_{-0.9}$  &$-2.6^{+0.8}_{-0.7}$  &$-2.7^{+0.8}_{-0.9}$  &$-3.0^{+0.6}_{-0.6}$ &$-2.1$ &$-5.6$ &$-$ &$-$  &$-2.5^{+0.8}_{-0.9}$  &$-2.8^{+0.8}_{-0.8}$  &$-2.7^{+0.9}_{-0.8}$  &$-3.1^{+0.6}_{-0.6}$  \\  
&$L($$\mathrm{O_3}$$)$  &$-7.4$ &$-5.9$ &$-5.0$ &$-5.4$  &$-5.6^{+0.8}_{-0.7}$  &$-5.7^{+0.7}_{-0.6}$  &$-5.8^{+0.8}_{-0.7}$  &$-6.0^{+0.5}_{-0.5}$ &$-7.3$ &$-6.0$ &$-5.0$ &$-5.4$ &$-5.8^{+0.8}_{-0.7}$  &$-5.8^{+0.6}_{-0.6}$  &$-5.7^{+0.8}_{-0.6}$  &$-6.0^{+0.5}_{-0.5}$  \\  
&$L($$\mathrm{CH_4}$$)$  &$-6.0$ &$-6.1$ &$-6.4$ &$-6.6$  &$-4.7^{+1.2}_{-1.3}$  &$-4.5^{+0.8}_{-0.7}$  &$-4.7^{+1.0}_{-0.9}$  &$-4.8^{+0.6}_{-0.6}$ &$-6.0$ &$-6.1$ &$-6.4$ &$-6.6$ &$-4.9^{+1.2}_{-1.3}$  &$-4.6^{+0.8}_{-0.7}$  &$-4.6^{+1.0}_{-0.8}$  &$-4.8^{+0.6}_{-0.5}$  \\  
&$\mathrm{T_{eq} [K]}$ &\multicolumn{4}{c|}{$256.5$}  &$262.7^{+10.0}_{-9.6}$  &$266.8^{+5.0}_{-5.8}$  &$266.0^{+6.6}_{-7.0}$  &$268.0^{+3.3}_{-3.4}$ &\multicolumn{4}{c|}{$259.5$} &$266.0^{+9.9}_{-9.9}$  &$269.5^{+5.4}_{-5.8}$  &$268.1^{+6.9}_{-7.6}$  &$270.2^{+3.4}_{-3.5}$  \\  
&$\mathrm{A_B}$ &\multicolumn{4}{c|}{$0.28$}   &$0.21^{+0.11}_{-0.13}$  &$0.16^{+0.07}_{-0.07}$  &$0.17^{+0.09}_{-0.09}$  &$0.14^{+0.05}_{-0.05}$ &\multicolumn{4}{c|}{$0.24$} &$0.17^{+0.12}_{-0.13}$  &$0.12^{+0.08}_{-0.07}$  &$0.14^{+0.10}_{-0.09}$  &$0.11^{+0.05}_{-0.05}$  \\\hline
\multirow{4}{*}{\rotatebox[origin=c]{90}{Figure~\ref{fig:Reduced Posteriors}}}&$L($$\mathrm{CO_2}$$)$ &\multicolumn{4}{c|}{$-3.4$}  &$-3.4^{+0.5}_{-0.5}$  &$-3.6^{+0.4}_{-0.4}$  &$-3.4^{+0.4}_{-0.4}$  &$-3.5^{+0.4}_{-0.4}$ &\multicolumn{4}{c|}{$-3.4$} &$-3.4^{+0.5}_{-0.5}$  &$-3.5^{+0.4}_{-0.4}$  &$-3.5^{+0.5}_{-0.4}$  &$-3.5^{+0.3}_{-0.3}$\\  
&$L($$\mathrm{H_2O}$$)$  &$-2.1$ &$-5.6$ &$-$ &$-$  &$-3.4^{+0.5}_{-0.5}$  &$-3.8^{+0.4}_{-0.4}$  &$-3.6^{+0.4}_{-0.4}$  &$-3.8^{+0.4}_{-0.4}$ &$-2.1$ &$-5.6$ &$-$ &$-$ &$-3.4^{+0.5}_{-0.6}$  &$-3.7^{+0.4}_{-0.4}$  &$-3.6^{+0.5}_{-0.4}$  &$-3.8^{+0.4}_{-0.4}$\\  
&$L($$\mathrm{O_3}$$)$ &$-7.4$ &$-5.9$ &$-5.0$ &$-5.4$  &$-6.6^{+0.4}_{-0.4}$  &$-6.7^{+0.3}_{-0.3}$  &$-6.5^{+0.4}_{-0.4}$  &$-6.6^{+0.4}_{-0.3}$ &$-7.3$ &$-6.0$ &$-5.0$ &$-5.4$  &$-6.5^{+0.4}_{-0.4}$  &$-6.5^{+0.4}_{-0.3}$  &$-6.5^{+0.4}_{-0.4}$  &$-6.5^{+0.3}_{-0.3}$\\  
&$L($$\mathrm{CH_4}$$)$  &$-6.0$ &$-6.1$ &$-6.4$ &$-6.6$  &$-5.8^{+0.7}_{-1.0}$  &$-5.6^{+0.4}_{-0.4}$  &$-5.6^{+0.5}_{-0.5}$  &$-5.6^{+0.4}_{-0.4}$ &$-6.0$ &$-6.1$ &$-6.4$ &$-6.6$ &$-5.8^{+0.7}_{-1.1}$  &$-5.5^{+0.4}_{-0.4}$  &$-5.6^{+0.5}_{-0.5}$  &$-5.5^{+0.4}_{-0.4}$\\\hline
\multirow{4}{*}{\rotatebox[origin=c]{90}{Figure~\ref{fig:Reduced Posteriors}}}&$L($$\mathrm{P_{0} [bar]}$$)$  &\multicolumn{4}{c|}{$0.0$}  &$-0.7^{+0.4}_{-0.4}$  &$-0.8^{+0.3}_{-0.3}$  &$-0.7^{+0.4}_{-0.4}$  &$-0.7^{+0.3}_{-0.3}$  &\multicolumn{4}{c|}{$0.0$} &$-0.6^{+0.4}_{-0.4}$  &$-0.7^{+0.3}_{-0.3}$  &$-0.7^{+0.3}_{-0.4}$  &$-0.6^{+0.3}_{-0.2}$  \\
&$\mathrm{T_{0} [K]}$  &\multicolumn{4}{c|}{$280.5$}  &$274.8^{+3.9}_{-2.5}$  &$274.5^{+2.0}_{-1.4}$  &$273.7^{+2.3}_{-1.6}$  &$272.3^{+0.9}_{-0.7}$ &\multicolumn{4}{c|}{$285.1$} &$278.5^{+3.3}_{-2.4}$  &$277.8^{+1.8}_{-1.3}$  &$278.4^{+2.7}_{-1.8}$  &$276.3^{+0.9}_{-0.8}$  \\
&$\mathrm{T_{eq} [K]}$  &\multicolumn{4}{c|}{$256.5$}  &$256.7^{+1.9}_{-2.1}$  &$255.8^{+1.0}_{-1.0}$  &$256.2^{+1.4}_{-1.4}$  &$255.8^{+0.6}_{-0.6}$ &\multicolumn{4}{c|}{$259.5$} &$259.4^{+2.1}_{-2.1}$  &$258.8^{+1.0}_{-1.0}$  &$259.3^{+1.4}_{-1.4}$  &$258.8^{+0.7}_{-0.7}$  \\
&$\mathrm{A_B}$  &\multicolumn{4}{c|}{$0.28$} &$0.27^{+0.03}_{-0.04}$  &$0.29^{+0.02}_{-0.02}$  &$0.29^{+0.03}_{-0.03}$  &$0.3^{+0.02}_{-0.02}$&\multicolumn{4}{c|}{$0.24$}  &$0.24^{+0.03}_{-0.04}$  &$0.26^{+0.02}_{-0.02}$  &$0.25^{+0.03}_{-0.03}$  &$0.26^{+0.02}_{-0.02}$ \\\hline
\multirow{6}{*}{\rotatebox[origin=c]{90}{Figure~\ref{fig:Posteriors_Rel_Abund}}}&$L\left(\frac{\mathrm{H_2O}} {\mathrm{CO_2}}\right)$ &$1.3$ &$-2.3$ &$-$ &$-$ &$-0.2^{+0.6}_{-0.6}$  &$-0.3^{+0.3}_{-0.3}$  &$-0.3^{+0.4}_{-0.4}$  &$-0.3^{+0.2}_{-0.2}$ &$1.3$ &$-2.2$ &$-$ &$-$  &$-0.2^{+0.5}_{-0.5}$  &$-0.3^{+0.3}_{-0.3}$  &$-0.3^{+0.4}_{-0.4}$  &$-0.3^{+0.2}_{-0.2}$\\  
&$L\left(\frac{\mathrm{O_3}}{\mathrm{CO_2}}\right)$ &$-4.0$ &$-2.5$ &$-1.6$ &$-2.0$ &$-3.4^{+0.4}_{-0.4}$  &$-3.3^{+0.2}_{-0.2}$  &$-3.3^{+0.3}_{-0.3}$  &$-3.2^{+0.2}_{-0.2}$ &$-3.9$ &$-2.6$ &$-1.6$ &$-2.0$ &$-3.3^{+0.4}_{-0.4}$  &$-3.2^{+0.2}_{-0.2}$  &$-3.3^{+0.3}_{-0.3}$  &$-3.2^{+0.2}_{-0.2}$\\  
&$L\left(\frac{\mathrm{CH_4}}{\mathrm{CO_2}}\right)$  &$-2.6$ &$-2.7$ &$-3.0$ &$-3.2$  &$-2.5^{+0.8}_{-1.5}$  &$-2.1^{+0.3}_{-0.4}$  &$-2.2^{+0.4}_{-0.6}$  &$-2.1^{+0.2}_{-0.2}$ &$-2.6$ &$-2.7$ &$-3.0$ &$-3.2$ &$-2.3^{+0.7}_{-1.2}$  &$-2.1^{+0.3}_{-0.4}$  &$-2.1^{+0.4}_{-0.5}$  &$-2.0^{+0.2}_{-0.2}$ \\  
&$L\left(\frac{\mathrm{O_3}}{\mathrm{H_2O}}\right)$  &$-5.3$ &$-0.3$ &$-$ &$-$  &$-2.8^{+0.5}_{-0.4}$  &$-2.7^{+0.3}_{-0.3}$ &$-2.7^{+0.4}_{-0.4}$  &$-2.6^{+0.2}_{-0.2}$ &$-5.2$ &$-0.4$ &$-$ &$-$  &$-3.2^{+0.4}_{-0.4}$  &$-3.1^{+0.3}_{-0.2}$  &$-3.2^{+0.3}_{-0.3}$  &$-3.0^{+0.2}_{-0.2}$ \\  
&$L\left(\frac{\mathrm{CH_4}}{\mathrm{H_2O}}\right)$  &$-3.9$ &$-0.4$ &$-$ &$-$  &$-2.0^{+0.9}_{-1.6}$  &$-1.7^{+0.4}_{-0.4}$  &$-1.7^{+0.5}_{-0.6}$  &$-1.6^{+0.3}_{-0.3}$ &$-4.0$ &$-0.5$ &$-$ &$-$  &$-2.3^{+0.7}_{-1.5}$  &$-2.0^{+0.4}_{-0.4}$  &$-2.1^{+0.5}_{-0.6}$  &$-1.9^{+0.2}_{-0.3}$ \\  
&$L\left(\frac{\mathrm{CH_4}}{\mathrm{O_3}}\right)$  &$1.4$ &$-0.2$ &$-1.4$ &$-1.2$  &$0.8^{+0.6}_{-1.3}$  &$1.0^{+0.3}_{-0.3}$ &$1.0^{+0.4}_{-0.4}$  &$1.0^{+0.2}_{-0.2}$ &$1.3$ &$-0.1$ &$-1.4$ &$-1.2$ &$0.9^{+0.6}_{-1.2}$  &$1.1^{+0.3}_{-0.3}$  &$1.1^{+0.3}_{-0.4}$  &$1.1^{+0.2}_{-0.2}$
\enddata
\addtolength{\leftskip} {-0cm}
\addtolength{\rightskip}{-4.8cm}
\tablecomments{Retrieved model parameter posteriors for the all combinations of spectral resolutions (\Rv{50, 100}) and noise levels (\SNv{10, 20}) of disk-integrated EqC Earth spectra. Here, $L(\cdot)$ abbreviates \lgrt{\cdot}. We provide the median of the retrieved posterior and indicate the $16\% - 84\%$ range via $+/-$ indices. We Further provide the ground truth values. If independent of the atmospheric pressure, we provide a single value. Otherwise, we provide the ground truths at pressures $1$~bar, $10^-1$~bar, $10^-2$~bar, and $10^-3$~bar if available.}
\label{tab:Rel_Abund_Values_EqC}
\end{deluxetable}
\end{rotatetable}

\clearpage
\FloatBarrier
\restartappendixnumbering
\section{Eliminating Biases in Retrieval Results By Reducing Posterior Distributions}\label{app:red_post_dist}

{As we motivated in Section~\ref{sec:results}, the underestimation of \Ps{} and \Rpl{} can be directly linked to biased retrieved estimates of \Teq{}, \Ab{}, and the atmospheric trace-gas abundances. Here, we provide further evidence for these correlations between parameter biases by reducing the retrieved posterior distribution to Earth's true \Ps{} and \Rpl{}.}

\subsection{Posterior Reduction Method}
\label{app:decorr_post}
\begin{figure}
  \centering
    \includegraphics[width=0.43\textwidth]{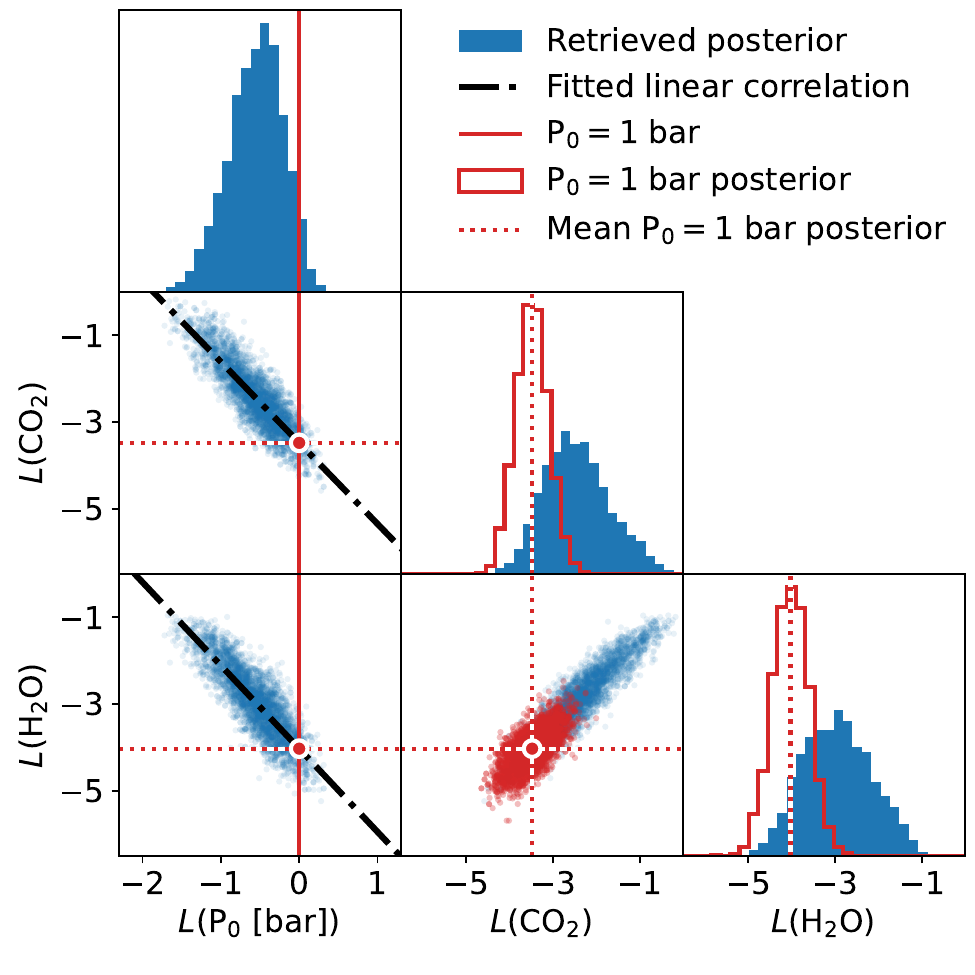}\qquad\qquad
    \includegraphics[width=0.43\textwidth]{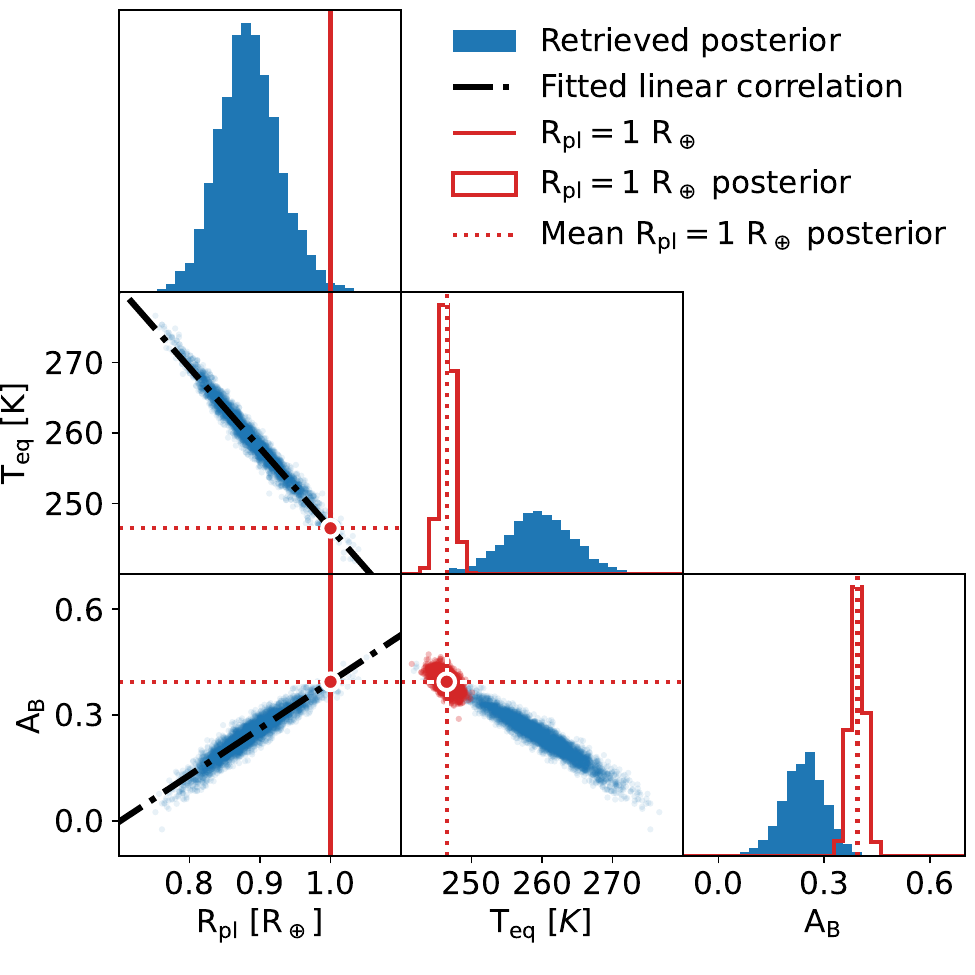}
  \caption{Reduction of selected retrieved posterior distributions with respect to \Ps{} (left plot) and \Rpl{} (right plot). The blue-filled histograms on the diagonal of the corner plots show the true retrieved parameter posteriors. Red-outlined histograms indicate the reduced posteriors. Below the diagonal, we show the 2D parameter posteriors in blue (true posteriors) and red (reduced posteriors) scatter plots. Black dashed-dotted lines represent the fitted linear correlation between parameters. Solid red lines mark the $\Ps{}=1$~bar and $\Rpl{}=1\Rearth{}$ position corresponding to the reduced posteriors, red-dotted lines indicate the median of the reduced posteriors.}
  \label{fig:post_red_proc}
\end{figure}

In Section~\ref{app:red_post_dist}, we reduced the retrieved posterior distributions to fixed values of $\Ps{}=1$~bar and $\Rpl{}=1\Rearth{}$ by assuming a linear correlation between \Ps{} or \Rpl{} and the remaining posteriors. Here, we outline the method used to reduce the posterior distributions. A schematic illustration showing both the true and the reduced posteriors can be found in Figure~\ref{fig:post_red_proc}. 

In the following, let us consider one point in the retrieved posterior distribution. We denote the value of \Ps{} or \Rpl{} (i.e., the parameter we want to reduce over) as $\theta_\mathrm{red, true}$ and values of the other model parameters as $\theta_\mathrm{param, true}$. If we assume a linear correlation between $\theta_\mathrm{red, true}$ and $\theta_\mathrm{param, true}$, we can make a prediction $\theta_\mathrm{param, pred}$ for $\theta_\mathrm{param, true}$ using $\theta_\mathrm{red, true}$ as follows:
\begin{equation}
    \theta_\mathrm{param, pred} = m\cdot\theta_\mathrm{red, true}+q.
\end{equation}
Here, $m$ is the slope and $q$ the offset with respect to the origin of the linear model. $\theta_\mathrm{param, pred}$ is the parameter value predicted by the linear model. We search for the best fit linear model by minimizing the square difference $\Delta$ between $\theta_\mathrm{param, pred}$ and $\theta_\mathrm{param, true}$:
\begin{equation}
    \Delta=\sum_\mathrm{Posterior}\left(\theta_\mathrm{param, pred}-\theta_\mathrm{param, true}\right)^2=\sum_\mathrm{Posterior}\left(m\cdot\theta_\mathrm{red, true}+q-\theta_\mathrm{param, true}\right)^2.
\end{equation}
The best fit linear models are indicated in Figure~\ref{fig:post_red_proc} as black dash-dotted lines. From the figure, we see that the correlations between the parameters considered here are well described by our linear model. In the next step, we fix the value of $\theta_\mathrm{red, true}$ to $\theta_\mathrm{red, fix}$ and calculate the corresponding reduced posterior values $\theta_\mathrm{param, red}$ of the other parameters as follows:
\begin{equation}
    \theta_\mathrm{param, red} = \theta_\mathrm{param, true} + m\cdot\left(\theta_\mathrm{red, fix} - \theta_\mathrm{red, true}\right).
\end{equation}
This yields the reduced posterior distribution of a parameter, which we plot in Figure~\ref{fig:post_red_proc}. This reduction method allows us to remove the effect of one parameter on the posterior distribution, and identify the origin of biases in the retrieval results.

\subsection{Reduction Relative to \Ps{}}

{To demonstrate the effects of underestimating \Ps{}, we reduce the abundance posteriors to Earth's true \Ps{} of 1~bar. The resulting reduced posteriors are shown in the left panel of Figure~\ref{fig:Reduced Posteriors} (numerical values in Tables~\ref{tab:Rel_Abund_Values_NP} to \ref{tab:Rel_Abund_Values_EqC}). The posterior reduction to the true \Ps{}, leads to significantly better estimates for \ce{CO2} and \ce{CH4}. For \ce{CO2}, the reduced posteriors are perfectly centered on the true value, while the \ce{CH4} abundances are significantly less overestimated. This demonstrates, that the shifts in the \ce{CO2} and \ce{CH4} posteriors in Figure~\ref{fig:PosteriorsM4} are directly linked to the inaccurately retrieved \Ps{}. For \ce{O3} and \ce{H2O}, the reduced posteriors are shifted to lower abundances and show a smaller variance between the individual retrievals. These findings suggest that the \Ps{} reduction also yields improved estimates for atmospheric \ce{O3} and \ce{H2O} abundances.}

\subsection{Reduction Relative to \Rpl{}}

\begin{figure}
  \centering
  \includegraphics[width=0.44\textwidth]{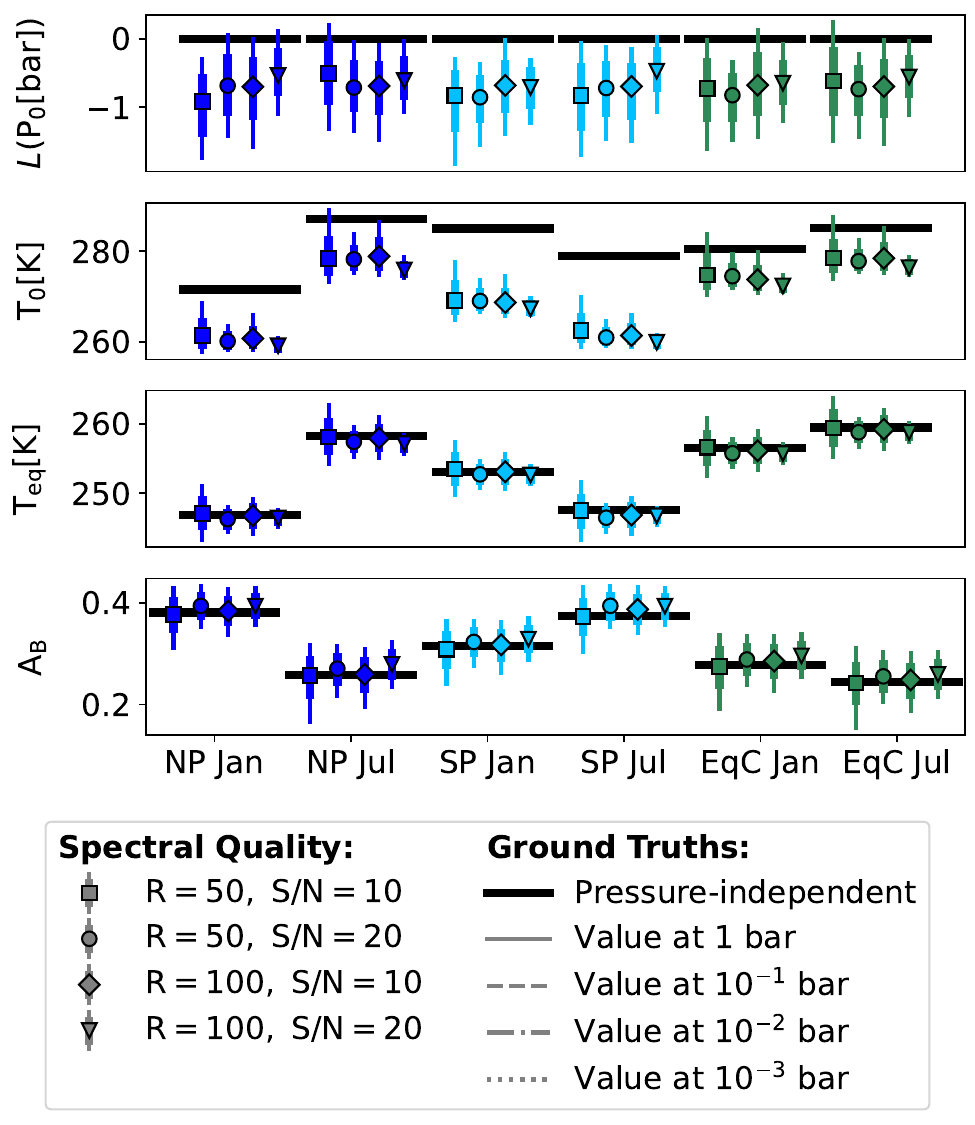}\qquad\qquad  
    \includegraphics[width=0.44\textwidth]{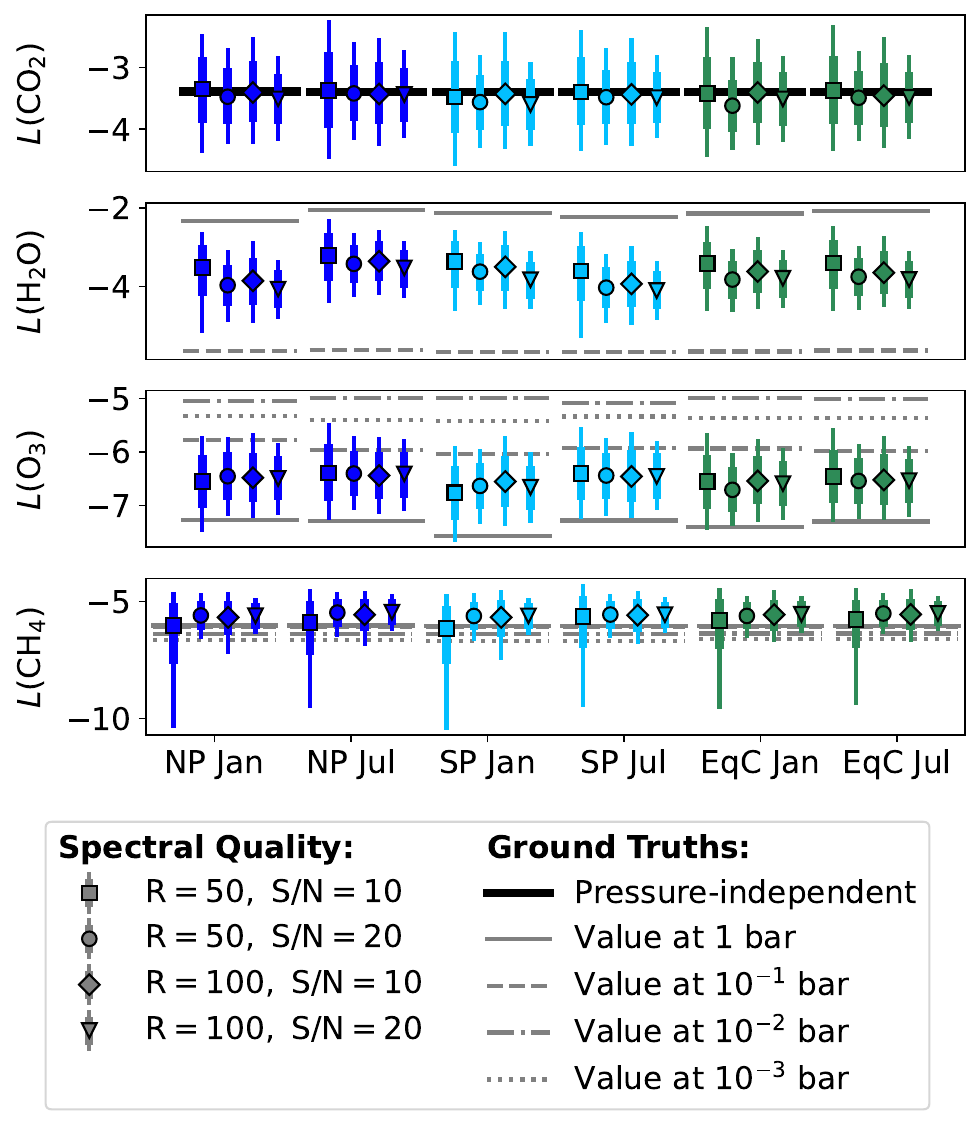}  
  \caption{Reduced posteriors for all combinations of \R{} (50, 100) and \SN{} (10, 20) for the six disk-integrated Earth spectra. Here, $L(\cdot)$ abbreviates $\lgrt{\cdot}$. The marker shape represents the spectral \R{} and \SN{}, the color shows the viewing geometry. Colored lines indicate posterior percentiles (thick: $16\%-84\%$; thin: $2\%-98\%$). Thick black lines indicate pressure-independent ground truths. Thin gray lines show the ground truth abundance at different atmospheric pressures (solid: 1~bar; dashed: $10^{-1}$~bar; dashed-dotted: $10^{-2}$~bar; dotted: $10^{-3}$~bar). Columns (left to right) show the results for: NP Jan, NP Jul, SP Jan, Sp Jul, EqC Jan, and EqC Jul. \textit{Left panel:} Selected posteriors reduced to the true \Rpl{} value of 1~\Rearth{}. \textit{Right panel:} Abundance posteriors reduced to the true \Ps{} value of 1~bar.}
  \label{fig:Reduced Posteriors}
\end{figure}

{In order to investigate how our underestimation of \Rpl{} affects the other posteriors, we reduce the \Rpl{} posterior to 1~\Rearth{}. We plot the reduced posteriors of \Ps{}, \Ts{}, \Teq{}, and \Ab{} in the right panel of Figure~\ref{fig:Reduced Posteriors} (numerical values in Tables~\ref{tab:Rel_Abund_Values_NP} to \ref{tab:Rel_Abund_Values_EqC}). First, we observe no significant differences between the reduced and the true \Ps{} posteriors from Figure~\ref{fig:PosteriorsM4}. Thus, no direct correlation between the \Rpl{} and the \Ps{} posterior exists. Second, for \Ts{}, which is accurately estimated in all retrievals, the reduced posteriors underestimate the truth by $\geq5$~K. This finding indicates that Earth's disk-integrated flux is smaller than what is expected for a cloud-free 1~\Rearth{} planet with surface temperature \Ts{}. This suggests that patchy clouds, which partially block the emission from the high pressure atmospheric layers and thereby reduce the total planet flux, are the likely cause of the \Rpl{} biases (see also Appendix~\ref{app:quant_radius_biases} for further evidence). Finally, the reduced \Teq{} and \Ab{} posteriors are unbiased and provide accurate truth estimates (uncertainties: $\leq\pm2$~K for \Teq{}; $\leq\pm0.1$ for \Ab{}), which demonstrates the correlations with \Rpl{}.}

\clearpage
\FloatBarrier

\section{Quantifying the Effect of Neglecting Clouds on Retrieved Planet Radius Estimates}\label{app:quant_radius_biases}

{Here, we use a simplified model for Earth's thermal emission to motivate that the magnitude of the biases on the retrieved \Rpl{} estimates (see Section~\ref{sec:results}) can be explained by an Earth-like patchy cloud coverage. In our cloud-free retrievals, we model Earth as a spherical Black Body (BB) with radius $R_\mathrm{pl,\,ret}$ and surface temperature $T_\mathrm{0,\,ret}$. Neglecting absorption and emission by Earth's atmosphere, the total power emitted ($P_\mathrm{cloud-free}$) is equivalent to the power emitted by a spherical BB (where $\sigma$ is the Stefan-Boltzmann constant):
\begin{equation}
    \label{eq:BB_emission_no_clouds}
    P_\mathrm{cloud-free}=4\pi R_\mathrm{pl,\,ret}^2\sigma T_\mathrm{0,\,ret}^4.
\end{equation}
However, we know that clouds are present in Earth's atmosphere. To obtain a first order approximation of the total power emitted by a partially cloudy Earth ($P_\mathrm{cloudy}$), we assume opaque clouds (i.e., the clouds block all thermal radiation from lower atmosphere layers) that emit BB radiation of temperature \Tct{} at the cloud-top. Using the cloud-coverage fraction ($f_\mathrm{cov}$; i.e. the percentage of Earth's surface covered by clouds), we can approximate $P_\mathrm{cloudy}$ as a weighted sum of the BB emission from cloudy and cloud-free regions:
\begin{equation}
    \label{eq:BB_emission_clouds}
    P_\mathrm{cloudy}=4\pi R_\mathrm{pl,\,true}^2\sigma\left(f_\mathrm{cov}\Tct^4 + \left(1-f_\mathrm{cov}\right) T_\mathrm{0,\,true}^4\right).
\end{equation}
Here, $R_\mathrm{pl,\,true}$ and $T_\mathrm{0,\,true}$ are Earth's true radius and average surface temperature respectively. The power emitted by Earth via its thermal emission is measurable and independent of the selected model. We thus set Equations~\ref{eq:BB_emission_no_clouds} and \ref{eq:BB_emission_clouds} equal to each other:
\begin{equation}
    \label{eq:BB_equation1}
    P_\mathrm{cloud-free}=4\pi R_\mathrm{pl,\,ret}^2\sigma T_\mathrm{0,\,ret}^4 \stackrel{!}{=}4\pi R_\mathrm{pl,\,true}^2\sigma\left(f_\mathrm{cov}\Tct^4 + \left(1-f_\mathrm{cov}\right)T_\mathrm{0,\,true}^4\right)=P_\mathrm{cloudy}.
\end{equation}
The results from Section~\ref{sec:results} suggest that \Ts{} is accurately estimated by our retrievals, despite not accounting for Earth's partial cloud coverage. Motivated by this finding, we substitute $T_\mathrm{0,\,ret}$ and $T_\mathrm{0,\,true}$ by $\Ts{}$ in Equation~\ref{eq:BB_equation1}. Subsequent rearranging yields:
\begin{equation}
    \label{eq:BB_equation2}
    \left(\frac{\Tct{}}{\Ts{}}\right)^4=\frac{1}{f_\mathrm{cov}}\left(\frac{R_\mathrm{pl,\,ret}}{R_\mathrm{pl,\,true}}\right)^2-\frac{1}{f_\mathrm{cov}}+1.
\end{equation}
Next, we assume that the temperature difference between \Tct{} and \Ts{} is $\Delta T$. We implement this assumption by replacing $\Tct{}$ with $\Ts{}-\Delta T$ in Equation~\ref{eq:BB_equation2}. Further, for Earth, $\Delta T$ is significantly smaller than \Ts{}. Thus, we can approximate as follows:
\begin{equation}
    \label{eq:BB_high_order_terms}
    \left(\frac{\Tct{}}{\Ts{}}\right)^4=\frac{\left(\Ts{}-\Delta T\right)^4}{\Ts{}^4}=1-4\frac{\Delta T}{\Ts{}}+\mathcal{O}\left(\frac{\Delta T^2}{\Ts^2}\right)\approx1-4\frac{\Delta T}{\Ts{}}.
\end{equation}
By inserting Equation~\ref{eq:BB_high_order_terms} into Equation~\ref{eq:BB_equation2} and simplifying the resulting expression, we obtain:
\begin{equation}
    \label{eq:BB_equation3}
    \Delta T = \frac{\Ts{}}{4f_\mathrm{cov}}\left(1-\left(\frac{R_\mathrm{pl,\,ret}}{R_\mathrm{pl,\,true}}\right)^2\right).
\end{equation}}

{By inserting numeric values into Equation~\ref{eq:BB_equation3}, we assess if the retrieved \Rpl{} biases are consistent with an Earth-like patchy cloud coverage. Motivated by our retrieval results (see, Section~\ref{sec:results} and Appendix~\ref{app:additional_retrieval_data}), we assume $R_\mathrm{pl,\,ret}/R_\mathrm{pl,\,true}=0.90\pm0.03$. For \Ts{}, we select the lowest and the highest retrieved values to cover the full \Ts{} range ($272\pm6$~K for NP~Jan; $287\pm6$~K for NP~Jul). Last, we assume an Earth-like $f_\mathrm{cov}$ of $0.67$ (see, Appendix~\ref{app:patchy_clouds}). Inserting these values into equation \ref{eq:BB_equation3} yields:
\begin{equation}
    \Delta T = \begin{cases}
\,\,19\pm5~\text{K} &\text{for NP Jan view.}\\
\,\,20\pm5~\text{K} &\text{for NP Jul view.}
\end{cases}
\end{equation}
This implies that \Tct{} must lie roughly $20$~K below \Ts{}, if the retrieved bias on \Rpl{} is evoked by Earth's patchy cloud coverage. Assuming a lower limit of $4$~K/km for Earth's moist adiabatic lapse rate, the $\Delta T$ requirement translates to an upper limit of $5.0\pm1.2$~km for the cloud-top altitude. Similarly, from Earth's dry adiabatic lapse rate ($\approx10$~K/km), we obtain a lower limit of $2.0\pm0.5$~km for the cloud-top position. Both altitudes lie well below the tropopause ($\approx9$~km at the poles to $\approx17$~km at the equator) and span the atmospheric layers where Earth's abundant low- to mid-level clouds form \citep{Houze2014}. This first-order approximation demonstrates that the magnitude of the retrieved \Rpl{} biases can be explained by the missing cloud treatment in our retrieval study. }

\clearpage
\FloatBarrier

\end{document}